\documentclass[fleqn,usenatbib]{mnras} 
\pdfoutput=1 
\usepackage{mathptmx}
\usepackage{txfonts}
\usepackage{float}
\usepackage{capt-of}
\usepackage{multicol}

\usepackage[T1]{fontenc}

\DeclareRobustCommand{\VAN}[3]{#2}
\let\VANthebibliography\thebibliography
\def\thebibliography{\DeclareRobustCommand{\VAN}[3]{##3}\VANthebibliography}

\usepackage{graphicx}
\usepackage{xcolor}
\usepackage{amssymb}
\usepackage{placeins}
\usepackage{tabularx}
\usepackage[inline]{enumitem}
\usepackage{tabulary}
\usepackage{booktabs}
\usepackage{multirow}
\usepackage{siunitx}
\usepackage{placeins}
\usepackage{listings}
\usepackage{subcaption}
\usepackage{orcidlink}
\usepackage[normalem]{ulem}
\DeclareUnicodeCharacter{2248}{$\approx$}
\DeclareUnicodeCharacter{0306}{$\check{c}$}
\usepackage{txfonts}

\usepackage{amsmath}
\usepackage{xspace}
\newcommand\HII{\ion{H}{II}\xspace} 
\newcommand\NII{[\ion{N}{II}]\xspace} 
\newcommand\OIII{[\ion{O}{III}]\xspace} 
\title[\texttt{TODDLERS}: UV--mm emission for star-forming regions]{\texttt{TODDLERS}: A new UV--mm emission library for star-forming regions. I. Integration with \texttt{SKIRT} and public release}
\author[A. U. Kapoor et al.]{Anand Utsav Kapoor\orcidlink{0000-0002-5187-1725},$^{1}$\thanks{E-mail: \href{anandutsav.kapoor@ugent.be}{anandutsav.kapoor@ugent.be}}
Maarten Baes\orcidlink{0000-0002-3930-2757},$^{1}$
Arjen van der Wel\orcidlink{0000-0002-5027-0135},$^{1}$
Andrea Gebek\orcidlink{0000-0002-0206-8231},$^{1}$
Peter Camps\orcidlink{0000-0002-4479-4119},$^{1}$
\newauthor
Angelos Nersesian,$^{1}$
Sharon E. Meidt,$^{1}$ 
Aaron Smith\orcidlink{0000-0002-2838-9033},$^{2}$
Sebastien Vicens,$^{3}$
Francesco D'Eugenio$^{4}$ $^{5}$,
\newauthor
Marco Martorano$^{1}$, 
Daniela Barrientos\orcidlink{0000-0002-1710-1460},$^{1}$
Nina Sanches Sartorio$^{1}$
\\
$^{1}$Sterrenkundig Observatorium, Universiteit Gent, Krijgslaan 281 S9, B-9000 Gent, Belgium\\
$^{2}$Center for Astrophysics | Harvard \& Smithsonian, 60 Garden St, Cambridge, MA 02138, USA \\
$^{3}$Facult\'e des sciences et de g\'enie - Universit\'e Laval, Pavillon Alexandre-Vachon, Qu\'ebec, G1V 0A6, Canada \\
$^{4}$Kavli Institute for Cosmology, University of Cambridge, Madingley Road, Cambridge CB3 0HA, UK \\
$^{5}$Cavendish Laboratory - Astrophysics Group, University of Cambridge, 19 JJ Thomson Avenue, Cambridge CB3 0HE, UK 
}

\date{Accepted 2023 September 27. Received 2023 September 27; in original form 2023 May 26}

\pubyear{2023}

\begin{document}
\label{firstpage}
\pagerange{\pageref{firstpage}--\pageref{lastpage}}
\maketitle

\begin{abstract}
We present and publicly release a new star-forming regions emission library \texttt{TODDLERS} ({T}ime evolution of {O}bservables including {D}ust {D}iagnostics and {L}ine {E}mission from {R}egions containing young {S}tars) for the publicly available radiative transfer code \texttt{SKIRT}. The library generation involves the spherical evolution of a homogeneous gas cloud around a young stellar cluster that accounts for stellar feedback processes including stellar winds, supernovae, and radiation pressure, as well as the gravitational forces on the gas. The semi-analytical evolution model is coupled with the photoionization code \texttt{Cloudy} to calculate time-dependent UV--mm spectral energy distributions (SEDs) from star-forming regions of varying metallicity, star-formation efficiency, birth-cloud density, and mass. 
The calculated SEDs include the stellar, nebular, and dust continuum emission along with a wide range of emission lines originating from \HII, photo-dissociation, and molecular gas regimes tabulated at high resolution. The SEDs incorporated in \texttt{SKIRT} are generated by calculating a stellar-mass normalized luminosity, which assumes that each emission source is composed of a power-law population of star-forming clouds.
When compared to the previous treatment of star-forming regions in \texttt{SKIRT}, \texttt{TODDLERS} shows a better agreement with low-redshift observational data in the IR wavelength range while offering a more comprehensive line-emission support. This paves the way for a variety of applications using simulated galaxies at low and high redshift.
\end{abstract}



\section{Introduction}
Galaxy formation and evolution is a complex problem involving multi-scale and multi-physics phenomena. Such complexity necessitates the use of numerical experiments to track the interplay of many involved processes \citep{2015ARA&A..53...51S, 2017ARA&A..55...59N, 2020NatRP...2...42V}. One of the key products of such numerical experiments is the distribution of baryonic mass and its associated properties, e.g. the distribution of dark matter, gas, metals, dust, stars, and black holes throughout the Universe over cosmic history.A fair comparison between the observable and numerical universes necessarily requires a conversion of mass to light, or vice-versa.
The generation of synthetic/mock observations utilizes the former methodology. In this forward modeling approach, not only the effects of the complex interplay of radiation, dust, and gas in realistic geometries are realized, but instrumental effects are also considered. This makes it possible to meaningfully compare observations with simulated data \citep{2015MNRAS.454.2381G, 2015MNRAS.447.2753T, 2019MNRAS.487.1529D, 
popping2021dustcontinuum, 2021MNRAS.506.5703K, 2022MNRAS.512.2728C, 2022MNRAS.516.3728T, 2023MNRAS.tmp..841G}. 
In this framework, multi-wavelength comparison of simulations with observations can provide increasingly comprehensive understanding of the properties and behavior of astronomical objects, as different wavelengths of radiation can reveal complementary aspects of an object's characteristics. 

The spectral energy distribution (SED) is the rate of energy emitted by luminous sources at different wavelengths of the electromagnetic spectrum. The SED encodes information about the physical processes occurring within a galaxy, such as its star formation rate (SFR), star formation history (SFH), the presence of an active galactic nucleus, or the amount and characteristics of dust and gas \cite[see, for example][]{2013ARA&A..51..393C, 2017ApJ...837..170L,  2018MNRAS.476.1705S, 2019ApJ...876....3L, 2019A&A...622A.103B}. The SED of a galaxy is significantly influenced by the presence of massive stars in the galaxy. The most massive of which live up to a few Myr and their radiation is highly/efficiently reprocessed by the dust and gas present in their natal environments (star-forming regions\footnote{We use the term star-forming region to refer to the gas in various phases i.e., ionized, neutral, and molecular gas surrounding a stellar cluster.}), making stellar clusters containing massive stars significant contributors to the UV and IR continuum \citep{1994ARA&A..32..227M, 2002ARA&A..40...27C, 2009PASP..121..213C, 2019PASJ...71....6H}. These wavelengths offer complementary approaches to accurately measure a galaxy's SFR, one of the most fundamental properties of a galaxy. 
Apart from this, massive young stars also serve as engines for the recombination and collisionally excited lines, which serve as diagnostic tools for determining SFR, densities, temperatures, chemical compositions, and ionization states \citep{2016MNRAS.461.3111B, 2019ApJ...887...80K, 2019ARA&A..57..511K, 2022ApJ...929..118G}. 
Synthetic observations involving line and continuum emission, in the UV, optical, and IR have thus become increasingly important in the current era of integral field unit (IFU) instruments such as JWST-NIRSpec, KMOS, MUSE at VLT, SINFONI. Here mock observations play a crucial role in interpreting the observational data and understanding various biases that affect the overall systematics and statistical scatter \citep{2022arXiv221202522H, 2022arXiv221100931J, 2023MNRAS.524..907B}. At the same time, mock observables shed light on the fidelity and limitations of the numerical simulations when it comes to the mass buildup and kinematics of galaxies.
Modeling the dust and gas emission from star-forming regions requires, among other things, the intrinsic spectra of the radiation sources, the dust/gas geometry, and a description of the physical properties of the interstellar medium (ISM). For large-box simulations with a sizeable galaxy sample, this sub-$\mathrm{pc}$ information is generally not available from the simulation snapshots due to the lack of resolution and the fact that the small-scale physics and the feedback processes are usually dealt with in a very approximate manner. Thus, sub-grid models describing the emission from young clusters are usually necessary in order to generate synthetic data products from simulated galaxies \citep[see, for example][]{2023arXiv230409261Y}. At the same time, galaxy formation simulations continue to become increasingly realistic due to improved physics and resolution \cite[see, for example, ][]{10.1093/mnras/staa3249, 2020MNRAS.492.2973T, 2022arXiv220515325F, 2022arXiv221104626K, 2023arXiv230107116S, 2023MNRAS.519.3154H} and there is an ongoing effort to produce synthetic observables in a self-consistent manner without employing sub-grid models \citep{2022MNRAS.517....1S, 2022MNRAS.513.2904T}. However, these efforts remain limited to isolated galaxy simulations in the current state of affairs.

The \texttt{SKIRT} radiative transfer code \citep{2011ApJS..196...22B, 2015A&C.....9...20C, 2020A&C....3100381C} is a Monte Carlo radiative transfer (RT) code that has been used extensively to generate multi-wavelength synthetic data for simulated galaxies. Apart from dust RT, the current version of the code is designed to perform Lyman-$\alpha$ RT \citep{2021ApJ...916...39C}, X-ray RT \citep{2023arXiv230410563V}, and non-LTE line RT \citep{Kosei_2023} without any constraints on geometrical complexity while accounting for polarization and kinematics.
 When generating synthetic UV--mm observations of galaxies with \texttt{SKIRT}, emission sources (particles from the parent simulation snapshot) are typically separated by age, with young stars (age below 10 Myr) assumed to be enshrouded by dust and gas. Such particles are assigned an SED from the library discussed in \cite{2008ApJS..176..438G} generated using the \texttt{MAPPINGS-III} code. These templates model emission from both the immediate \HII region and the surrounding photodissociation region (PDR), including the dust contained within each of these regions. We refer to the version of this library implemented in \texttt{SKIRT} \citep{2010MNRAS.403...17J} as \texttt{HiiM3}\footnote{Documented online at \url{https://skirt.ugent.be/skirt9/class_mappings_s_e_d_family.html}}
 throughout this work. 
 \texttt{HiiM3} has been used to generate synthetic broadband fluxes for simulated galaxies from the EAGLE  simulation suite \citep{2016MNRAS.462.1057C, 2019MNRAS.484.4069B, 2020MNRAS.494.2823T}. More recently it has been applied in a similar fashion to TNG-50 galaxies \citep[][Gebek et al. in preparation]{2022MNRAS.516.3728T}. It has also been used to generate synthetic high-resolution multi-wavelength images for zoom-in simulations, Auriga \citep{2021MNRAS.506.5703K} and Artemis \citep{2022MNRAS.512.2728C}.  
 However, it has been recognized by the aforementioned authors that the use of \texttt{HiiM3} could be partly responsible for the high FUV and MIR, and tension in MIR--FIR colors of the simulated galaxies when compared with their observational counterparts. This motivates the development of updated options for the treatment of emission from star-forming regions in \texttt{SKIRT}.
 
The aim of this work is to construct a physically motivated, time-resolved model for the UV--mm emission from star-forming regions. In order to post-process simulated galaxies, we need a model that encompasses a large parameter space while leveraging the simulation's information. The model should incorporate relevant physics and remain computationally feasible for parameter sweeping \footnote{parameters may include age, metallicity, and sub-grid physics' parameters.}.
To this end, two relevant state-of-the-art models currently available include \texttt{HiiM3} and the recent model presented in \citet[][WARPFIELD-EMP]{2020MNRAS.496..339P}. The former model is not time-resolved and is designed for modeling the integrated spectra of starburst galaxies by luminosity-weighted averaging young clusters of different ages ($<10$\,Myr). The latter model couples the evolution of gas clouds under stellar feedback \citep[][WARPFIELD]{2017MNRAS.470.4453R} with a photoionization code to generate time-dependent observables.
Given its suitability for our work, we adopt an approach similar to the second model mentioned above, i.e., a semi-analytical calculation for the evolution of a spherical, homogeneous gas cloud exposed to stellar feedback coupled with the photoionization code \texttt{Cloudy} to generate observables. Our approach expands on WARFIELD/WARPFIELD-EMP by covering a broader range of metallicities and incorporating an additional feedback channel into our semi-analytical calculations, namely radiation pressure from the resonant scattering of Lyman-$\alpha$ (Ly$\alpha$) photons. The resulting emission spectra span the UV--mm electromagnetic spectrum, including features like the sub--mm CO lines.
\begin{figure*}
\centering
\includegraphics[width=.8\textwidth]{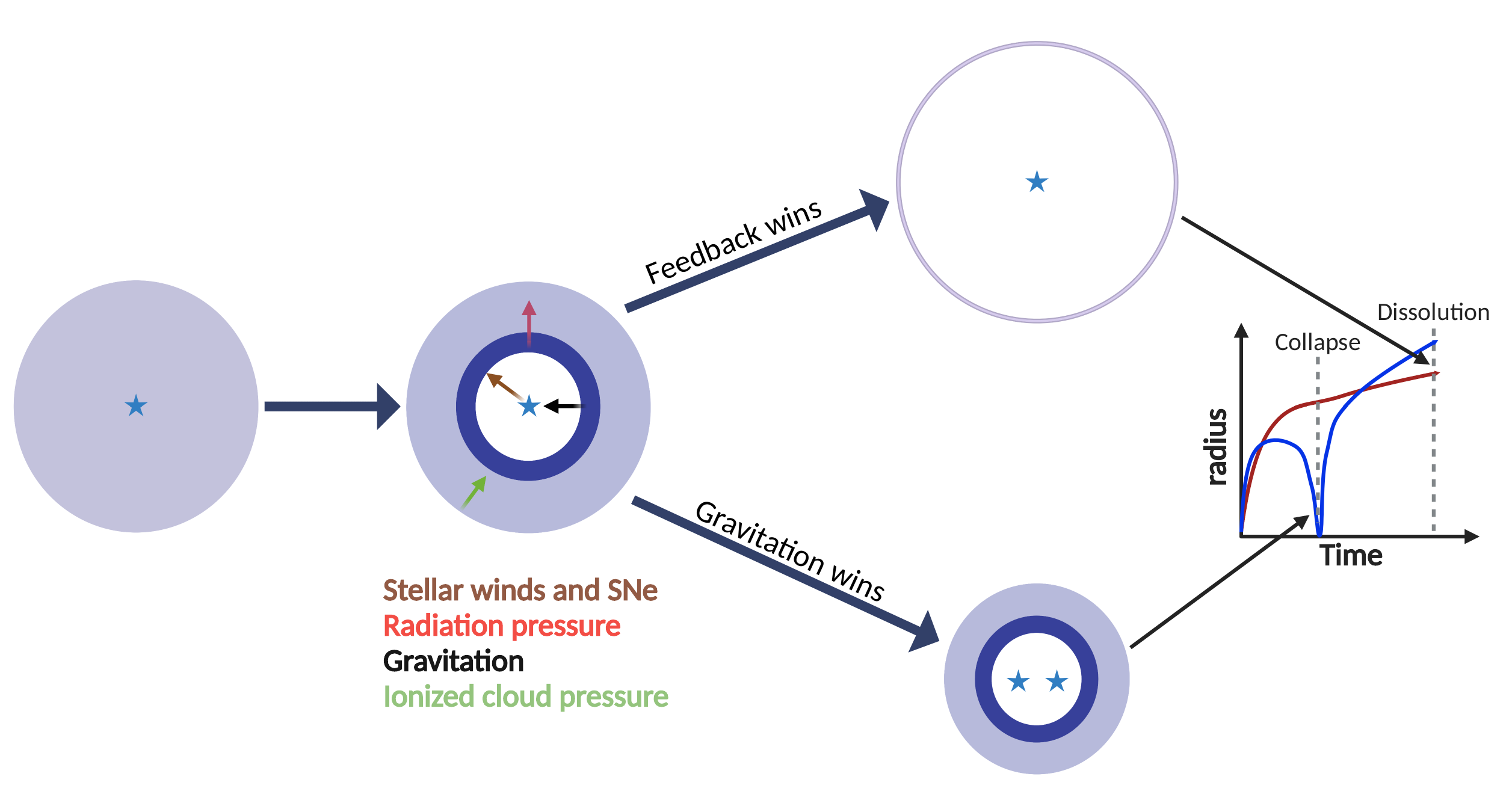}
\caption{Schematic of the key ingredients of the shell-evolution model setup consisting of the rapid formation of a shell (dark blue) under stellar feedback. The stellar feedback and radiation pressure push the shell, while the gravitational force due to the central cluster and the shell's self-gravity oppose the expansion. If the shell is optically thin to ionizing photons, then any remaining unswept gas cloud also opposes the expansion. These forces lead to two possible outcomes for this system: either the feedback is strong enough to push the gas, so that the cloud is assumed to be dissolved, or the gravitational force dominates and leads to a secondary collapse episode that forms another stellar cluster. In the latter case, the model equations are solved again assuming the presence of the younger cluster formed with the same $\epsilon_{\mathrm{SF}}$.
}
\label{fig:model_setup_schematic}
\end{figure*}

Our custom-made model is designed to seamlessly integrate into \texttt{SKIRT}. We consider the fact that young stellar particles in simulations do not represent a single star-forming region but rather a population with a range of ages. This custom model also facilitates the incorporation of additional modifications, such as substituting the stellar library, initial mass function (IMF), dust models, and so on.

This is the first of a two-part series of papers with the main goal of presenting the new library, \texttt{TODDLERS}\footnote{A toddler is a child aged 1--3 years old. The word is derived from ``to toddle'', which means to walk unsteadily, like a child of this age. The toddler years are a time of great cognitive, emotional, and social development.}.
The paper organizes its contents as follows:
In Sec.~\ref{sect:evolutionary_model}, we describe the semi-analytic evolution model and its output.
Sec.~\ref{cloudy_methods_library_gen.sec} discusses the methodology used for generating the observables using \texttt{Cloudy}.
Sec.~\ref{sect:observables from individual models} showcases key diagnostics resulting from the coupling of the evolution model and \texttt{Cloudy} post-processing.
In Sec.~\ref{sect:integration_with_SKIRT}, we integrate the \texttt{TODDLERS}' observables within \texttt{SKIRT}.
Sec.~\ref{sect:comparison_with_HiiM3} focuses on comparing \texttt{TODDLERS} and \texttt{HiiM3}, particularly the IR colors resulting from the application of these two libraries without any other changes.
Finally, in Sec.~\ref{sect:conclusions}, we summarize and conclude.

\section{Evolution of gas cloud under feedback from young stars}
\label{sect:evolutionary_model}
\begin{figure}
\centering
\includegraphics[width=.9\columnwidth]{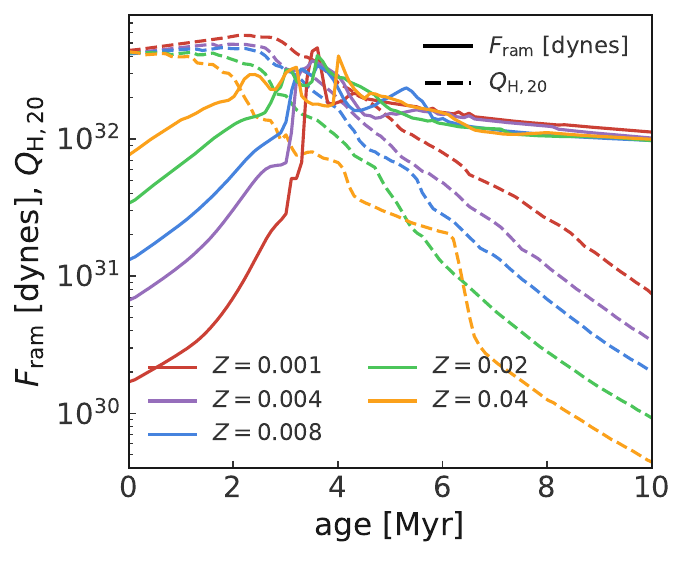}
\caption{Evolution of two key stellar feedback properties used in this work shown as a function of metallicity. This assumes a coeval population of stellar mass $10^{6}~M_{\odot}$ and a well sampled IMF (see Sec.~\ref{subsect:stellar_data}). The solid curves represent the force due to mass loss, $F_{\mathrm{ram}}$ as a function of time and metallicity for the employed stellar library.
The dashed curves are the production rate of Hydrogen ionizing photons in units of $10^{20}~\mathrm{s^{-1}}$.}\label{fig:stellar_feedback_Fram_and_Qion}
\end{figure}
We model the evolution of a homogeneous gas cloud around a young stellar cluster under stellar feedback in spherical symmetry inspired by the work presented by \citet{2019MNRAS.483.2547R}. The semi-analytical model calculates the evolution of a finite gas cloud under stellar feedback, accounting for stellar winds, supernovae (SNe), and radiation pressure due to ionizing radiation and dust. The bubble expansion is initially mediated by the shocked stellar wind and is pressure-driven, but a switch to momentum-driven evolution takes place based on prescriptions for instabilities that could lead to a rapid loss of bubble pressure.
The gravitational force on the gas, both due to the self-gravity of the gas and due to the central stellar cluster is taken into consideration. This allows for multiple star-formation events to take place in the event that the gravitational force overpowers the feedback of the cluster. In contrast, if the stellar feedback is strong enough, it could eventually lead to the dissolution of the cloud. 
This is schematically shown in Fig.~\ref{fig:model_setup_schematic}. We refer to the semi-analytical treatment as the shell-evolution model throughout this work. 

The equations of motion for the shell during various evolutionary phases are discussed briefly next.
The stellar feedback data (quantities such as mass-loss rates, terminal velocities of the stellar ejecta, ionizing/non-ionizing luminosities, etc.) used in this work come from \texttt{STARBURST99} models whose details are given in Sec.~\ref{subsect:stellar_data}. The evolution of two such quantities, the force due to stellar ejecta ($F_{\mathrm{ram}}$) and the rate of production of Hydrogen ionizing photons by the cluster ($Q_{\mathrm{H}}$) are shown as a function of cluster metallicity in Fig.~\ref{fig:stellar_feedback_Fram_and_Qion}. Initially ($t<3$Myr), $F_{\mathrm{ram}}$ comes almost entirely from the stellar winds of OB stars, which increases with metallicity. W-R stars can contribute significantly to $F_{\mathrm{ram}}$ starting around $3-4$~Myr, lasting for a period of $0.25-2$~Myr depending on the metallicity. Once the W-R phase is over, $F_{\mathrm{ram}}$ comes mostly from the SNe. The production rate of Hydrogen ionizing photons drops as the massive stars die. Increasing the metallicity leads to increasing line blanketing, lowering the production rate of ionizing photons.

\subsection{Dynamics of the shell}
\label{sect:shellDynamics}
It is assumed that feedback from a central star cluster interacts with a finite, massive cloud of number density $n_{\text{cl}}$ surrounding it. The central source is an instantaneously born, young stellar cluster.
The amount of stellar mass ($M_{\star}$) and the cloud gas mass susceptible to the stellar feedback are dictated by the star formation efficiency parameter ($\epsilon_{\mathrm{\textsc{SF}}}$).
\begin{equation}
    M_{\star} = \epsilon_{\mathrm{\textsc{SF}}}M_{\mathrm{cl}} \quad \text{and} \quad
    M_{\mathrm{cl,r}} = (1 - \epsilon_{\mathrm{\textsc{SF}}})M_{\mathrm{cl}} \, ,
\label{eqn:definition_SFE}   
\end{equation}
where $M_{\mathrm{cl}}$ is the initial mass of the gas cloud, while $M_{\mathrm{cl,r}}$ is the remaining cloud mass around the cluster.
The cloud and the central stellar cluster/s are assumed to have the same metallicity. The gas has a mean mass per nucleus $\mu_{\mathrm{n}} = (14/11)~ m_{\mathrm{H}}$  and the mean mass per particle $\mu_{\mathrm{p}} = (14/23)~m_{\mathrm{H}}$, where $m_{\mathrm{H}}$ is the proton mass. The cloud's density is given as $\rho_{\text{cl}} = \mu_{\mathrm{n}}n_{\text{cl}}$, {where $n_{\text{cl}}$ is the Hydrogen number density of the cloud}. {We note that while we have adopted $Z=0.02$ values for $\mu_{\mathrm{n}}$ and $\mu_{\mathrm{p}}$ throughout, this choice is expected to have minimal impact on the shell dynamics.}

In order to solve for the shell's dynamics, we solve the equations of the conservation of mass, momentum, and energy.  For the conservation of mass, it is assumed that as the shell expands, the unswept cloud quickly settles and becomes a part of the shell.
This allows us to write the mass conservation as:
\begin{equation}
   \frac{dM_{\text{sh}}}{dt}=\left\{\begin{array}{ll} \rho_{\text{cl}} A_{\text{sh}} v_{\text{sh}}  & \text { if } M_{\text{sh}} < M_{\text{cl,r}} ~\text{and}~ v_{\text{sh}} > 0 \\ 0 & \text { if } M_{\text{sh}} = M_{\text{cl,r}} ~\text{or}~ v_{\text{sh}} \leq 0 \end{array}\right . \, .
\label{eqn:mass_conservation}
\end{equation}
In Eqn.~\eqref{eqn:mass_conservation}, $A_{\text{sh}}$, $v_{\text{sh}}$ refer to the shell's surface area, and velocity, respectively. We consider only homogeneous clouds in this work, hence, $\rho_{\text{cl}}$ is a constant for a given cloud in this work. Note that during infall (see Sec.~\ref{subsect:shell_collapse}), no mass change occurs. The conservation of momentum is written by considering the forces due to the mechanical luminosity and the radiation pressure due to the stellar cluster, gravitational forces on the shell, and the external pressure of the cloud in which the shell is expanding. 
The general form of the momentum equation is as follows:
\begin{equation}
\frac{\mathrm{d}}{\mathrm{d} t}\left(M_{\mathrm{sh}} v_{\text{sh}}\right)= F_{\mathrm{w,sn}}- F_{\mathrm{grav}} + F^{\mathrm{UV,IR}}_{\mathrm{rad}} + F^{\mathrm{Ly\alpha}}_{\mathrm{rad}} - F_{{\mathrm{ext}}} \, .
\label{eqn:momentum_conservation_general}
\end{equation}
In Eqn.~\eqref{eqn:momentum_conservation_general}, $F_{\mathrm{w,sn}}$ is the term attributed to the stellar winds and SNe, and its exact form depends on the evolutionary phase (Sec.~\ref{sect:pressure_driven_phase}, \ref{sect:momentum_phase}) and is discussed along with the specific phase.  $F_{\mathrm{grav}}$ is the gravitational force on a thin shell of radius $r_{\text{sh}}$ due to the star cluster and its self-gravity. $F_{\mathrm{grav}}$ can be written as:
\begin{equation} \label{eqn:F_grav_shell}
F_{\mathrm{grav}}=\frac{G M_{\mathrm{sh}}}{r_{\mathrm{sh}}^{2}}\left(M_{*}+\frac{M_{\mathrm{sh}}}{2}\right) \, .
\end{equation}
The last three terms in Eqn.~\eqref{eqn:momentum_conservation_general} are dependent on the shell structure. The third and the fourth terms are the forces due to radiation pressure on the shell. We have two components of the radiation pressure acting on the shell. The first radiation pressure term, $F^{\mathrm{UV,IR}}_{\mathrm{rad}}$, is due to photoionization and dust, and includes the additional momentum provided by dust scattering. The second radiation pressure term,  $F^{\mathrm{Ly\alpha}}_{\mathrm{rad}}$ is due to the resonant scattering of the Ly$\alpha$ photons by neutral hydrogen. This component becomes increasingly important as the metallicity of the system decreases. We describe the methodology to calculate this force in Sec.~\ref{subsect:Lyman_alpha_method}.
The fifth term is the force due to the ISM (cloud or diffused) outside of the shell. These terms are described along with the shell structure in Sec.~\ref{sect:shell_structure}.  

\subsubsection{Pressure driven phase}
\label{sect:pressure_driven_phase}

The initial bubble evolution is dominated by the shocked stellar ejecta from the massive stars in the cluster. This evolutionary phase continues till the shell fragments and the shell loses pressure support due to the hot gas in the bubble. We refer to this evolutionary phase as the pressure-driven phase.
During this phase, the stellar winds and/or the supernovae are shocked and their energy feeds the hot bubble interior. The bubble pressure then pushes the shell. Due to the high temperatures and short sound crossing time within the hot bubble, the bubble interior is assumed to be isobaric. The energy equation can be written as follows:
\begin{equation}
\frac{{dE}_{\mathrm{b}}}{dt}=L_{\text {mech }}-L_{\text {cool }}-  P_{\mathrm{b}} A_{\mathrm{sh}} v_{\mathrm{sh}} \, .
\label{eqn:energy_equation}
\end{equation}
The mechanical luminosity from the wind and the SNe is given as:
\begin{equation}
L_{\mathrm{mech}}= L_{\mathrm{w}} + L_{\mathrm{sn}} = \frac{1}{2} \left( \dot{M}_{\mathrm{w}} v_{\mathrm{w}}^{2} + \dot{ M}_{\mathrm{sn}} v_{\mathrm{sn}}^{2} \right) \, , 
\label{eqn:mechanical_lum_def}
\end{equation}
where $\dot{M}_{\mathrm{w}}$ and $\dot{M}_{\mathrm{sn}}$ are the mass loss rates due to stellar winds and supernovae (SNe), respectively, and $v_{\mathrm{w}}$ and $v_{\mathrm{sn}}$ are the terminal velocities of the winds and SNe ejecta, respectively. The bubble pressure, $P_{\mathrm{b}}$ is given as:
\begin{equation}
P_{\mathrm{b}}=(\gamma-1) \frac{E_{\mathrm{b}}}{\frac{4 \pi}{3}\left(r_{\mathrm{sh}}^{3}- r_{\mathrm{w}}^{3}\right)} \, ,
\end{equation}
with $\gamma = 5/3$ being the adiabatic index for an ideal gas. Here $r_{\mathrm{w}},~r_{\mathrm{sh}}$ are the free-streaming radius and the overall bubble radius, respectively. The region between these two radii contains the shocked stellar wind \citep[see Fig.~1 in][]{1977ApJ...218..377W}. $r_{\mathrm{w}}$ can be found by equating $F_{\mathrm{ram}}$ (see Eqn.\eqref{eqn:ram_force_def}) and the force due to the bubble pressure.
The first term in Eqn.~\eqref{eqn:momentum_conservation_general} for this phase is given as: $F_{\mathrm{w,sn}} = P_{\mathrm{b}} A_{\mathrm{sh}}$.
\cite{2022MNRAS.509.4498G} find that the radiative cooling rate from the wind bubble is $\leq~1\%$ of the wind luminosity. Thus, for the pressure-driven phase, we set $L_{\text {cool }}=0$. We do note that there could be other channels for cooling the hot bubble, which could slow down its radial expansion. For example, the presence of a turbulent mixing layer at the contact discontinuity at the hot bubble-shell interface could serve as a means of efficient radiative cooling \cite[see, for example, ][]{2019MNRAS.490.1961E, 2020ApJ...894L..24F, 2021MNRAS.502.3179T, 2021ApJ...914...89L, 2021ApJ...914...90L}. This cooling could be added as an additional contribution to the cooling in the energy equation following \citet{2019MNRAS.490.1961E}. However, we do not address this complexity in the present work in order to limit the number of free variables in the model.

The equation for the conservation of energy is coupled to the system only when the shocked stellar ejecta drives the expansion of the shell. In this case, the energy of the stellar ejecta feeds the bubble pressure which pushes the shell. The coupling of the energy equation is, therefore, limited to the period when the hot stellar ejecta is strongly confined within the bubble. If the shell fragments, which we describe next, the hot gas is assumed to escape at a time scale determined by the sound crossing time. Once the bubble is devoid of hot gases, the expansion is due to the direct impingement of the stellar ejecta onto the shell. Since no energy build-up takes place in the bubble, the energy conservation equation is no longer coupled to the rest of the equations. 

\subsubsection{Momentum driven phase}
\label{sect:momentum_phase}

We terminate the pressure-driven phase {assuming efficient cooling at the contact discontinuity and/or shell fragmentation when the conditions for the Rayleigh-Taylor (RT) instability or the gravitational instability are met. The RT instability occurs when a dense fluid is accelerated with respect to a lighter fluid. Thus, an accelerating dense shell in the cloud/ISM would be RT unstable. RT instability causes disruption of sharp density jumps at contact discontinuities and promotes turbulent mixing \citep{1978ApJ...219..994C, 2016ApJ...821...76D}. Similarly, the system is expected to be gravitationally unstable when the thermal and kinetic energy of a gas parcel are overcome by its local gravitational binding energy \citep{1981ApJ...243L.127O, 2011EAS....51...45E}. Additionally, it is assumed that shell fragmentation occurs when the entire cloud is swept by the shell during the pressure-driven phase.}

{In order to identify the time of onset of the aforementioned instabilities ($t_{\mathrm{frag}}$), and consequently, the end of the pressure-driven phase, we use the prescriptions given in \cite{2019MNRAS.483.2547R} and the references therein. For the RT instability, this is simply an acceleration condition for the dense shell, i.e., $\ddot{r}_{\mathrm{sh}}>0$, while the gravitational instability is assumed to occur when
\begin{equation}
0.67 \frac{3 \, G \, M_{\mathrm{sh}}} {4 \pi \, v_{\rm{sh}} r_{\rm{sh}} c_{\mathrm{s}, \mathrm{sh}}}>1,
\label{eqn:gravitational_instability}
\end{equation}
where, $c_{\mathrm{s}, \mathrm{sh}}$ is the minimum sound speed in the shell. Based on Eqn.~\eqref{eqn:gravitational_instability}, it is clear that neutral shells (characterized by a lower sound speed) with high mass and low velocities are vulnerable to gravitational instability.}

The termination of the pressure-driven phase is associated with a rapid loss of the bubble's pressure{, leading to a rapid decrease of density at the inner face of the shell}. In this case, the cooling term in Eqn.~\eqref{eqn:energy_equation} is: 
\begin{equation}
    L_{\mathrm{cool}} = \frac{E_{\mathrm{b}}(t_{\mathrm{frag}})}{\Delta t_{c}} \, ,
\end{equation}

$E_{\mathrm{b}}(t_{\mathrm{frag}})$ is the bubble energy at the moment of shell fragmentation. $\Delta t_{c}$ is the sound crossing time scale assuming a volume averaged bubble temperature of $10^{6}~\mathrm{K}$ and the radius of the bubble at the time of the fragmentation, $\Delta t_{c} = r_{\mathrm{sh}}(t_{\mathrm{frag}})/c_{\mathrm{s,b}}$. Here $c_{\mathrm{s,b}}$ is the volume averaged speed of sound in the bubble. This transition lasts till all of the bubble's energy is lost, and generally lasts less than a Myr. 
Once the pressure-driven phase is over, it is assumed that there are no intervening media in the interior of the bubble, leading to direct impingement of the cluster's ejecta on the shell. Since there is no energy buildup in the bubble, the evolution is determined by the momentum conservation equation alone. 
The first term in Eqn.~\eqref{eqn:momentum_conservation_general} for this phase is given as: $F_{\mathrm{w,sn}} = F_{\mathrm{ram}}$, where $F_{\mathrm{ram}}$ is given as:
\begin{equation} \label{eqn:ram_force_def}
F_{\mathrm{ram}}= \dot{M}_{\mathrm{w}} v_{\mathrm{w}} + \dot{M}_{\mathrm{sn}} v_{\mathrm{sn}} \, .
\end{equation}
Momentum-driven evolution is weaker in comparison to pressure-driven evolution. At constant stellar feedback, the momentum-driven bubble radius scales as $\propto t^{1/2}$, while the pressure-driven scaling is $\propto t^{3/5}$ \citep{1977ApJ...218..377W}. 
Momentum-driven \HII regions are a plausible explanation for weak X-ray emission \citep{2009ApJ...693.1696H, 2011ApJ...731...91L, 2013ApJ...769...12V} and low shell expansion velocities of observed sources \citep{2021ApJ...914...89L}.
Generally, when the evolution switches to the momentum-driven phase, the shell is massive enough to have a significant amount of gravitational force. Thus, the bubble either dissolves if the feedback is strong enough (Sec.~\ref{subsect:shell_dissolution}), or collapses under gravity (Sec.~\ref{subsect:shell_collapse}).

\subsubsection{Shell structure}
\label{sect:shell_structure}
\begin{figure}
\centering    \hspace{-0.75cm}\includegraphics[width=.8\columnwidth]{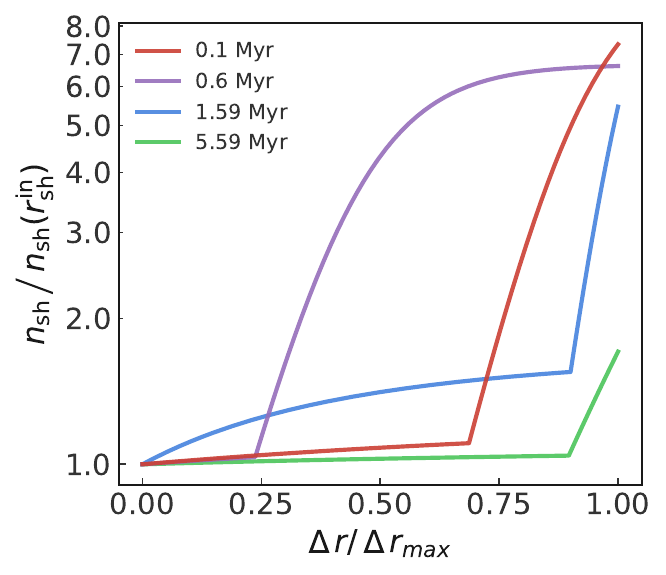}
\caption{The shell density normalized by the density at its inner edge as a function of the normalized shell depth (i.e., shell depth divided by shell thickness) at selected times for a system with $Z=0.02$,  $\epsilon_{\mathrm{SF}}=5\%$, $n_{\mathrm{cl}}=160~\mathrm{cm^{-3}}$ and $\log{M_{\mathrm{cl}}}=5.75$.}
\label{fig:density_profile_examples}
\end{figure}
To calculate the shell's structure, we use the model presented in \citet{2011ApJ...732..100D}.
The model couples the number density in the shell, $n_{\mathrm{sh}}(r)$, the attenuation function for the ionizing radiation, $\phi(r)$, and the optical depth of the dust, $\tau_{\mathrm{d}}$.
These equations are written in two energy regimes, ionizing radiation (photons with energies above $13.6~\mathrm{eV}$) which is absorbed by hydrogen and dust, and non-ionizing radiation which is absorbed by dust alone. Thus for the ionized region of the shell, we have:
\begin{equation}
\frac{\mathrm{d}}{\mathrm{d} r}\left(\frac{\mu_{\mathrm{n}}}{\mu_{\mathrm{p}}} n_{\mathrm{sh}} k T_{\mathrm{i}}\right)=-\frac{1}{4 \pi r^{2} c} \frac{\mathrm{d}}{\mathrm{d} r}\left(L_{\mathrm{n}} e^{-\tau_{\mathrm{d}}}+L_{\mathrm{i}} \phi\right) \, ,
\label{eqn:Shell_structure_density_ionized}
\end{equation}
\begin{equation}
\frac{\mathrm{d} \phi}{\mathrm{d} r}=-\frac{4 \pi r^{2}}{Q_{\mathrm{H}}} \alpha_{\mathrm{B}} n_{\mathrm{sh}}^{2}-n_{\mathrm{sh}} \sigma_{\mathrm{d}} \phi \, , \quad \text{and}
\label{eqn:Shell_structure_ionization_param}
\end{equation}
\begin{equation}
    \frac{\mathrm{d} \tau_{\mathrm{d}}}{\mathrm{d} r}=n_{\mathrm{sh}} \sigma_{\mathrm{d}} \, .
    \label{eqn:Shell_structure_dust_optical_depth_ionized}
\end{equation}

Eqn.~\eqref{eqn:Shell_structure_density_ionized} is an equation of hydrostatic equilibrium of the shell in the limit of low magnetic and turbulent pressures. 
The two components on the right side of Eqn.~\eqref{eqn:Shell_structure_density_ionized} quantify the absorption of neutral radiation due to the dust in the shell and the ionizing radiation due to the gas, $L_{\mathrm{n}}(t)$ and $L_{\mathrm{i}}(t)$ is the neutral and ionizing luminosities, respectively.
Eqn.~\eqref{eqn:Shell_structure_ionization_param}
describes the two ways to attenuate ionizing radiation in the shell, i.e., by ionizing the gas (which can be written in terms of the recombination rate), and by dust. Here, $Q_{\mathrm{H}}(t)$ is the rate of hydrogen ionizing photons emitted by the cluster, $\alpha_{\mathrm{B}}$ is the case B recombination coefficient $= 2.59 \times 10^{-13}~\mathrm{cm}^3 \mathrm{s}^{-1}$ at $\mathrm{T} = 10^4 \mathrm{K}$ \citep{2006agna.book.....O}, $\sigma_{\mathrm{d}}$ is the dust cross-section, and  $c$ is the speed of light.
It is assumed that the quantity of dust scales linearly with metallicity, hence,  $\sigma_{\mathrm{d}} = \sigma_{0}{Z/Z_{\odot}}$, where $\sigma_{0} = 1.5 \times 10^{-21}~\mathrm{cm^2H^{-1}}$ \citep{2011ApJ...732..100D}. 
Eqn.~\eqref{eqn:Shell_structure_dust_optical_depth_ionized} calculates the dust optical depth. 

{The density values at the inner edge of the shell result from the assumption of hydrostatic equilibrium between the forces resulting from winds/SNe and the thermal pressure at the inner edge of the shell. Depending on whether the shell is pressure-driven or momentum-driven, the winds/SNe force at the shell's inner edge differ and result in 
different density initial conditions for the coupled differential equations, given as}:
\begin{equation}
  n_{\mathrm{sh}}(r^\text{in}_{\mathrm{sh}}) = \left\{\begin{array}{ll} \frac{\mu_{\mathrm{p}}}{\mu_{\mathrm{n}} k T_{\mathrm{i}}}P_{\mathrm{b}} & \text { Energy driven } \\ 
\frac{\mu_{\mathrm{p}}}{\mu_{\mathrm{n}} k T_{\mathrm{i}}}  \frac{F_{\mathrm{ram}}}{A_{\mathrm{sh}}} & \text { Momentum driven } \end{array}\right. \, .
\label{eqn:shell_IC_density}
\end{equation}
In Eqn.~\eqref{eqn:shell_IC_density}, as the inner edge of the shell is part of an \HII region, its temperature is assumed to be $T_{\mathrm{i}}=10^{4} \mathrm{K}$.
{We noted in Sec.~\ref{sect:momentum_phase}, a switch from pressure-driven to momentum-driven shells is expected based on the weak X-ray emission and low shell expansion velocities. Another important point about the switch to momentum-driven can be made based on the inner edge density of the shell. This density value, along with the flux of ionizing photons is a key parameter that determines the emission line ratios from \HII regions. 
As noted by \citet{2005ApJ...619..755D}, pressure-driven shells typically exhibit high inner shell density and expand to large radii, resulting in a low ratio of ionizing photon flux to the inner edge density of the shell. This is inconsistent with observations. Transitioning to the momentum-driven phase helps resolve this issue, a point elaborated upon in Sec.~\ref{sect:BPT diagram and parameters}.}

The initial conditions for the other two variables follow from zero initial attenuation:
\begin{equation}
        \phi(r^\text{in}_{\mathrm{sh}}) = 1, ~~
         \tau_{\mathrm{d}}(r^\text{in}_{\mathrm{sh}}) = 0 \, .
\label{eqn: shell_ICs_ionization_attenuation}     
\end{equation}
The solution process is terminated either if the attenuation function drops to zero, or if the entire shell's mass is accounted.
If the former happens, a modified set of equations is then solved using the $n_{\mathrm{sh}}$ and $\tau_{\mathrm{d}}$ at the termination radius as initial conditions.
{Following \citet{2014ApJ...785..164M}, we have:}
\begin{equation}
\frac{\mathrm{d}}{\mathrm{d} r}\left(n_{\mathrm{sh}} k T_{\mathrm{n}}\right)=-\frac{1}{4 \pi r^{2} c} \frac{\mathrm{d}}{\mathrm{d} r}\left(L_{\mathrm{n}} e^{-\tau_{\mathrm{d}}}\right) \, ,
\label{eqn:Shell_structure_density_neutral}
\end{equation}
\begin{equation}
\frac{\mathrm{d} \tau_{\mathrm{d}}}{\mathrm{d} r}=n_{\mathrm{sh}} \sigma_{\mathrm{d}} \, .
\label{eqn:Shell_structure_dust_optical_depth_neutral}
\end{equation}
These are essentially the same equations as those for the ionized shell, except that the terms related to ionizing radiation have been dropped (as $\phi(r)=0$), and the neutral shell temperature, $T_{\mathrm{n}} = 100~\mathrm{K}$ is employed.

These equations are terminated at a radius where the computed mass (integrating the shell density structure) equals the mass of the shell {determined by integrating Eqn.~\eqref{eqn:mass_conservation},} allowing us to calculate $F^{\mathrm{UV,IR}}$ as
\begin{equation}
F^{\mathrm{UV,IR}}_{\text {rad }} \approx f_{\text {abs }} \frac{L_{\text {bol }}}{c}\left(1+\tau_{\mathrm{IR}}\right) \, .
\label{eqn:radiation_pressure_fabs}
\end{equation}
Here $\tau_{\mathrm{IR}}$ is the IR optical depth of the shell given as
\begin{equation}
\tau_{\mathrm{IR}}=\kappa_{\mathrm{IR}} \int_{r^\text{in}_{\mathrm{sh}}}^{r^\text{out}_{\mathrm{sh}}} \mu_{\mathrm{n}} n_{\mathrm{sh}} \mathrm{d} r \, ,
\label{eqn:IR_optical_depth}
\end{equation}
where $\kappa_{\mathrm{IR}} = (Z/Z_{\odot})\kappa_{\mathrm{IR,0}}$ with $\kappa_{\mathrm{IR,0}}=4\,\mathrm{cm^2g^{-1}}$, {noting that for gas-to-dust ratio $\sim100$, values of $\kappa_{\mathrm{IR}}$ are likely to be in the range $1-5 ~ \mathrm{cm^{2} \, g^{-1}}$ \citep{2003A&A...410..611S, 2015ApJ...809..187S}.}

Only the first term in Eqn.~\eqref{eqn:radiation_pressure_fabs} would show up if each photon interacts just once with the medium and then escapes the system. The presence of an additional term allows for momentum exchange between trapped IR radiation and the shell, assuming that the gas and the dust are dynamically well coupled.
Here, $L_{\mathrm{bol}} = L_{\mathrm{i}} + L_{\mathrm{n}}$.
The absorption fraction $f_{\text {abs}}$ is a luminosity-weighted average  of absorption fractions in the ionizing and neutral wavebands at the outer edge of the shell, $r_{\mathrm{sh}}^\text{out}$,  given as:
\begin{equation}
f_{\mathrm{abs}}=\frac{f_{\mathrm{abs}, \mathrm{i}} L_{\mathrm{i}}+f_{\mathrm{abs}, \mathrm{n}} L_{\mathrm{n}}}{L_{\mathrm{bol}}} \, ,
\label{eqn:fabs_definition}
\end{equation}
where, $f_{\mathrm{abs}, \mathrm{i}}=1-\phi(r_{\mathrm{sh}}^\text{out})$, 
and
    $f_{\mathrm{abs}, \mathrm{n}}=1-e^{-\tau_{\mathrm{d}}(r_{\mathrm{sh}}^\text{out})}$
. 

{Fig.~\ref{fig:density_profile_examples} shows examples of the density profile calculated using the method discussed here. The notable increase in density signifies the transition from the ionized to the neutral regions of the shell. The decline in the maximum relative density with age is largely due to the thinning of the shell as the gas is pushed out.}

{We note that the density structure in the \HII regions of the shell results from the use of Eqns.~\ref{eqn:Shell_structure_density_ionized}, \ref{eqn:Shell_structure_ionization_param}, and \ref{eqn:Shell_structure_dust_optical_depth_ionized} is fairly consistent with those obtained using detailed calculations in \texttt{Cloudy}. We explicitly compared these profiles for a subset of the parameter space and found the difference in the mass-weighted density to be within $10\%$, implying a similar position of the ionization front when using the approximate calculations.
On the other hand, neutral parts of the shell can show significant deviations when calculated using \texttt{Cloudy}, and those calculated using Eqns.~\eqref{eqn:Shell_structure_density_neutral}, \eqref{eqn:Shell_structure_dust_optical_depth_neutral}. The differences originate from the approximate shell density profiles using a constant grain cross-section, assuming a constant temperature of $100\,\rm{K}$, and ignoring the absorption of Lyman-Werner band radiation by $\rm{H}_{2}$ when the shell is dense enough to form molecules. The deviations in the density profiles of the neutral shells are unlikely to affect the calculation of $F^{\mathrm{UV,IR}}_{\text {rad }}$ as the
density profiles differ the most in optically thick cases where molecules form. In such cases, the absorption fraction associated with neutral radiation, $f_{\rm{abs, n}}$ tends to unity whether or not we use \texttt{Cloudy}.}

\begin{figure}
\centering    \hspace{-0.75cm}\includegraphics[width=.8\columnwidth]{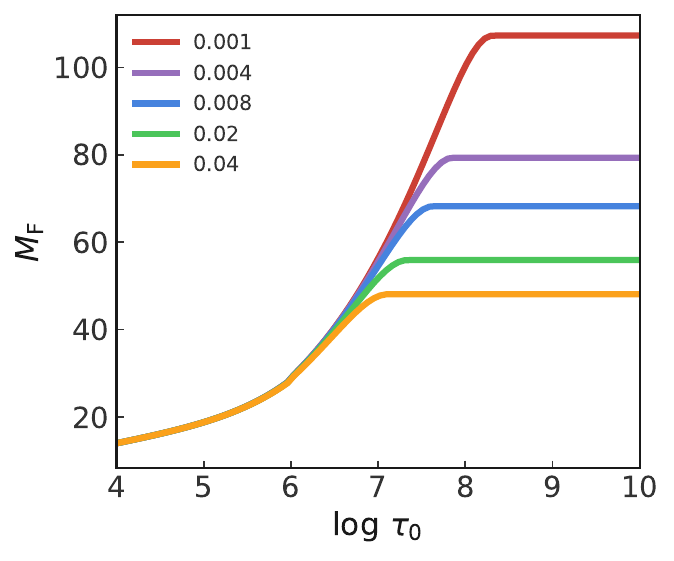}
\caption{$M_\mathrm{F}$ in the case of dusty gas as a function of the Ly$\alpha$ optical depth at line center. The solid lines are the fits provided in \citet{2018MNRAS.475.4617K} at the metallicities considered in this work.}
\label{fig:M_f_Lya}
\end{figure}

\subsubsection{External pressure}
If the shell is completely ionized, ionizing radiation leaks out to ionize the cloud behind it, photo-heating it to a temperature of $\sim 10^{4}~\mathrm{K}$. This represents a significant increase in external force 
in Eqn.~\eqref{eqn:momentum_conservation_general} in comparison to the force due to the cloud when the shell is neutral and the cloud has a temperature of $\approx 10^{2}~\mathrm{K}$. The different cases can be written as:
\begin{equation}
   F_{\mathrm{ext}} = \left\{\begin{array}{ll} \frac{\mu_{\mathrm{n}}}{\mu_{\mathrm{p}}} n_{\mathrm{cl}} k T_{\mathrm{i}} A_{\mathrm{sh}}  & \text { if } M_{\text{sh}} < M_{\text{cl}} \text { and } \phi(r_{\mathrm{sh}}^\text{out}) > 0  \\ n_{\mathrm{cl}} k T_{\mathrm{n}} A_{\mathrm{sh}}   & \text { if } M_{\text{sh}} < M_{\text{cl}} \text { and } \phi(r_{\mathrm{sh}}^\text{out}) = 0 \\ 0  & \text { if } v_{\mathrm{sh}} < 0
   \end{array}\right. \, .
\label{eqn:external_force_conditions}
\end{equation}
{Additionally, once the shell has swept through the entire cloud, we assume that an external pressure from the diffuse ISM, given by \( P_{\rm{ext}}/k = 10^{3}\,\rm{cm^{-3}\,K} \), acts on the expanding shell.}

{We remark that we assume that the cloud is in virial equilibrium, which implies that we do not allow for the natal cloud to infall. Including this would lead to a significant difference in the evolution of the system for the highest density and cloud mass cases. This is exemplified by the change in the amount of minimum stellar mass required to unbind the cloud where this additional force is taken into account, as in \citet{2023MNRAS.521.5686K}.} We additionally note that the cloud itself is assumed to be not affected by the ionized gas pressure to simplify the calculations.

\subsection{Ly\texorpdfstring{$\alpha$}{Lg} radiation pressure}
\label{subsect:Lyman_alpha_method}

Ly$\alpha$ photons resonantly scatter in optically thick regions due to the large absorption cross-section of neutral hydrogen before they escape or get absorbed by dust.
As the metallicity of the system decreases, the central cluster's population exhibits weakened line-driven winds and undergoes lower mass loss, which impacts their main sequence lifetimes. As compared to their metal-rich counterparts, the low-metallicity clusters exhibit a more gradual decrease in the rate of production of ionizing photons. 
Therefore, there is a shift in the mode of stellar feedback. Decreasing the metallicity shifts the dominant mode of energy removal from the central cluster from stellar winds to radiative.
At the same time, the gas around the lower metallicity clusters is increasingly devoid of dust grains in our models. This translates to a lower coupling to radiation by absorption and scattering by dust grains. On the other hand, the decreased presence of dust grains ensures reduced destruction of Ly$\alpha$ photons in the neutral medium, opening another feedback channel. Avoiding destruction, the trapping of Ly$\alpha$ photons could represent a significant radiative force \citep{2008MNRAS.391..457D, 2009MNRAS.396..377D, 2017MNRAS.464.2963S, 2018MNRAS.475.4617K}. 

\cite{2021MNRAS.504...89T} argue for the implementation of Ly$\alpha$ pressure in galaxy formation due to its greater influence compared to photoionization and UV radiation pressure in initiating gas acceleration around bright sources. Their conclusions bracket a broad range of gas columns and metallicities  ($16 < \log N_{\mathrm{HI}} < 23$ $;~ -4 < \log~Z/Z_{\odot} < 0$).
The trapping of the Ly$\alpha$ results in force multiplication. The multiplication factor ($M_{\mathrm{F}}$) represents the number of times, on average, a photon contributes to the momentum deposition, with respect to the case in which only a single scattering takes place.

In this work, we use the approach presented in \cite{2018MNRAS.475.4617K} to calculate $M_{\mathrm{F}}$ in dusty media. They provide fitting formulas for $M_{\mathrm{F}}$ in the dusty case based on calculations performed using 3D Monte Carlo Ly$\alpha$ radiative transfer assuming a central source in a uniform medium. The fitting formulas use the multiplication factor in the dust-free case, $M_\text{F,\,no\ dust}$, and the escape fraction that mimics the destruction of Ly$\alpha$ by dust, $f_{\mathrm{esc,\,dust}}^{\mathrm{Ly} \alpha}$. They are written as follows:
\begin{equation}
\begin{split}
&M_\text{F,\,no\ dust} \approx 29\left(\frac{\tau_0}{10^6}\right)^{0.29} \quad \left(\tau_0 \geq 10^6\right) \, , \\
&\begin{aligned}
\log M_\text{F,\,no\ dust} \approx &-0.433+0.874 \log \tau_0 -0.173\left(\log \tau_0\right)^2 \\
&+0.0133\left(\log \tau_0\right)^3 \quad \left(\tau_0<10^6\right) \, ,
\end{aligned}
\end{split}
\end{equation}
\begin{equation}
\begin{aligned}
\tau_0^{\text {peak }} & = 4.06 \times 10^6 ~T_4^{-1 / 4}\left(\frac{\sigma_{\mathrm{d},-21}}{3}\right)^{-3 / 4} \, ,
\end{aligned}
\end{equation}
\begin{equation}
\tau_0^{\prime}=\min \left(\tau_0, \tau_0^{\text {peak }}\right) \, , \quad N^{\prime}_{\mathrm{HI}}=\min \left(N_{\mathrm{HI}}, N_{\mathrm{HI}}^{\text {peak }}\right) \, ,
\end{equation}
\begin{equation}
f_{\mathrm{esc,\,dust}}^{\mathrm{Ly} \alpha}=1 / \cosh \left\{\frac{\sqrt{3}}{\pi^{5 / 12} \xi}\left[\left(a_{\mathrm{V}} \tau_0\right)^{1 / 3} \tau_{\mathrm{da}}\right]^{1 / 2}\right\} \, ,
\end{equation}
\begin{equation}
\tau_{\mathrm{da}}=N_{\mathrm{HI}} \sigma_{\mathrm{d}}\left(1-\mathcal{A}_{\mathrm{b}}\right) \, ,
\end{equation}
\begin{equation}
M_{\mathrm{F}} = M_\text{F,\,no\ dust} \left(\tau_0^{\prime}\right) \times f_{\mathrm{esc,\,dust}}^{\text {Ly} \alpha}\left(N_{\mathrm{HI}}^{\prime}, \xi\right) \, .
\end{equation}

In the equations above, $\tau_0$ is the Ly$\alpha$ optical depth at the line center using the cross-section $\sigma_0=5.88 \times 10^{-14} \mathrm{~cm}^2 T_4^{-1 / 2}$ ($T_{4}\equiv T_{\mathrm{n}}/10^{4}\mathrm{K}$), $N_{\mathrm{HI}}$ is the atomic Hydrogen column density, $Z^{\prime}$ is the metallicity in units of solar metallicity, $\sigma_{\mathrm{d},-21} \equiv \sigma_{\mathrm{d}} / 10^{-21} \mathrm{~cm}^2 / \mathrm{H}$\footnote{Note that the dust cross section is defined as $\sigma_{\mathrm{d}} = \sigma_{0}{Z/Z_{\odot}}$.}, $f_{\mathrm{d} / \mathrm{m}}$ is the dust to metal ratio, $\mathcal{A}_{\mathrm{b}}=0.46$ is the dust albedo,  $a_{\mathrm{V}} = 4.7 \times 10^{-4} T_{4}^{-1/2}$ is the Voigt parameter in the optically thick regime, and $\xi=1.78$ is a fitting parameter.

{Note that $\tau_0$ is calculated using the shell profile given in Sec.~\ref{sect:shell_structure}. We use Eqn.~\eqref{eqn:Shell_structure_dust_optical_depth_neutral} to estimate $N_{\mathrm{HI}}$ after removing the contribution of the ionized column. 
In order to not overestimate $N_{\mathrm{HI}}$, we use the model presented in \citet{2013MNRAS.436.2747K} to get an approximate value of the molecular hydrogen fraction in the shell as a function of depth. We compute $ N_{\mathrm{HI}} $ by considering the neutral gas column up to the point in the shell where the molecular hydrogen fraction becomes non-zero. The calculation method for the molecular fraction is provided in Appendix~\ref{appendix:krumholz2013}, while examples of the atomic column density can be found in Appendix~\ref{appendix:Atomic_column_trends}. In practice, the Ly$\alpha$ multiplication factor at all metallicities saturates at neutral column depths lower than where any molecular hydrogen could be present.}
The $M_{\mathrm{F}}$ values as a function of the Ly$\alpha$ optical depth in the case of the five metallicities considered in this work are shown in Fig.~\ref{fig:M_f_Lya}.

At each time step during the shell evolution, we calculate the Ly$\alpha$ luminosity, $L_{\mathrm{Ly}\alpha}$  as:
\begin{equation}
    L_{\mathrm{Ly}\alpha} = 4 \pi \int_{r(\phi=1)}^{r(\phi=0)} \mathcal{P}_{\mathrm{B}} \, E_{\mathrm{Ly\alpha}} \, \alpha_{\mathrm{B}} \, n_{\mathrm{sh}}^2  r^2 \text{d}r \, . 
\label{eqn:LymanAlphaLuminosity}
\end{equation}
\begin{figure}
\centering
\hspace{-0.75cm}\includegraphics[width=.8\columnwidth]{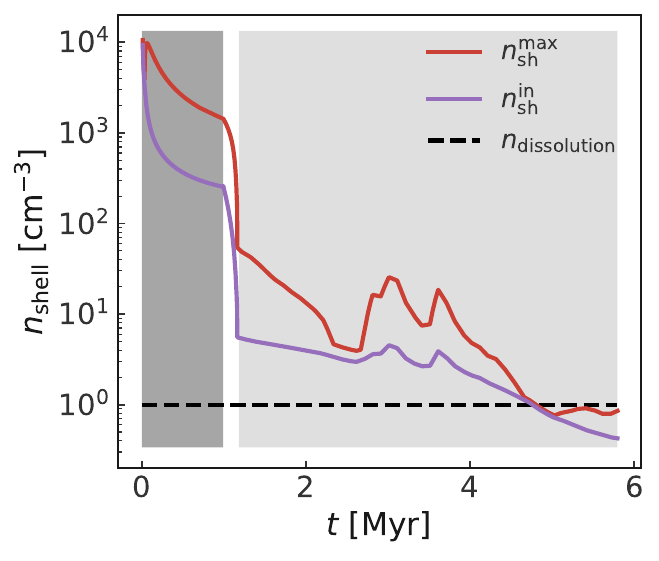}
    \caption{Evolution of the shell density for a system with $Z=0.02$,  $\epsilon_{\mathrm{SF}}=5\%$, $n_{\mathrm{cl}}=160~\mathrm{cm^{-3}}$ and $\log{M_{\mathrm{cl}}}=5.75$.
    Density values at the inner edge and the maximum in the shell are shown, {obtained using the profiles such as those in Fig.~\ref{fig:density_profile_examples}}.
    The shell is considered dissolved when the maximum density falls below the threshold value (dashed black line) for a period of more than $1$~Myr. The darker shaded area represents the period when the shell is expanding due to the pressure of the shocked gases in the bubble, while the lighter one is that when the expansion is momentum-driven. The non-shaded region between the two is the transition between the two regimes.}    \label{fig:shell_dissolution_example}
\end{figure}
The integral is carried out from the inner edge of the shell, where the attenuation function for the ionizing radiation ($\phi$) is unity, till the point where the shell turns neutral (details of the shell structure are given in Sec.~\ref{sect:shell_structure}). This takes into account the absorption of ionizing photons by the dust present in the system.
{The probability $\mathcal{P}_{\mathrm{B}}$ of an absorbed Lyman continuum photon resulting in the emission of a Ly$\alpha$ photon is taken to be a fixed, Case-B value of $0.68$ at $T=10^{4}\,\mathrm{K}$. Similarly, $\alpha_{\mathrm{B}}$ represents the Case-B recombination coefficient and is assigned a value of $2.59\times 10^{-13} \, \mathrm{cm}^3 \mathrm{s}^{-1}$ \citep{2006agna.book.....O}}. $E_{\mathrm{Ly\alpha}}$ is the energy of individual Ly$\alpha$ photon ($10.16$ eV).
Given this, we calculate $F_{\mathrm{rad}}^{\mathrm{Ly\alpha}}$ as:
\begin{equation}
    F_{\mathrm{rad}}^{\mathrm{Ly\alpha}} = f_{\mathrm{esc,\,v}}^{\text {Ly} \alpha} \, M_{\mathrm{F}} \, \frac{L_{\mathrm{Ly}\alpha}}{c} \, .
\end{equation}
Here we have added a shell velocity-dependent escape fraction, $f_{\mathrm{esc,\,v}}^{\text {Ly} \alpha}$, to mimic the reduction in the opacity of moving shells to Ly$\alpha$ photons. This is based on the scaling given in \cite{2008MNRAS.391..457D}) (refer to Fig.~6 in that paper), which suggests a drop in the multiplication factor by an order of magnitude if the absolute velocity of the shells increases to 100 km/s. {In practice,  $f_{\mathrm{esc,\,v}}^{\text {Ly} \alpha}$ is a linear function that drops from 1 to 0.1 as the absolute value of the shell velocity increases from 0 to 100 km/s. We note that the shell velocity rarely exceeds 100 km/s in our parameter space, in which case, $f_{\mathrm{esc,\,v}}^{\text {Ly} \alpha}$ is assumed to be zero.}

Apart from the caveats highlighted in \cite{2018MNRAS.475.4617K}, we note additional caveats of our approach here. We calculate $N_{\mathrm{HI}}$ assuming the ionized and neutral parts of the shell to be in hydrostatic equilibrium, not accounting for any force gradients that would necessarily be present. We have also ignored the force on the cloud if the ionization front lies outside the shell. {Another aspect we have not considered here is the leakage of Ly$\alpha$ photons once the shell fragments.
This is difficult to handle in our model without introducing another free parameter, for example, the escape fraction of Lyman Continuum, which would propagate to Ly$\alpha$ pressure. Additionally, fragmentation of the shell could alter the Ly$\alpha$ radiation pressure by changing the geometry to one with escape channels \cite[see, for example][for Ly$\alpha$ escape mechanisms through clumpy gas.]{2017A&A...607A..71G, 2019MNRAS.484...39S}}

To address these limitations, a comprehensive approach using Monte Carlo radiation hydrodynamics is required \cite[see, for example,][]{2020ApJ...905...27S}. However, this is outside the purview of this study, which necessitates a broad exploration of the parameter space.

\subsection{Shell dissolution}\label{subsect:shell_dissolution}

Following \cite{2017MNRAS.470.4453R}, it is assumed that the shell is dissolved if the entire cloud has been swept and the maximum density in the shell falls below $1~\mathrm{cm^{-3}}$ for a period greater than $1~\mathrm{Myr}$.

The maximum density in the shell is determined by the density at the inner edge and the gas/dust column density {through the shell-structure equations given in Sec.~\ref{sect:shell_structure}. The former is determined by the winds/SNe feedback intensity, and the latter by the shell's mass and the impinging radiation intensity}. Both of these quantities are affected by the shell's radius. As the shell expands, the density at the inner edge generally decreases due to the increasing radius and the aging of the stellar cluster, {except during the Supernovae (SNe) and Wolf-Rayet (W-R) phases, when the feedback intensity experiences a significant increase. This is shown in Fig.~\ref{fig:shell_dissolution_example}, with additional examples given in Figs.~\ref{fig:evolution_vary_all_params_Z02} and \ref{fig:evolution_vary_all_params_Z004}. Keeping in mind that the models presented here employ finite clouds, the shell begins to thin and the column density decreases once the entire cloud is swept up and no more mass is added to the shell. These cumulative effects result in a decrease in the maximum shell density as the shell expands beyond the cloud, which is the case when the stellar feedback provides sufficient outward momentum to overcome the cloud's binding energy. Fig.~\ref{fig:shell_dissolution_example} also shows the declining maximum number density in the shell for a case where the shell eventually dissolves.}

{In Sec.~\ref{sect:shell_structure}, we emphasized that the impact of utilizing approximate profiles for the shell structure is expected to be limited. Additionally, we observe that as the shells become thinner, the discrepancies in the density profiles in the neutral regions of the shell, calculated using the approximate method of the evolutionary model, tend to decrease compared to the results from \texttt{Cloudy}. This suggests that the criteria for shell dissolution would also not be significantly affected even if we were to transition to more sophisticated calculations.}
 
\subsection{Shell collapse and multi-generational star formation}\label{subsect:shell_collapse}
During the momentum-driven phase, it is possible that the feedback is not strong enough to dissolve the shell. In such a scenario, the shell starts to shrink under the gravitational force. We follow the evolution of such a system till the moment the shell radius becomes equal to the initial radius given by Eqn.~\eqref{eqn:initial_conds_W77}. At this point, we restart the evolution (initially pressure-driven and later momentum-driven) with an additional star cluster. The new cluster is formed with the same $\epsilon_{\mathrm{SF}}$ out of the available cloud mass.
This process can take place multiple times during the total evolution time, $t_{\mathrm{evo}} = \mathrm{min}\left( t_{\mathrm{diss.}}, 30~\mathrm{Myr} \right)$.
The remaining cloud mass surrounding a system containing $N$ generations of stars is, $M_{\mathrm{cl,r},N} = (1 - \epsilon_{\mathrm{\textsc{SF}}})^{N} M_{\text{cl}}$.
Similarly, the mass of the $N^{\mathrm{th}}$ generational cluster of stars is, $M_{\star,N} = \epsilon_{\mathrm{\textsc{SF}}} M_{{\mathrm{cl}, N-1}}$.
The stellar feedback is dependent on the ages of the stellar clusters present in the system. The total stellar feedback $\mathcal{L}$ at time $t$, with $N$ generations of star clusters is given as:
\begin{equation}
\mathcal{L}(t)=\sum_{i=1}^{N(t)} \mathcal{L}_{i}\left(M_{*, i},~ t-t_{\mathrm{\textsc{SF}}, i}\right) \, ,
\end{equation}
where $t_{\mathrm{\textsc{SF}}, i}$ is the time at which the $i^{\mathrm{th}}$ star formation event takes place. 
Although this prescription is very simplistic and ignores many intricate processes that regulate star formation, it could be thought of as a way to incorporate the complexity of ``collect and collapse'' star formation \cite[][N49 \HII bubble]{2010A&A...518L.101Z}
 and multiple generations of stars present in stellar nurseries
 \cite[][Tarantula nebula]{2018MNRAS.473L..11R}. 

 {We note that the assumption that stellar clusters fully sample the IMF may not be entirely accurate for smaller clusters. This limitation is even more significant in the recollapse scenario, where the remaining cloud mass, and thus the resulting stellar mass, is even smaller. Averaging over multiple populations, as discussed in Sec.~\ref{sect:integration_with_SKIRT}, can help mitigate this issue to some extent. We plan to address the effects of stochastically sampling the IMF in the future.}

\subsection{Initial conditions and the parameter space}

\begin{figure}
    \centering
\hspace{-0.75cm}\includegraphics[width=.8\columnwidth]{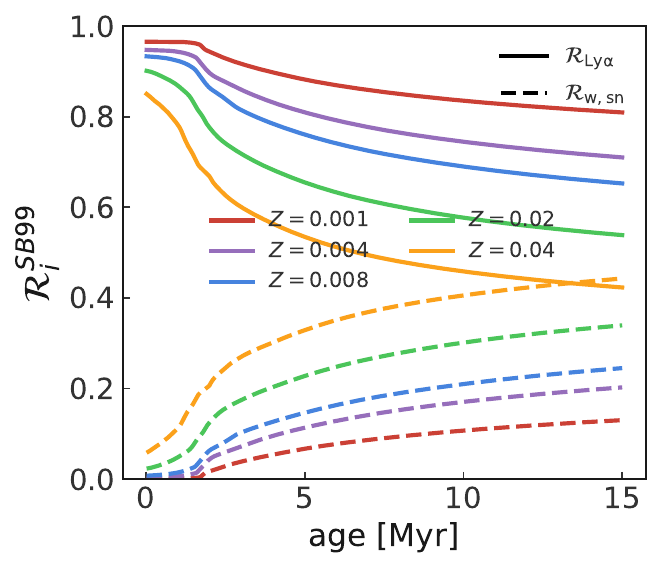}
    \caption{Upper limit on the {relative} contribution of Ly$\alpha$ radiation pressure to the {total} outward momentum deposition, $\mathcal{R}_{\mathrm{Ly\alpha}}$ (solid) and the associated lower limit on the $\mathcal{R}_{\mathrm{w,sn}}$ (dashed) as a function of metallicity and age based on the stellar template data. See the discussion in Sec.~\ref{sect:trends_in_shell_evolution.ssec}.}
    \label{fig:template_R_Lya}
\end{figure}
\begin{figure*}
    \centering
    \includegraphics[width=.8\textwidth]{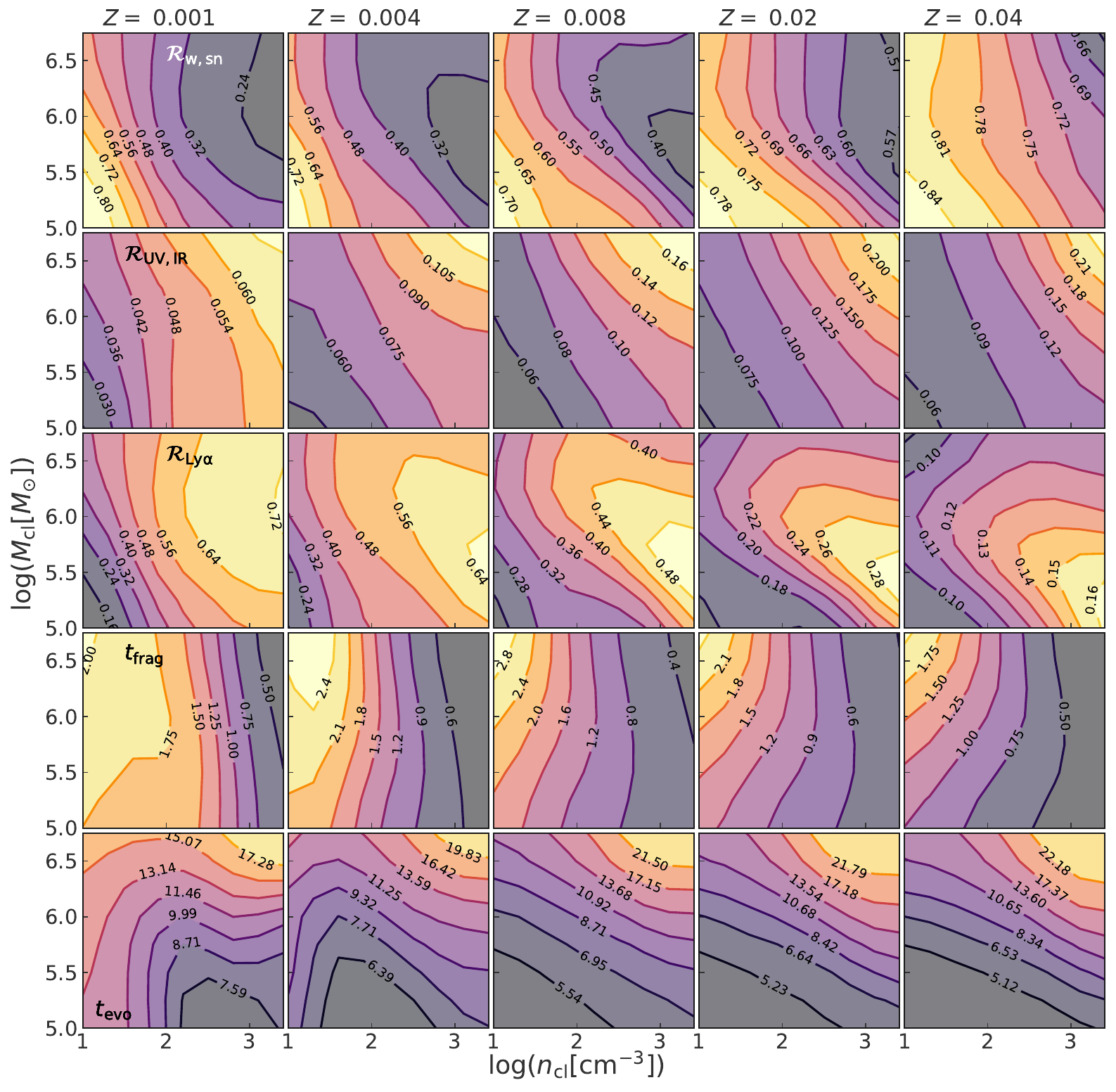}
    \caption{{The top three rows depict contours of $\mathcal{R}_{i}$, which quantifies the relative total outward momentum deposition by channel $i$ at the end of the evolution period. Specifically, the rows show $\mathcal{R}_{\mathrm{w, sn}}$ (top row), $\mathcal{R}_{\mathrm{UV,IR}}$ (second row), and $\mathcal{R}_{\mathrm{Ly\alpha}}$ (third row) as functions of cloud density $n_{\mathrm{cl}}$ and mass $M_{\mathrm{cl}}$. The fourth row illustrates the fragmentation time, $t_{\mathrm{frag}}$, marking the switch to momentum-driven evolution for the first generation of star formation (if multiple star formation events take place), and the bottom row displays the total evolution time, $t_{\mathrm{evo}} = \mathrm{min}\left( t_{\mathrm{diss.}}, 30~\mathrm{Myr} \right)$. All panels assume $\epsilon_{\mathrm{SF}}=5$ per cent.}}
    \label{fig:R_contours_fix_SFE}
\end{figure*}

\begin{figure*}
    \centering
    \includegraphics[width=.8\textwidth]{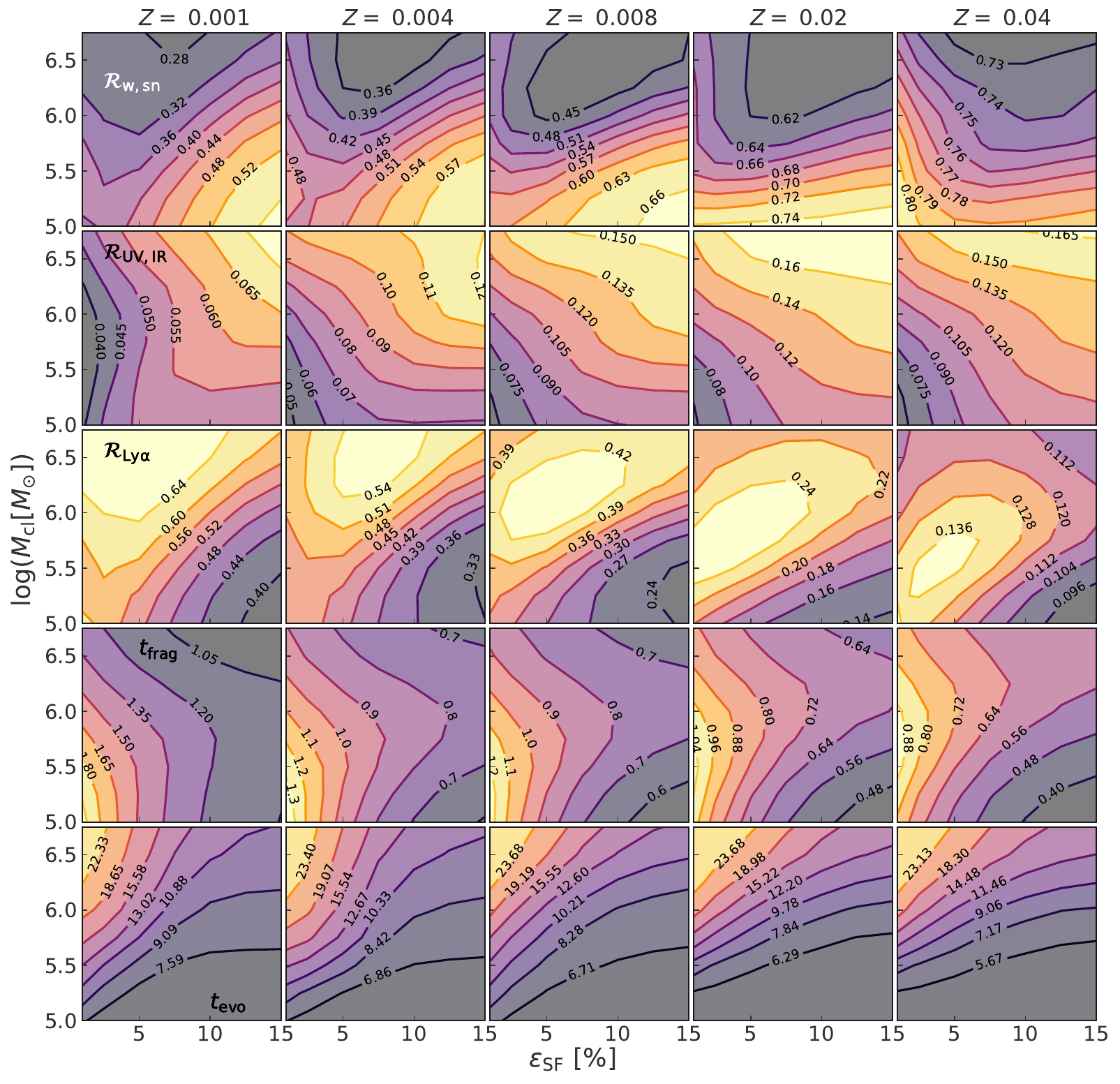}
    \caption{Same as Fig.~\ref{fig:R_contours_fix_SFE}, but here the x-axis represents $\epsilon_{\mathrm{SF}}$. The cloud density is fixed at $320~\mathrm{cm^{-3}}$.}
    \label{fig:R_contours_fix_n}
\end{figure*}

The initial conditions for the mass, momentum, and energy equations are generated using the initial data point of the stellar feedback library ($t_{\mathrm{0}}$, $L_{\mathrm{mech, 0}}$). Assuming a constant mechanical luminosity, and an adiabatic bubble for the initial  period of $t_{0}$, we use the relations in \citet{1977ApJ...218..377W}:
\begin{equation}
\begin{aligned}
& E_{0} =\frac{5}{11} L_{\mathrm{mech, 0}} \, t_{\mathrm{0}}
\hspace{.5em} \text{ and}
\hspace{.5em} 
r_{\mathrm{0}} =\left(\frac{250}{308 \pi}\right)^{1/5} \,  L_{\mathrm{mech, 0}}^{1 / 5} \, \rho_{\mathrm{cl}}^{-1 / 5} \,  t_{\mathrm{0}}^{3 / 5} \, .
\end{aligned}
\label{eqn:initial_conds_W77}
\end{equation}
The initial mass of the shell is calculated using the initial shell radius and the density of the cloud, the shell velocity follows from the expression of the radius. The initial mechanical luminosity comes from all generations of stellar clusters in the system (see Sec.~\ref{subsect:shell_collapse}).

The parameter space explored here is based on typically observed values.
The cloud mass range is based on the observational data suggesting that most of the molecular mass in the Milky Way resides in clouds of mass greater than $10^{5}~M_{\odot}$ and there is an upper limit of $5\times10^{6}~M_{\odot}$ \citep[see][and references therein]{2015ARA&A..53..583H}.
The cloud number density values used are based on the observationally reported mean surface density values \citep[e.g.,][]{2009ApJ...699.1092H, 2012ApJ...761...37M, 2014ApJ...784....3C, 2017ApJ...834...57M}.
We use all five metallicities available for the \texttt{STARBURST99} templates. 
Finally, the star formation efficiency parameter runs from $1-15\%$ \citep{1994ApJ...436..795F, 2016ApJ...819..137K}. 
The parameter space is summarized in Tab.~\ref{Table:template_parameter_space} and represents a total of $2520$ models that were evolved for the library. 
The model ODEs are written in \texttt{Python} and solved using the stiff ODE solver support in the \texttt{scipy.integrate.solve\_ivp}\footnote{Documented online at \url{https://docs.scipy.org/doc/scipy/reference/generated/scipy.integrate.solve_ivp.html}} function.

\begin{table}
\caption{The parameter space used for the evolutionary models in this work, which includes the cloud density $n_{\mathrm{cl}}$, cloud mass $M_{\mathrm{cl}}$, star-formation efficiency $\epsilon_{\mathrm{\textsc{SF}}}$, and stellar/gas metallicity $Z$.}
\centering
\renewcommand{\arraystretch}{1.25}
\begin{tabular}{lc}
\hline
Parameter & Values \\
\hline $n_{\mathrm{cl}}~[\mathrm{~cm}^{-3}]$ & $10, ~20, ~40, ~80, ~160, ~320, ~640, ~1280, ~2560$ \\
$\log M_{\mathrm{cl}}~[M_{\odot}]$ & $5.00, ~5.25, ~5.50, ~5.75, ~6.00, ~6.25, ~6.50, ~6.75$ \\
$\epsilon_{\mathrm{\textsc{SF}}}~[\%]$ & $1.0, ~2.5, ~5.0, ~7.5, ~10.0, ~12.5, ~15.0$ \\
$Z$ & $0.001, ~0.004, ~0.008, ~0.02, ~0.04$ \\
\hline
\label{Table:template_parameter_space}
\end{tabular}
\renewcommand{\arraystretch}{0.8}
\end{table}

\subsection{Trends in the feedback channels driving the shell evolution}
\label{sect:trends_in_shell_evolution.ssec}
\begin{figure*}
\centering
\includegraphics[width=.9\textwidth]{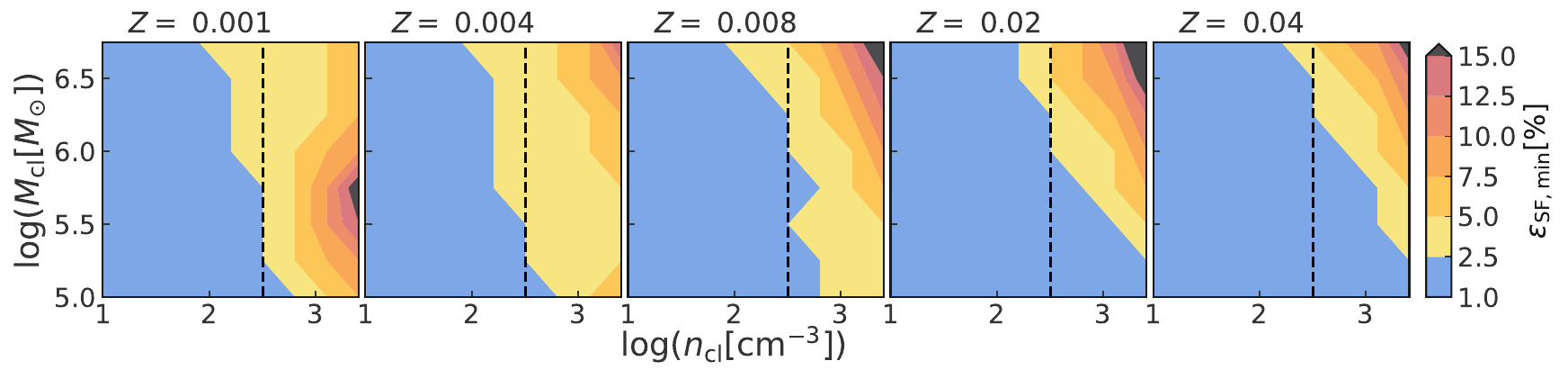}
\includegraphics[width=.9\textwidth]{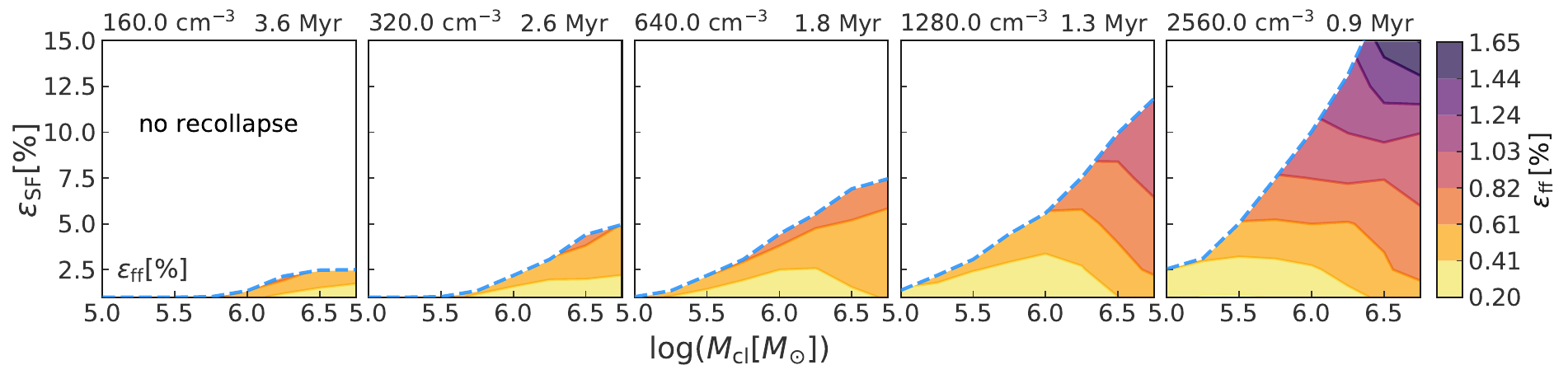}

    \caption{Top: Contours of the minimum single burst efficiency ($\epsilon_{\mathrm{SF,min}}$) required to disrupt the clouds as a function of the cloud mass, density, and metallicity. At all $\epsilon_{\mathrm{SF}}$ values smaller than this value, more than one generation of stars is present at the end of the $t_{\mathrm{evo}}$. The dashed line corresponds to $n_\mathrm{cl}=320~\mathrm{cm^{-3}}$, which is the value used in  Fig.~\ref{fig:R_contours_fix_n}.
    Bottom: Contours of the star formation efficiency per free-fall time ($\epsilon_{\mathrm{ff}}$) given as percentage values for $Z=0.02$ and a selection of $n_{\mathrm{cl}}$ values listed above each panel along with the associated free-fall time $\tau_{\mathrm{ff}}$. The contours are calculated according to Eqn.~\eqref{eqn:freefall_SFE}. The blue dashed line represents the minimum $\epsilon_{\mathrm{SF}}$ required to disrupt the cloud.  {The white region represents the models not exhibiting recollapse.}}
\label{fig:SFE_contours}
\end{figure*}
To elucidate the impact of changing the model parameters on the evolution of the system, 
We define the {relative total outward momentum deposition} on the shell by a force due to a given feedback channel $i$ at the end of the calculation period, $t_{\mathrm{evo}} = \mathrm{min}\left( t_{\mathrm{diss.}}, 30~\mathrm{Myr} \right)$. This could be written as:
\begin{equation}
    \mathcal{R}_{i} = \frac{\int_{0}^{t_{\mathrm{evo}}} F_{i} ~\text{d}t} {\int_{0}^{t_{\mathrm{evo}}} \big( F_{\mathrm{w,sn}} + F^{\mathrm{UV,IR}}_{\mathrm{rad}} + F^{\mathrm{Ly\alpha}}_{\mathrm{rad}}\big)~\text{d}t} \, .
\label{eqn:R}
\end{equation}
Before we look at the results from the models, it is worth looking at the expected values of $\mathcal{R}_{i}$ as a function of the cluster age and metallicity resulting directly from the template library. Fig.~\ref{fig:template_R_Lya} shows the values of $\mathcal{R}_{\mathrm{Ly\alpha}}$ and $\mathcal{R}_{\mathrm{w,sn}}$ assuming, \begin{enumerate*}
    \item the maximum value of $M_{\mathrm{F}}$ (see Fig.~\ref{fig:M_f_Lya}) and conversion of all ionizing photons to Ly$\alpha$, 
    \item $F^{\mathrm{w,sn}}_{\mathrm{rad}} = F_{\mathrm{ram}}$, which assumes that the shell evolution is momentum-driven throughout (cf. Sec.~\ref{sect:momentum_phase}), and
    \item $F^{\mathrm{UV,IR}}_{\mathrm{rad}} = L_{\mathrm{bolo}}/c$.
\end{enumerate*} 
As momentum deposition by the shocked gases during the pressure-driven phase tends to be higher in comparison to the direct deposition by $F_{\mathrm{ram}}$\footnote{In Fig.~\ref{fig:shell_dissolution_example}, the reduction of force acting on the shell can be inferred by comparing the densities at the shell's inner edge as the switch to momentum-driven phase takes place.}, the values of $\mathcal{R}_{\mathrm{Ly\alpha}}$ in Fig.~\ref{fig:template_R_Lya} serve as an upper limit on the value expected in our models. Additional factors, such as the lack of neutral gas, the absorption of ionizing photons by dust, and lower opacity to Ly$\alpha$ photons due to shell thinning and/or high shell velocity would lower the values of $\mathcal{R}_{\mathrm{Ly\alpha}}$ depending upon the model parameters. The metallicity based reduction in $\mathcal{R}_{\mathrm{Ly\alpha}}$ is clear in Fig.~\ref{fig:template_R_Lya}. It can also be inferred that shells that dissolve slowly are expected to exhibit lower $\mathcal{R}_{\mathrm{Ly\alpha}}$ or higher $\mathcal{R}_{\mathrm{w,sn}}$.

In the following, we examine $\mathcal{R}_{i}$ and discuss how the variations in the model parameters affect the feedback mechanisms and the dissolution and fragmentation times of the shells. The top three rows in Figs.~\ref{fig:R_contours_fix_SFE} and \ref{fig:R_contours_fix_n} show the contours of $\mathcal{R}_{i}$ due to the forces attributed to winds and supernovae ($\mathcal{R}_{\mathrm{w, sn}}$), the absorption of UV and  trapped IR ($\mathcal{R}_{\mathrm{UV, IR}}$), and the Ly$\alpha$ radiation pressure ($\mathcal{R}_{\mathrm{Ly\alpha}}$), respectively. 
The contours for $t_{\mathrm{frag}}$ (row 4) and those for $t_{\mathrm{evo}}$ (row 5) are also shown.
In Fig.~\ref{fig:R_contours_fix_SFE}, the contours are shown as a function of $n_{\mathrm{cl}}$ and $M_{\mathrm{cl}}$ at a fixed $\epsilon_{\mathrm{SF}} = 5 \%$, while in Fig.~\ref{fig:R_contours_fix_n}, they are shown as a function of $\epsilon_{\mathrm{SF}}$ and $M_{\mathrm{cl}}$ at a fixed $n_{\mathrm{cl}}=320~\mathrm{cm^{-3}}$.
Some additional examples of the time evolution of key shell properties as a function of the model parameters are provided in the Appendix, in Figs.~\ref{fig:evolution_vary_all_params_Z02} and \ref{fig:evolution_vary_all_params_Z004}.

    

\subsubsection{Metallicity}
It can be inferred from the $\mathcal{R}_{i}$ values in Figs.~\ref{fig:R_contours_fix_SFE} and \ref{fig:R_contours_fix_n} that at the lower end of the metallicity range considered in this work ($Z\leq0.004$), a significant amount of the total momentum imparted to the shells can come from Ly$\alpha$ scattering pressure. At the higher metallicity end, the main drivers are the winds and the supernovae. This departure from Ly$\alpha$ driven shells with increasing metallicity is a result of both increased energy carried by the stellar winds relative to radiative energy, as well as the increased destruction of the Ly$\alpha$ photons by dust. 
As the Ly$\alpha$ radiation pressure only affects the shell dynamics only if neutral Hydrogen is present, even at the lowest metallicity values, the impact of Ly$\alpha$ can be subdominant when considering the lower end of cloud densities as we discuss below along with the effects of changing the cloud density. 

$\mathcal{R}_{\mathrm{UV, IR}}$ remains subdominant across the parameter space considered in this work, although its values rise as the cloud density and mass are increased owing to increasing dust optical depths. For the metallicities in the range $Z\leq0.008$ shown in Fig.~\ref{fig:R_contours_fix_SFE}, the ratio $\mathcal{R}_{\mathrm{Ly\alpha}}/\mathcal{R}_{\mathrm{UV,IR}}$ tends to fall in the range $5-20$, while this ratio is around  $0.5-3$ at the higher metallicity end.

\subsubsection{Cloud density}
Cloud density plays a role in determining when the switch to the momentum-driven phase takes place (fourth row, Fig.~\ref{fig:R_contours_fix_SFE}). The pressure-driven phase lasts longer for low-density clouds as it takes longer for them to become susceptible to either gravitational fragmentation or the Rayleigh-Taylor instability, which leads to shell fragmentation when the entire cloud has been swept and the shell expands in the low-density ISM (see Sec.~\ref{sect:momentum_phase}). As mentioned before, the deposition of momentum by winds and SNe in the pressure-driven phase is significantly higher. 
While this is true at all metallicities, at the lower end of the metallicities, the shells of low-density clouds ($n_{\mathrm{cl}}<100~\mathrm{cm^{-3}}$) tend to be optically thin to ionizing photons for prolonged periods and couple less efficiently with Ly$\alpha$ radiation {due to a lack of neutral columns}. Thus, $\mathcal{R}_{\mathrm{w, sn}}$ remains high at lower cloud densities in general. The contribution of $F^{\mathrm{UV,IR}}_{\mathrm{rad}}$ towards overall momentum transfer to the shell also tends to increase as the cloud density is increased.

Increasing $n_{\mathrm{cl}}$ while keeping the $M_{\mathrm{cl}}$ fixed increases the gravitational binding energy, thus a higher amount of outward momentum is required to dissolve them. 
The dominant feedback mechanisms in the case of high metallicity systems (winds and SNe) scale with the stellar mass present in the system, thus increasing the gravitational binding energy while keeping the stellar mass fixed 
leads to slower dissolution of the shell as reflected by the $t_{\mathrm{evo}}$ contours. For the lower metallicity cases, Ly$\alpha$ radiation pressure and the external pressure due to the ionized cloud can play a significant role in shaping the expansion of the shell. Both of these quantities depend on the density structure of the shell. 
The contribution of the Ly$\alpha$ radiation pressure to the total momentum deposition increases with cloud density at a fixed $M_{\mathrm{cl}}$ as shown in Fig.~\ref{fig:R_contours_fix_SFE} as the shells formed out of dense clouds have higher column densities of neutral gas and expand slowly. These factors favor the presence of neutral gas and an increase in the effective $M_{\mathrm{F}}$ in our models. {This can be understood by considering the fact that the shells carved out of denser clouds have smaller radii, higher inner-edge densities, and are more massive during the initial period of expansion. For a purely wind-driven shell in the pressure-driven phase $r_{\rm{sh}}\,\propto n^{-0.2}_{\rm{cl}}$, $n_{\mathrm{sh}}(r^\text{in}_{\mathrm{sh}})\,\propto\,n^{0.6}_{\rm{cl}}$, and $M_{\rm{sh}}\,\propto n^{0.4}_{\rm{cl}}$ \citep[see, for example][]{1977ApJ...218..377W} promoting higher neutral column densities, as can be inferred from Figs.~\ref{fig:NHI_Z02} and \ref{fig:NHI_Z004}. This difference in neutral column densities during the initial period of expansion is particularly important as it plays out when the ionizing radiation is the strongest.}

At the lower cloud-density end, {other parameters fixed, shells have lower inner-edge densities and larger radii compared to their higher-density counterparts, both these effects can push the ionization front deeper in the shell, or make the shell density bounded}. Shells that are optically thin to ionizing radiation while they are still in the natal cloud can lead to strong ambient force which reduces the effective outward momentum deposition, slowing the shell expansion and delaying its dissolution. 
{The effects of shell structure, which manifest distinctly in different density regimes – retardation due to external pressure in low-density cases and Ly$\alpha$ radiation pressure in higher-density cases – are especially pronounced for lower metallicity models. This results in non-monotonic trends in the evolutionary time, $t_{\mathrm{evo}}$, with cloud density.}

\subsubsection{Cloud mass}
Increasing $M_{\mathrm{cl}}$ at a fixed $n_{\mathrm{cl}}$ increases its binding energy, which varies as $M_{\mathrm{cl}}^{2}$. 
Generally, at a fixed $\epsilon_{\mathrm{SF}}$ and $n_{\mathrm{cl}}$, more massive clouds generally take longer to dissolve. This effect is clear in the contours of $t_{\mathrm{evo}}$.

For the low metallicity, low-density models, increasing the cloud mass generally increases $\mathcal{R}_{\mathrm{Ly\alpha}}$. This is due to the increasing {atomic} gas columns with increasing cloud mass ({c.f. Fig.~\ref{fig:NHI_Z004}}) and the increase in $M_{\mathrm{F}}$ whose saturation values tend to lie at higher $N_{\mathrm{H}}$ than those encountered in this regime.
At high cloud densities, increasing the cloud mass tends to have a non-monotonic effect on $\mathcal{R}_{\mathrm{Ly\alpha}}$. At high cloud densities, the shell neutral gas columns can exceed those at which $M_\mathrm{F}$ saturates. As noted above, massive clouds take longer to dissolve or in some cases contain multiple generations. The late dissolution leads to higher contributions by the $F_{\mathrm{w,sn}}$ increasing $\mathcal{R}_{\mathrm{w, sn}}$, as can be gauged from Fig.~\ref{fig:template_R_Lya}.
Increasing the metallicity decreases the $N_\mathrm{H}$ value at which $M_\mathrm{F}$ saturates, leading to the downward shift of the peak of $\mathcal{R}_{\mathrm{Ly\alpha}}$. 

Increasing $M_{\mathrm{cl}}$ at a fixed $n_{\mathrm{cl}}$ generally tends to increase the contribution of $F^{\mathrm{UV,IR}}_{\mathrm{rad}}$ to the overall outward momentum deposition due to increasingly large dust columns encountered with higher $M_{\mathrm{cl}}$.  

\subsubsection{Star formation efficiency}

Fig.~\ref{fig:R_contours_fix_n} shows the effect of changing $\epsilon_{\mathrm{SF}}$ on the quantities discussed here while keeping the cloud density fixed at $320~\mathrm{cm^{-3}}$. As expected, the increase in the amount of stellar mass relative to the cloud mass leads to a faster dissolution of the clouds across all metallicities. 

The quantities that scale with stellar mass increase with increasing $\epsilon_{\mathrm{SF}}$ at fixed $M_{\mathrm{cl}}$. These include the force due to winds and SNe, and the rate of production of ionizing photons. At the lower end of $M_{\mathrm{cl}}$, increasing $\epsilon_{\mathrm{SF}}$ results in faster-moving, rapidly thinning shells. This results in an increase in the relative contribution of the momentum deposition by the winds and the SNe with respect to the Ly$\alpha$ radiation pressure which is reflected by the increase in $\mathcal{R}_{\mathrm{w,sn}}$, and a decrease in $\mathcal{R}_{\mathrm{Ly\alpha}}$ at low cloud masses with increasing $\epsilon_{\mathrm{SF}}$. 
At the higher metallicity end, the above-mentioned increase in $\mathcal{R}_{\mathrm{w,sn}}$ is slightly offset by increased contribution by $F^{\mathrm{UV,IR}}_{\mathrm{rad}}$.
At the higher cloud mass end, the increase in $\epsilon_{\mathrm{SF}}$ does not lead to an appreciable change in the relative amount of momentum deposition by the various feedback channels in Fig.~\ref{fig:R_contours_fix_n}, this is reflective of the various forces showing a similar scaling with stellar mass. This is expected in this particular example where massive clouds of relatively dense gas are considered. In such cases, the shells are likely to possess the highest $M_{\mathrm{F}}$ values due to large gas columns, making $F^{\mathrm{Ly\alpha}}_{\mathrm{rad}}$ scale with the stellar mass. Note that the models at $\epsilon_{\mathrm{SF}} \leq 2.5\%$ harbour multiple generations of star clusters at the end of $t_{\mathrm{evo}}$ (cf. Fig.~\ref{fig:SFE_contours}, top panel).

The top panel in Fig.~\ref{fig:SFE_contours} shows the contours of the minimum star formation efficiency, $\epsilon_{\mathrm{SF, min}}$, required to dissolve a cloud of given mass and density. While similar values are obtained for the low-mass, low-density clouds across all metallicities, we find that the low-metallicity systems are able to destroy high-mass, high-density clouds with a lower burst stellar mass due to the Ly$\alpha$ radiation pressure. {At the higher density, intermediate cloud mass end of the low-metallicity systems, a distinct sequence of events emerges. The momentum-driven phase is initiated early when the shell is still within the cloud. Once the inner edge's density diminishes, shells with lower mass become ionized, having insufficient mass to stay radiation-bound. Confronted by the cloud's elevated external pressure and gravitational force, these shells decelerate and recollapse. Conversely, the more massive shells, which remain neutral during the phase transition, are propelled by the Ly$\alpha$ radiation pressure, resisting recollapse. This leads to a contour pattern seen at the highest cloud density and intermediate cloud mass values.}

In order to compare our results to those of \cite{2019MNRAS.483.2547R}, we ran tests {with} Ly$\alpha$ radiation feedback turned off at solar metallicity using a similar parameter space. We find a very good agreement between the $\epsilon_{\mathrm{SF, min}}$ values derived in that work and our results, the details of the comparison are given in Appendix~\ref{appendix:comparison_warpfield}. 

\begin{figure*}
\centering
   \includegraphics[width=.65\textwidth]{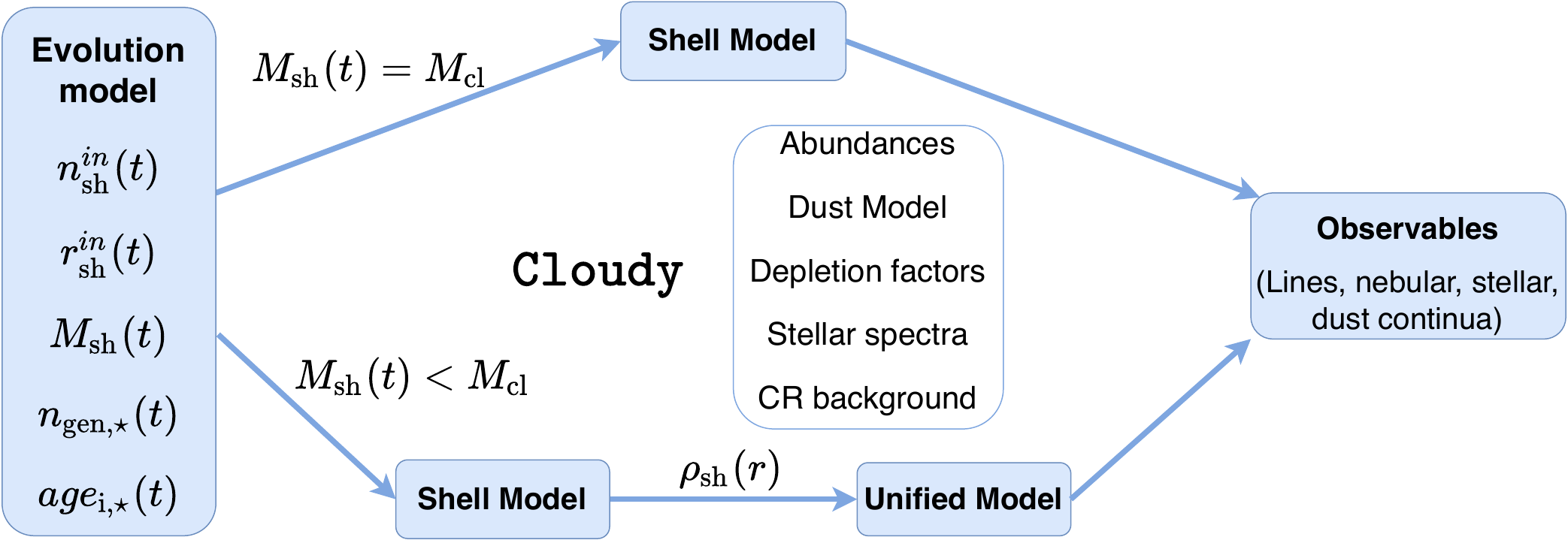} 
\caption{A schematic describing the kinds and ingredients of the \texttt{Cloudy} models generated for this work. The models' input parameters describing the density at the inner face, the radius of the shell, the mass of the shell, and the age and number of stellar clusters present come directly from the shell-evolution model.
Other parameters, such as the elemental abundances, metal depletion factors,  the dust model, and cosmic-ray spectra are also required for these models. Depending on the mass of the shell, one or two models are run in order to generate the observables. In the case where the shell has not overtaken the entire gas cloud, a second model which includes both the shell density profile and a constant density profile of the cloud is run.}
\label{fig:cloudy_flow_general}
\end{figure*}

Apart from this, we also calculate the star formation efficiency per free-fall time ($\epsilon_{\mathrm{ff}}$) for {recollapsing} models at $t_{\mathrm{evo}}$ as:

\begin{equation}
\epsilon_{\mathrm{ff}} = \frac{M_{\star} \left/ t_{\mathrm{evo}} \right.}{M_{\mathrm{cl}} \left/ \tau_{\mathrm{ff}} \right.}
, \quad \text{where} \quad
\tau_{\mathrm{ff}}=\sqrt{\frac{3 \pi}{32 , G , \rho_{\mathrm{cl}}}}
\end{equation}
\label{eqn:freefall_SFE}

is the gravitational free-fall time,  {and $M_{\star} (t_{\mathrm{evo}})$ is the total stellar mass in the system at $t_{\mathrm{evo}}$. This quantity 
is a measure of the star formation rate relative to the maximum rate dictated by gravity, thus representing the opposition offered by stellar feedback}.
The contours of this quantity for $Z=0.02$ and a selected set of $n_{\mathrm{cl}}$ are shown in the bottom panel of Fig.~\ref{fig:SFE_contours}.

The recollapsing models exhibit $0.20 \% \leq \epsilon_{\mathrm{ff}} \leq 1.65 \%$, which encompasses the range of values exhibited by star formation in giant molecular clouds of nearby galaxies \cite[][$0.40 \% \leq \epsilon_{\mathrm{ff}} \leq 1.10 \%$]{2018ApJ...861L..18U}.

\section{Post processing and library generation }
\label{cloudy_methods_library_gen.sec}

The output from the evolutionary model is passed on to the photo-ionization code, \texttt{Cloudy}\footnote{Available at \url{https://gitlab.nublado.org/cloudy/cloudy/-/tree/master}, \\  Commit SHA: \texttt{69c3fa5871da3262341910e37c6ed2e5fb76dd3c}} \citep{2017RMxAA..53..385F} in the second step. This allows us to produce various observables following the transition from $\mathrm{\HII}~\text{to}~\mathrm{H\textsc{i}}~\text{to}~\mathrm{H_{2}}$ regions while self-consistently accounting for gas, dust, and molecular microphysics.

We use a closed, spherical geometry for all the models generated using \texttt{Cloudy}. The data required for running \texttt{Cloudy} models is summarized in Fig.~\ref{fig:cloudy_flow_general}.
 As the shell expands and sweeps its birth cloud, two gas configurations arise: 
\begin{enumerate*}
\item Shell embedded within the cloud, or,  
\item shell has swept the entire cloud.
\end{enumerate*} 
During this post-processing, the shell's density profile is derived within \texttt{Cloudy} assuming a hydrostatic equilibrium. The density calculation starts from the inner face of the shell and, therefore, requires the initial density condition. This is given using Eqn.~\eqref{eqn:shell_IC_density}.
{It's worth noting that
We rely on \texttt{Cloudy} to compute the density profile with detailed chemical and thermal calculations instead of using the approximate profiles computed in Sec.~\ref{sect:shell_structure}, although, in both cases, hydrostatic equilibrium is assumed.} 

In order to limit the parameter space of our models, we do not account for the turbulent and magnetic pressures in the shell models. In the cases where the unswept cloud is present, the stellar spectra go through initial processing due to the shell, followed by subsequent processing of the shell output due to the unswept cloud beyond the shell. 
We use the \texttt{dlaw table} command in \texttt{Cloudy}, which allows the code to use arbitrary density values as a function of radius. Each of the cases where the shell is embedded in the birth cloud involves two \texttt{Cloudy} simulations:
\begin{enumerate*}
\item Shell only simulation: to get the density structure of the shell, 
\item Unified shell and cloud simulation: We use the \texttt{dlaw table} command to input an overall density structure for the shell-cloud system.  In such cases, the density structure obtained from the shell-only simulation is augmented with a constant cloud density profile at radii beyond the shell. We assume a transition length equal to $10\%$ of the shell depth.
\end{enumerate*} 
 {We note one could model the unswept cloud using the transmitted continuum from the shell as input SED; however, this is problematic as \texttt{Cloudy} lacks knowledge of the shell model's optical depth effects, which can result in inaccuracies \citep[][private communication]{Priv_Comm_2Shell_Model}.}

In the case where the entire cloud has been swept into the shell, only a single simulation is required.
The stellar data is consistent with that used for the shell-evolution model and is discussed in Sec.~\ref{subsect:stellar_data}. 
Only a mass-based stopping criterion is employed for all models, i.e., the radial extent of the model is determined by the density structure and the total mass of the gas.

\subsection{Stellar evolutionary tracks and spectra}\label{subsect:stellar_data}

The current work makes use of the high mass-loss Geneva tracks \citep{1994A&AS..103...97M} in \texttt{STARBURST99} population synthesis code \citep{1999ApJS..123....3L}. These do not consider binary population or stellar rotation.
The reasoning behind this choice is two-fold, \begin{enumerate*}

 \item They offer a better sampling of the metallicity. The newer models, including the ones considering stellar rotation \citep[see][]{2014ApJS..212...14L} are only available for two metallicities, $Z=0.002 \mathrm{~and~} 0.014$ ($Z_{\odot} = 0.014$), whereas, the older ones are available for five metallicities, $Z=0.001,~0.004,~0.008,~0.02,\mathrm{~and~} 0.04$ ($Z_{\odot} = 0.02$). 
 
    \item For the instantaneous burst models considered in this work, the high mass-loss rates produce a better agreement with observational data when comparing emission-line diagnostics \citep{2010AJ....139..712L}. Thus, the other set of the ``standard'' mass-loss tracks is not used.
\end{enumerate*}  

\begin{figure*}
    \centering 
    \subfloat{\includegraphics[width=.35\textwidth]{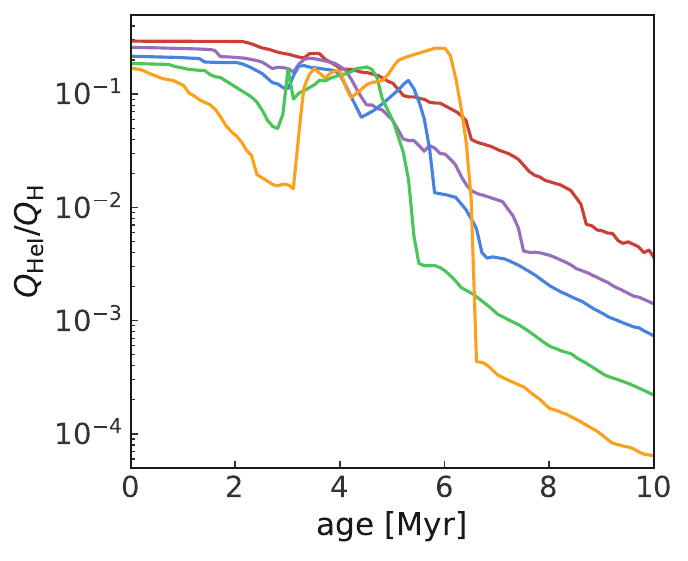}}
    \quad  
    \subfloat{\includegraphics[width=.35\textwidth]{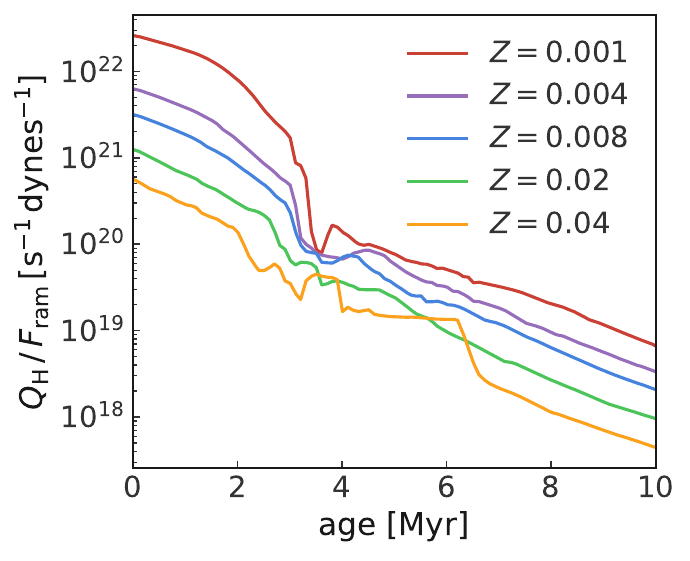}} 
    \caption{Evolution of stellar cluster properties that affect the emission from the surrounding gas as a function of metallicity. The properties assume a coeval population of stars having a well-sampled IMF (see Sec.~\ref{subsect:stellar_data}).
    Left: The ratio of the rate of helium ionizing photons to that of hydrogen ionizing photons. This serves as a proxy for spectral hardness. This value decreases as the OB stars die but it 
    is affected by the presence of W-R stars. The extreme mass loss suffered by W-R stars can expose their stellar cores and 
    significantly harden the ionizing spectrum of the stellar populations. The highest metallicity systems shown here show the highest spectral hardening driven by W-R stars. Right: The ratio of the ionizing photons to the compressive force at the inner face of the shell, this ratio scales with the ionization parameter during the momentum-driven phase due to the density condition given by Eqn.~\eqref{eqn:shell_IC_density}.}
    \label{fig:stellar_feedback_spectralHardness_logU}
\end{figure*}

 The spectra used for \texttt{Cloudy} models employ \texttt{STARBURST99}’s Pauldrach/ Hillier model atmospheres, which use the WMBASIC wind models of \cite{2001A&A...375..161P} for younger ages when O stars dominate the luminosity ($<3$ Myr), and the CMFGEN \cite{1998ApJ...496..407H} atmospheres for later ages when W–R stars are dominant. 
A Kroupa initial mass function (IMF) between $0.1-100~M_{\odot}$, with a power law break at $0.5~M_{\odot}$ has been employed. The power-law exponent for the lower mass end is $1.3$, while for the higher mass end is $2.3$.
The higher mass threshold is consistent with the Auriga \citep{2017MNRAS.467..179G}, EAGLE \citep{2015MNRAS.446..521S} , and Illustris-TNG \citep{2018MNRAS.473.4077P} models, although they rely on the Chabrier IMF \citep{2003PASP..115..763C}.
The results would not change appreciably if the IMFs were interchanged as we are only concerned with the early evolution driven by massive stars, whose number does not differ between these IMFs.
All other parameters are set to the default values recommended on the \texttt{STARBURST99} webpage\footnote{\href{https://www.stsci.edu/science/starburst99/docs/default.htm}{\texttt{www.stsci.edu/science/starburst99/docs/default.htm}}}.

The left panel in Fig.~\ref{fig:stellar_feedback_spectralHardness_logU} shows the spectral hardness by means of the ratio of the rate of helium ionizing photons (first ionization, 24.6 eV) to the Hydrogen ionizing photons. The right panel is the ratio of the rate of hydrogen ionizing photons to that of the compressive force on the shell, $F_{\mathrm{ram}}$. This ratio scales with the ionization parameter, $U$, during the momentum-driven phase, as discussed in Sec.~\ref{sect:BPT diagram and parameters}.

\subsection{Chemical and dust abundances}
\label{ChemicalAbund.subsect}
We use the solar abundance set from \cite{2010Ap&SS.328..179G} (The \texttt{GASS} abundance set in \texttt{Cloudy}). 

We scale the abundances according to the metallicity value of the stellar templates using the \texttt{GASS} value of $Z_{\odot}=0.014$.
The abundance of some specific elements is modified. For helium, we use the relation given in \cite{2002ApJ...572..753D}
\begin{equation}
    \text{He}/\text{H} = 0.0737 + 0.024\,(Z/Z_{\odot}) \, .
    \label{He_abund.eqn}
\end{equation}
For carbon and nitrogen, we use the prescription described in \cite{2013ApJS..208...10D}, interpolating between the values in Tab.~3 of that work.
The depletion factors used here are the ``classic'' \texttt{Cloudy} set (refer to the \texttt{Cloudy} documentation).
We note that the depletion factors employed are independent of the metallicity of the system. Finally, we note that the aforementioned modifications to the abundances are done before applying the depletion factors.
 
The dust model employed in this work specifies the graphite and silicate grains with size distribution and abundance appropriate for those along the line of sight to the Trapezium stars in Orion. The grain population is modeled by a  power law of index $-3.5$, resolved by ten bins running from $0.03-0.25~\micron$ 
The Orion size distribution is deficient in small particles and so produces relatively grey extinction. 
The polycyclic aromatic hydrocarbons (PAH) are added to the dust model as a fixed fraction of the dust mass, scaled by the H1 abundance at a given location in the nebula. This is based on the assumption that PAHs are present only in PDRs, i.e., assuming PAH destruction in ionized parts of the gas cloud, while depleting onto larger grains in molecular regions. The maximum possible value for the PAH to dust mass fraction ($q_{\mathrm{PAH}}$) is taken to be the same as the Galactic diffuse dust value of $4.6\%$ \citep{2001ApJ...554..778L, 2001ApJ...548..296W}.
The PAH population is a power law of index $-3.5$, resolved by ten bins in the range of $4.3-11\,\text{\AA}~(\mathrm{30-500~C~atoms})$.
The grain abundances are scaled along with the stellar template metallicity.
Further details about the dust modeling in \texttt{Cloudy} can be found in \citet{2004ASPC..313..380V}, \citet{2008ApJ...686.1125A}, and references therein. {We note that while there is evidence of a decline in dust-to-metal ratio at low metallicities on galaxy-wide scales and on smaller scales \citep{2014A&A...563A..31R, 2018ApJ...865..117C}, we adopt a constant dust-to-metal ratio in our study. This simplification is supported by the idea that regions of active star formation, where our study primarily focuses, should have a dust content that is relatively enhanced compared to the broader galactic environment \citep{2022MNRAS.509L...6P}. Considering the limitations of current observations, it's reasonable to assume a constant dust-to-metal ratio in these star-forming regions.}

Finally, for the highest metallicity models ($Z = 0.02,~0.04$), we employ a cosmic ray abundance of $2\times,~5\times$ the Galactic background value when the shell has not fully swept the natal cloud. This is done for the stability of the chemical network in the cases where the simulation goes deep into the molecular region.

\subsection{Time sampling}
\label{subsect:time_sampling_cloudy_runs}
The temporal sampling of the library represents a balance between the number of \texttt{Cloudy} models and resolving major changes in the physical conditions of the system. For the sake of simplicity,  we employ a single time grid for the generation of a look-up table used by \texttt{SKIRT}.
About $15\%$ of models in our parameter space do not dissolve till the end of the $t_{\mathrm{evo}}=30~\mathrm{Myr}$. Based on this we fix the endpoint of the template library, $t_{\mathrm{lib}}=30~\mathrm{Myr}$. 
On the other hand, nearly $80\%$ of the collapse events in our models occur before $15~\mathrm{Myr}$. Keeping that in mind, we deploy a higher number of the points in the period between $0-15~\mathrm{Myr}$.
For each of the models, we use $90$ time points between $0-30~\mathrm{Myr}$ to resolve the evolution of the system, where $60$ points are uniformly deployed in the first $15~\mathrm{Myr}$, and the rest are distributed uniformly in the last $15~\mathrm{Myr}$. This gives us a temporal resolution of $0.25~\mathrm{Myr}$ in the first half of $t_{\mathrm{lib}}$, and $0.5~\mathrm{Myr}$ in the subsequent half.
At each of these time steps, we run the \texttt{Cloudy} model/s as previously described to get the observables.
For all the time steps where the shells have dissolved, stellar spectra without any gas/dust reprocessing are added to the look-up table.

\section{\HII region diagnostics: Emission line ratios and IR colors}
\label{sect:observables from individual models}

\begin{figure*}
    \centering    \includegraphics[width=.7\textwidth]{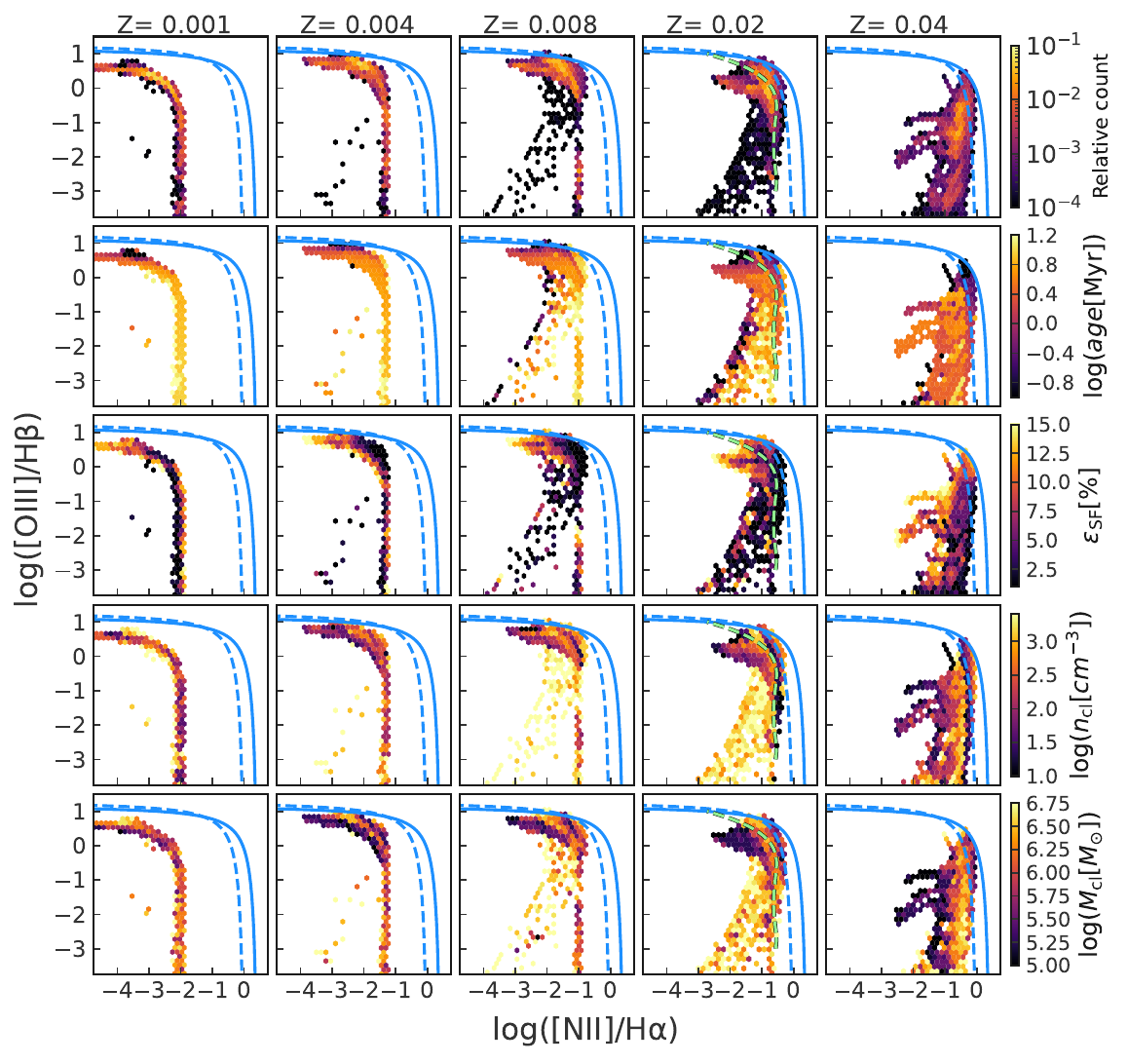}
    \caption{The BPT diagram with the colormaps representing the relative number of models in a given bin (top row) and median values of the primary variables (second row onward). The columns represent the different metallicity of the models as listed in the first row. In all panels, the solid black curve shows the demarcation between starburst galaxies and AGN defined in \citet{2001ApJ...556..121K}, the dashed black curve shows the revised demarcation from \citet{2003MNRAS.346.1055K} and the green curve {shown at $Z=0.02$} is a polynomial fit to the \HII regions' line ratios obtained by \citet{2018MNRAS.477.4152R} for the nearby galaxy NGC-628.}
\label{fig:Hii_region_BPTratios_Nii_Oiii_primary}
\end{figure*}

As a means to highlight the parameter space of the model observables, we focus on two \HII region diagnostics. Firstly, the BPT diagram using the $\text \NII\lambda 6584 / \mathrm{H} \alpha \text {~and~\OIII}\lambda 5007 / \mathrm{H} \beta  $ emission line ratios, and secondly, the dust emission continuum arising from the \HII regions by using color-color diagram making use of the four IRAS bands centered at $12,~25,~60,~100~\mu m$. 
For each of these diagnostics, we discuss the impact of the evolutionary model's free parameters (primary variables) and the connection with other parameters that result from either the evolutionary model, like the shell's radius or velocity, or as a result of the \texttt{Cloudy} post-processing, like PAH to gas fraction or dust temperature. We refer to the latter as secondary variables.

\subsection{BPT diagram}
\label{sect:BPT diagram and parameters}
\begin{figure*}
    \centering
\includegraphics[width=.7\textwidth]{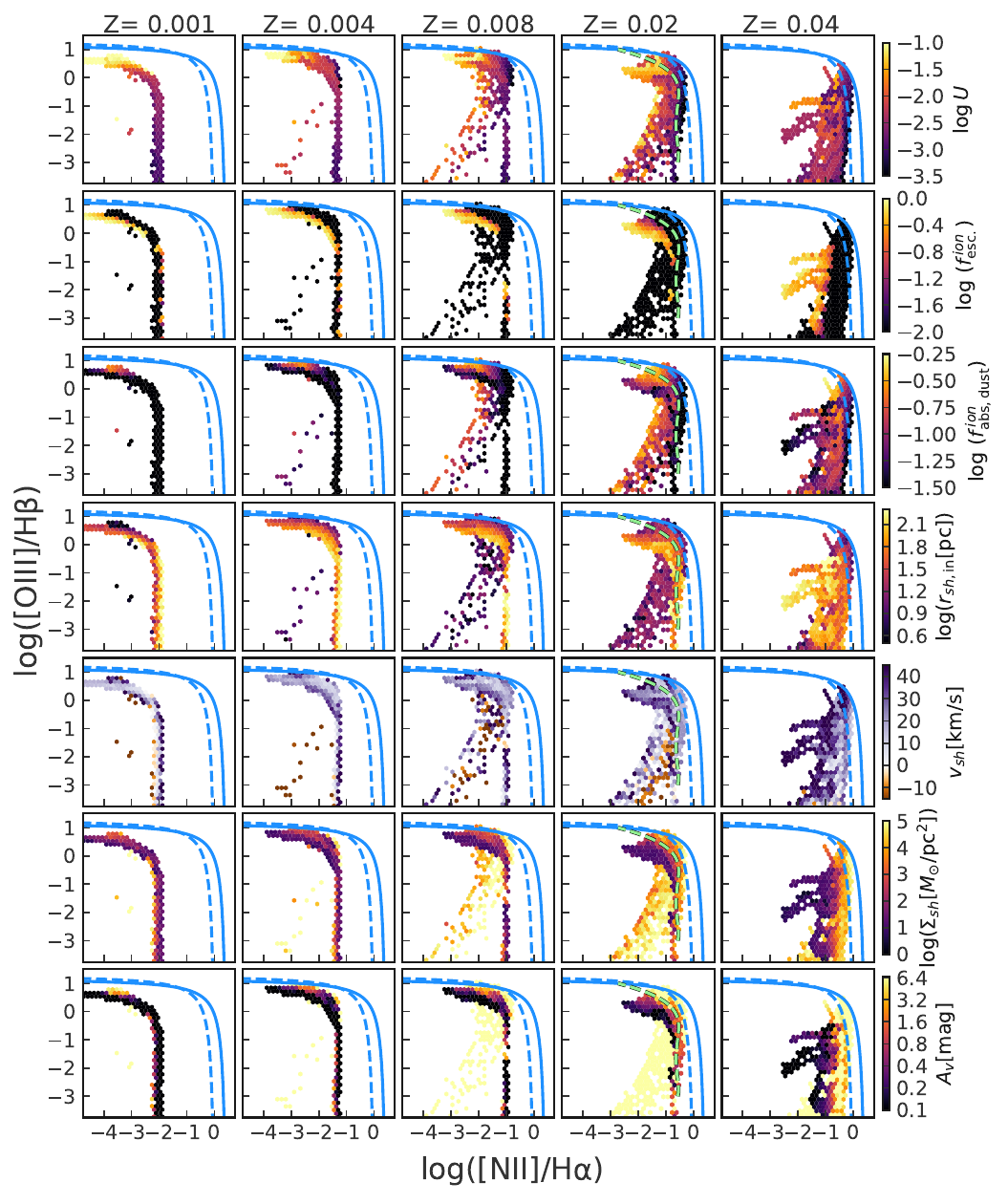}
    \caption{The BPT diagram with the colormaps representing the median values of the secondary variables. The columns represent the different metallicity of the models as listed in the first row. }
    \label{fig:Hii_region_BPTratios_Nii_Oiii_secondary}
\end{figure*}

The classic BPT diagram \citep{1981PASP...93....5B} has been extensively used to classify objects based on the emission excitation mechanism. The $\NII / \mathrm{H} \alpha \text {--}\OIII / \mathrm{H} \beta$ BPT diagram exploits the fact that N and O have similar second ionization potentials, and the proximity of \NII$\lambda 6584$ and \OIII$\lambda 5007$ to the hydrogen recombination lines $\mathrm{H} \alpha$ and $\mathrm{H} \beta$.  This allows $\mathrm{O}^{++}$ to serve as a proxy for $\mathrm{N}^{++}$, implying that any increase in the abundance of $\mathrm{O}^{++}$ levels must come at the expense of $\mathrm{N}^{+}$ abundance, thus shedding light on the photo-ionizing source's spectral properties. 
The spectral proximity of \OIII and \NII lines to the hydrogen recombination lines makes these ratios almost completely unaffected by the dust effects exterior to the ionized region.

Traditionally, the BPT diagram is populated by grids of \HII models with varying metallicity, stellar cluster age, and ionization parameter. The ionization parameter is generally defined as the value at the inner edge of the nebula and is  given as: 
\begin{equation}
U \equiv \frac{Q_{\mathrm{H}}}{4 \pi R^2  ~n_{\mathrm{H}}~c} \, ,
\end{equation}
where $Q_{\mathrm{H}}$ is the total ionizing photon rate incident on the inner edge. For a given density, $U$ acts as a normalization for a given ionizing spectrum shape and combines the intensity of the ionizing source, the density, and the geometry of the gas cloud.
As $U$ folds three physical parameters in it, each can be modified by keeping the rest constant, e.g., \cite{2001A&A...365..347M, 2010AJ....139..712L, 2017ApJ...840...44B}.
The grids generally assume no correlation among the grid parameters. In contrast, the current work introduces relations between the quantities listed above by connecting cluster evolution and the state of the gas around it through stellar feedback and gravity.

Fig.~{\ref{fig:Hii_region_BPTratios_Nii_Oiii_primary}} shows how the models populate the $\NII / \mathrm{H} \alpha \text {--}\OIII / \mathrm{H} \beta$ plane as a function of the primary variables.
The top row in this figure shows the histogram for our models in the BPT diagram space.  Also shown as the green dashed curve is the polynomial fit to H$\alpha$ spaxels found by \cite{2018MNRAS.477.4152R}.
Examining the locus of the most frequent data points in this plot as a function of metallicity allows us to see the dual-valued nature of the $\OIII / \mathrm{H} \beta$ ratio \citep{2005ApJ...631..231P, 2008ApJ...681.1183K, 2017ApJ...840...44B}, which is driven by the fact that this ratio is a function of the oxygen abundance, but also of the temperature of the gas and the spectral hardness of the ionizing source. While increasing the metallicity of the system boosts the amount of oxygen, it also lowers the equilibrium temperature. 
This leads to the lower $\OIII / \mathrm{H} \beta$ ratio at the higher end of the metallicity.
In contrast, on the lower end, oxygen abundance is the limiting factor when it comes to oxygen emission. On the higher end of the metallicity, it is noteworthy that the amount of ionizing photons is lower due to line-blanketing in stellar atmospheres and the spectra are softer except for the W-R phase (see fig.~\ref{fig:stellar_feedback_Fram_and_Qion}). 
$\NII / \mathrm{H} \alpha$, on the other hand, 
does not show a double-valued trend with respect to the metallicity, and the locus of the most frequent data points tends to shift rightwards with increasing metallicity.

\begin{figure*}
    \centering
 \includegraphics[width=.9\textwidth]{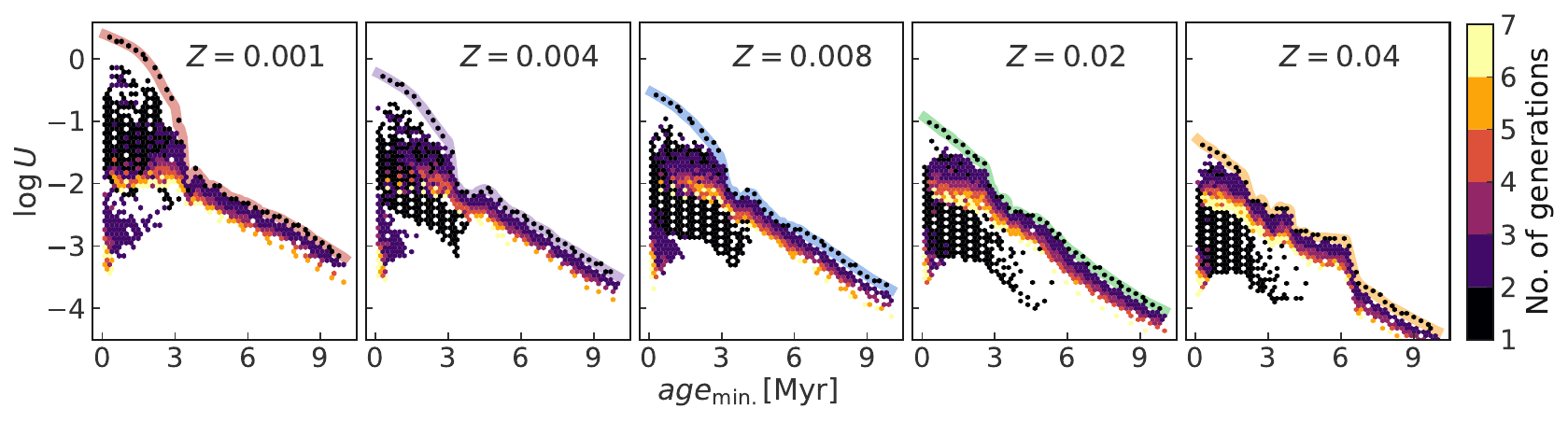}
    \caption{$U$ at the {shell inner edge} as a function of the age of the youngest cluster for all the models in our parameter space. The shape of the curve enveloping the hexbin plot in each panel and its color is the same as that shown in the right panel of Fig.~\ref{fig:stellar_feedback_spectralHardness_logU}, but scaled as $U=\frac{\mu_{\mathrm{n}} k T_{\mathrm{i}}}{\mu_{\mathrm{p}}  c} \frac{Q_{\mathrm{H}}} {F_{\mathrm{ram}}} $ using the shell's inner face density condition during the momentum driven phase (Eqn.~\eqref{eqn:shell_IC_density}).} \label{fig:logU_vs_age_of_the_youngest}
\end{figure*}

The second row in Fig.~{\ref{fig:Hii_region_BPTratios_Nii_Oiii_primary}} shows the line ratios for our models as a function of the age of the system. Broadly speaking, the systems move from high to low values of the $\OIII / \mathrm{H} \beta$ ratio with the fall in $U$ \footnote{{In this work, $U$ is defined at the shell's inner edge. We don't use the spherical ionization parameter, evaluated at the Str{\"o}mgren radius. In \texttt{Cloudy}, this corresponds to where the neutral Hydrogen fraction is 0.5 for radiation-bounded gas or the cloud's edge for density-bounded gas, requiring photoionization calculations.}} with age. The BPT diagram is populated as a function of $U$ in the first row of Fig.~{\ref{fig:Hii_region_BPTratios_Nii_Oiii_secondary}}. The decreasing $U$ leads to an increase in the $\NII / \mathrm{H} \alpha$ with a saturation based on the metallicity.
As mentioned previously, $U$ has three parameters in it, the flux of the ionizing photons, the shell's radius, and the gas density. Given Eqn.~\eqref{eqn:shell_IC_density}, $U$ for our models wraps together the gas compression due to the stellar feedback and the cluster's ionizing flux. The gas compression depends on whether the expansion is pressure-driven or momentum-driven.
Following \cite{Dopita_2006}, one could write a scaling for $U$ during the pressure-driven phase as:
\begin{equation}
    U(t) \propto~\delta(t)^{3/2} \frac{\sum Q_{\mathrm{H}}(t)}{\sum {{{L}}}_{\mathrm{mech}}(t)} \, ,
\label{eqn:U_pressure_driven}
\end{equation}
$\delta(t)$ is the instantaneous ratio of the shell's inner face density to the cloud density. The presence of this factor in the above equation leads to a coupling between $U$ and both the cloud density and the mass of the central cluster/s.
In comparison, $U$ during the momentum-driven phase is given as: 
\begin{equation}
U(t) \propto \frac{\sum Q_{\mathrm{H}}(t)} {\sum F_{\mathrm{ram}}(t)} \, .
\label{eqn:U_momentum_driven}
\end{equation}
The summation in Eqn.~{\eqref{eqn:U_pressure_driven}} and {\eqref{eqn:U_momentum_driven}} are due to the possibility of shell recollapse in our models, accounting for the feedback and ionizing flux from all the generations present. 
In the case of systems containing only a single generation of stars, Eqn.~\eqref{eqn:U_pressure_driven} shows a $U\propto {n_{\mathrm{cl}}}^{-1/5}{(\epsilon_{\mathrm{SF}}M_{\mathrm{cl}})}^{1/5}$ scaling.
On the other hand, $U$ in the case of the momentum-driven phase and for a system containing a single stellar population is free from additional dependencies on stellar mass and cloud density at a given time and metallicity. {$U$ also shows a positive correlation with the fraction of ionizing photons absorbed by dust ($f^{\rm{ion}}_{\rm{abs,\,dust}}$) in \HII regions \citep[for example,][]{2002ApJ...570..688I, 2009A&A...507.1327H}. This quantity is shown in the third row of Fig.~\ref{fig:Hii_region_BPTratios_Nii_Oiii_secondary}. 
In our models, at the higher end of \( U \) for metallicity values \( Z=0.02 \) and \( Z=0.04 \), up to \( 52\,\% \) of the ionizing photons can be absorbed by dust. We mention that \texttt{Cloudy} does not directly report \( f^{\rm{ion}}_{\rm{abs,\,dust}} \). To estimate this quantity, we adopt the method outlined by \cite{2011ApJ...732..100D}. An explanation of this method can be found in Appendix~\ref{appendix:Draine_fion}.}

The third row in Fig.~{\ref{fig:Hii_region_BPTratios_Nii_Oiii_primary}} shows the line ratios as $\epsilon_{\mathrm{SF}}$ is varied. This parameter controls the amount of stellar mass relative to the cloud mass that needs to be pushed by the stellar feedback. Due to the dependence on stellar mass, $U\propto M^{1/5}_{*}$ during the initial pressure-dominated phase, all other parameters kept the same, systems with higher $\epsilon_{\mathrm{SF}}$ possess somewhat higher $U$ values and thus higher values of $\OIII / \mathrm{H} \beta$. Furthermore, $\epsilon_{\mathrm{SF}}$ determines the feedback strength and, therefore, the number of generations present in the system at a given time. We discuss the impact of the presence of multiple stellar generations on $U$ along with the discussion on cloud density and mass below.


Another interesting impact of varying $\epsilon_{\mathrm{SF}}$ arises due to the presence of leaky \HII regions. Models with low density ($ < 320~\mathrm{cm^{-3}}$), low mass ($<10^{5.75}M_{\odot}$), and high $\epsilon_{\mathrm{SF}}$ ($\geq10\%$) exhibit a tendency to sweep the entire cloud and lower the shell column densities rapidly ($t < 5\mathrm{Myr}$ for $Z=0.02$) while the central cluster continues producing significant ionizing photons. This results in shells with low surface density, susceptible to ionizing radiation leakage, leading to reduced $\NII / \mathrm{H} \alpha$ values. The absence of intervening gas to soften the spectra diminishes \NII production. These models occupy a distinct region on the BPT diagram, especially prominent for higher metallicity systems when $Z \geq 0.02$. {The Hydrogen ionizing radiation's escape fraction\footnote{This is calculated using incident and transmitted source SED in the range 1--1.8 $E_{\rm{H}}$, where $E_{\rm{H}}=13.6\,\rm{eV}$.} and the shell surface densities, as demonstrated in Fig.~\ref{fig:Hii_region_BPTratios_Nii_Oiii_secondary}, further elucidate this behavior.}
For $Z=0.02$, these models partially span the region defined by $-2.5<\log(\NII / \mathrm{H} \alpha)<-1$ and $-0.5<\log(\OIII/ \mathrm{H} \beta)<0.5$. For $Z=0.04$, this region is $-2.5<\log(\NII / \mathrm{H} \alpha)<-1$ and $-2.4<\log(\OIII/ \mathrm{H} \beta)<-0.6$. Analogous models with lower metallicity reside approximately within the regions given by $\log(\NII / \mathrm{H} \alpha)<-2, -2.5, -3$ and $\log(\OIII / \mathrm{H} \beta )> 0$ for $Z=0.008,~0.004,~0.001$ respectively. It's noteworthy that the rapid thinning of the shell, driven by the intensity of stellar feedback, offers an additional avenue by which stellar feedback modifies the line ratios, beyond the influence exerted by $U$.

In contrast, low $\epsilon_{\mathrm{SF}}$ systems with elevated density and mass tend to harbor multiple generations of stars with their shells still embedded within the unswept cloud. This characteristic manifests in notably high $A_{\mathrm{v}}$ values for these models, as highlighted in the last row of Fig.~\ref{fig:Hii_region_BPTratios_Nii_Oiii_secondary}.

The influence of dust on line ratios in high $A_{\mathrm{v}}$ nebulae merits attention. Although the BPT diagram remains uninfluenced by a dust screen, additional effects are present in the case of high surface density shells that exhibit high $A_{\mathrm{v}}$ values. {Models with these characteristics are located in a distinct region of the BPT diagram. The primary reason for this positioning is the dominance of a secondary H$\alpha$ emission, originating from non-recombination contributions outside the \HII region, over the highly attenuated emission from within the \HII region itself. Consequently, there is an increase in the denominator of the line ratios under discussion. Further details are given in Appendix~\ref{appendix:line_ratios_of_highest_Av_objects}.} The high $A_{\mathrm{v}}$ branch seen prominently in the case of $Z=0.008,~0.02$ around $-3.0<\log (\NII / \mathrm{H} \alpha)<-1.5$ and $0.0<\log(\OIII / \mathrm{H} \beta)<-3.0$. At lower metallicities, the overall $A_{\mathrm{v}}$ is lower, while in the case of $Z=0.04$, they fall outside the range of the BPT diagram shown here. It is important to keep in mind that these embedded systems are unlikely to be observed in the optical due to the very high $A_{\mathrm{v}}$.

\begin{figure*}
    \centering
\includegraphics[width=.7\textwidth]{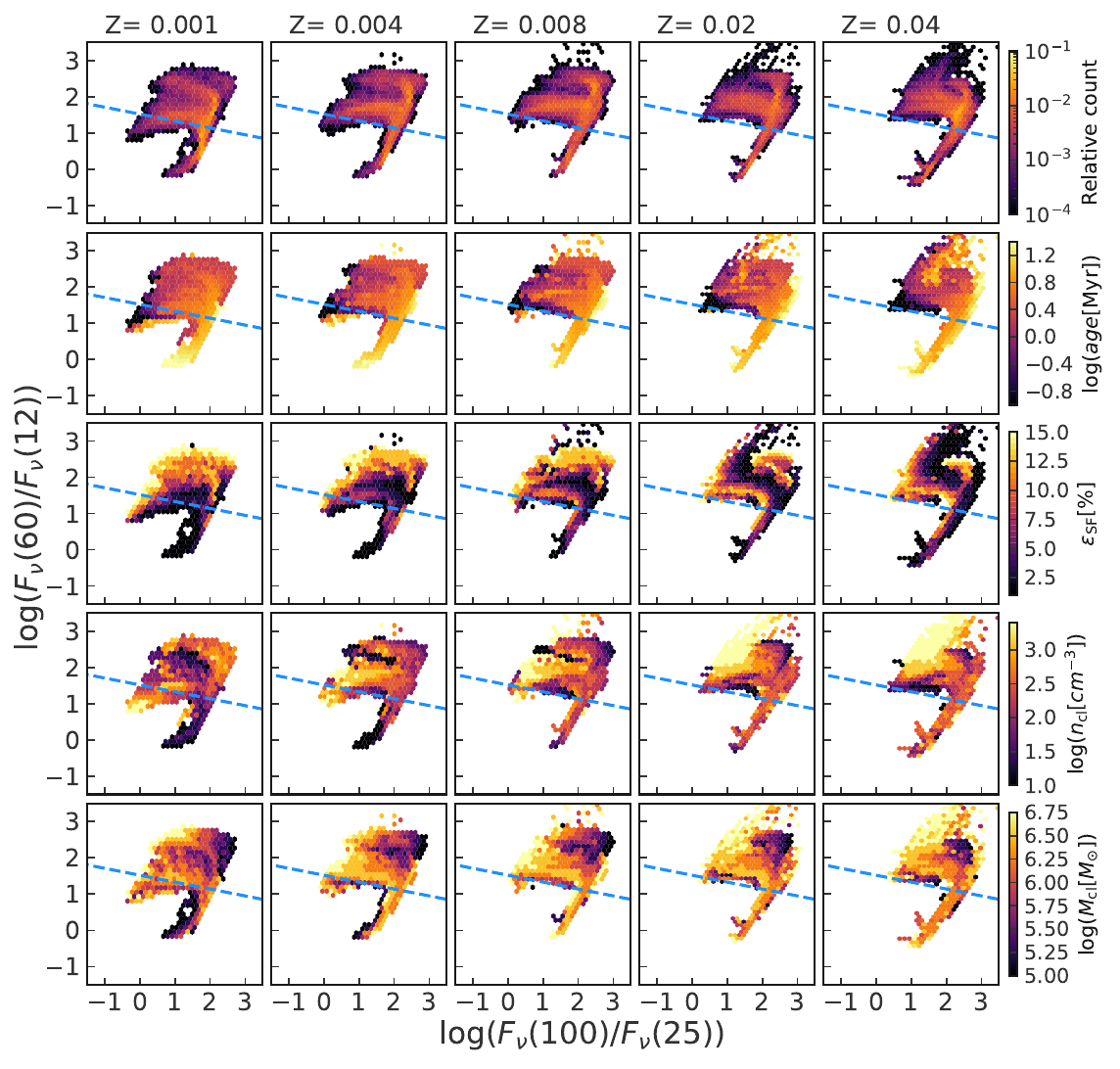}
    \caption{Same as Fig.~\ref{fig:Hii_region_BPTratios_Nii_Oiii_primary}, but using the IRAS colors. The dashed blue line in each panel marks the boundary derived by \citet{2018MNRAS.476.3981Y} distinguishing Milky-Way \HII regions from other sources.}
\label{fig:Hii_region_IRAScolors_primary}
\end{figure*}

\begin{figure*}
    \centering    \includegraphics[width=.7\textwidth]{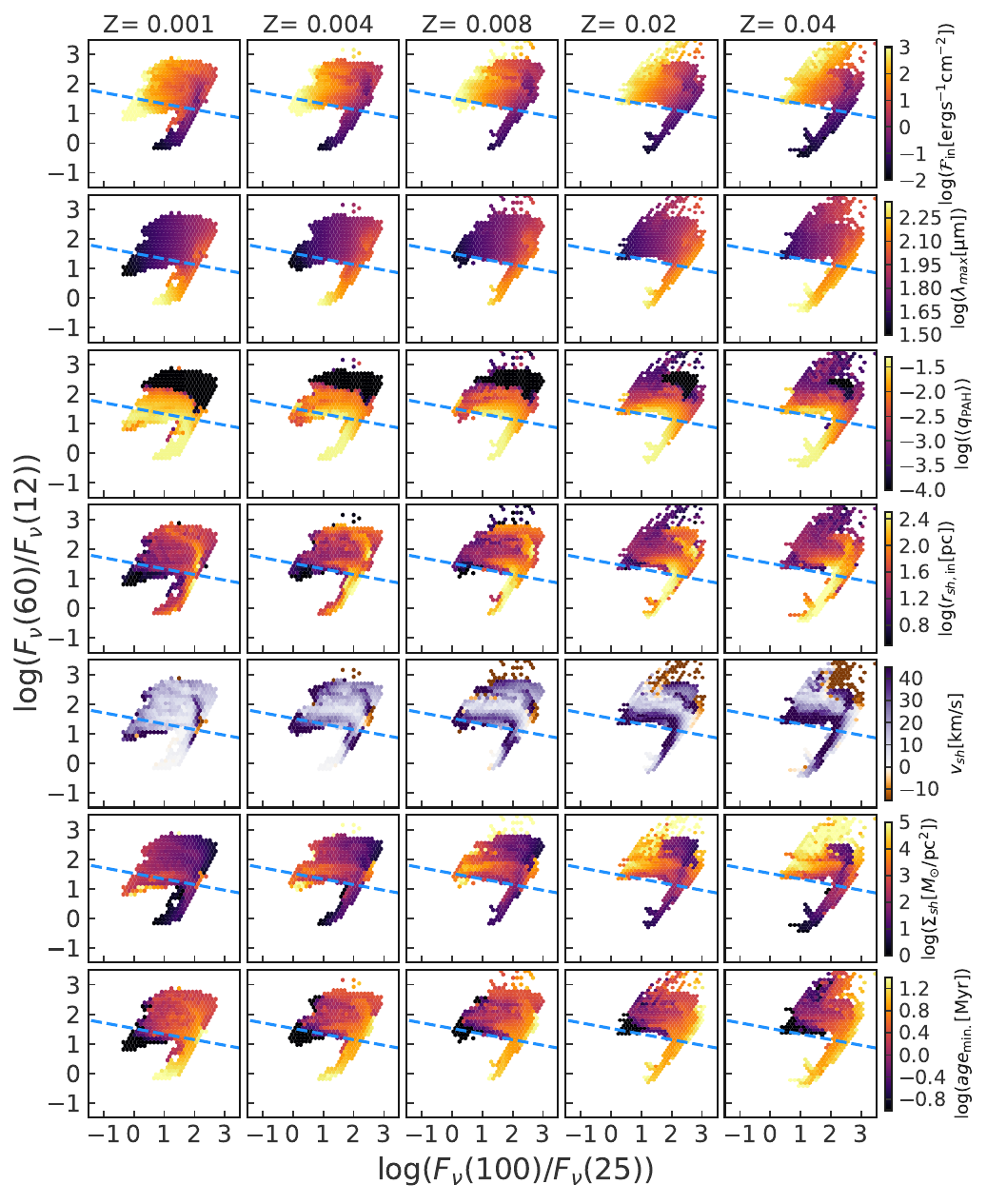}
    \caption{Similar to Fig.~\ref{fig:Hii_region_BPTratios_Nii_Oiii_secondary}, but using the IRAS colors and a different set of secondary variables. The top three rows are variables relevant to IR emission, with colormaps representing the median values of $\mathcal{F_{\mathrm{in}}}$, peak wavelength in the $20-1000~\micron$ range, and the model mass-weighted average value $q_{\mathrm{PAH}}$. The next four rows show the IRAS plane with colormaps for shell radius, velocity, mean surface density, and the age of the youngest stellar cluster.}
    \label{fig:Hii_region_IRAScolors_secondary}
\end{figure*}

The BPT diagram as a function of the cloud density and the cloud mass is shown in the fourth and fifth row of Fig.~{\ref{fig:Hii_region_BPTratios_Nii_Oiii_primary}}, respectively. Both of these parameters play a role in determining the gravitational binding energy of the system.
The cloud mass and density impact the position on the BPT diagram by determining the number of stellar generations present in the system and the time difference between star formation events.
The impact of the presence of multiple generations on $U$ could be understood by considering that the rate of production of ionizing photons for an instantaneous burst of stars decreases as the most massive stars die, while the mechanical luminosity and the ram force decrease less rapidly as they are sustained by the SNe (cf. Fig.~\ref{fig:stellar_feedback_Fram_and_Qion}). Thus, in the case of a system with multiple generations (if the generations have a sufficient difference in age), the youngest one contributes the most to the ionizing photons, while the contribution to the denominator can come from all generations, resulting in a lower $U$ in comparison to systems with a single population of the same age. 
The evolution of $U$ as a function of the number of generations and the age of the youngest cluster in the system is shown in Fig.~\ref{fig:logU_vs_age_of_the_youngest}. Systems with multiple generations tend to exhibit a lower $U$ based on the argument above.
The larger range of $U$ encountered during the first few years in  Fig.~\ref{fig:logU_vs_age_of_the_youngest} is due to the aforementioned dependence of $U$ on the stellar mass and cloud mass during the pressure-driven phase. On the other hand, the later evolution is momentum-driven where the range of $U$ is significantly reduced. In general, $U$ during the pressure-dominated phase is lower than that during the momentum-driven phase in our models. The highest possible $U$ is shown by the green curve enveloping the hexbin plots in Fig.~\ref{fig:logU_vs_age_of_the_youngest} and is given by 
$\frac{\mu_{\mathrm{n}} k T_{\mathrm{i}}}{\mu_{\mathrm{p}}  c} \frac{Q_{\mathrm{H}}(t)} {F_{\mathrm{ram}}(t)}$.
Given that the switch to the momentum-driven phase is faster for the densest clouds (cf. Fig.~\ref{fig:R_contours_fix_SFE}), they exhibit high $U$ values in our models.

{Based on Fig.~\ref{fig:logU_vs_age_of_the_youngest}, we also note that the $U$ values of our models are consistent with the range of $3\times10^{-4} \leq U \leq 3\times10^{-3}$ reported by \citet{2002ApJS..142...35K}. As mentioned in Sec.~\ref{sect:shell_structure}, this range of $U$ is not reproduced if only a pressure-driven expansion is employed \citep{2005ApJ...619..755D}.}

Finally, a few remarks could be made on the basis of a comparison between our models and the polynomial fit to NGC-628's \HII regions. The fit (going from high to low $\OIII / \mathrm{H} \beta$) follows an increase in the gas metallicity. We note that our models within the metallicity range $Z=0.008$ to $0.04$ appear to be consistent with this fit. The shell expansion velocity of these models falls in the range $5-15~\mathrm{km\,s}^{-1}$ along with an $A_{\mathrm{v}} < 3$.

\subsection{IRAS colours}
\label{sect:IRAS plane and parameters}

\begin{figure*}
    \centering
    \includegraphics[width=.9\textwidth]{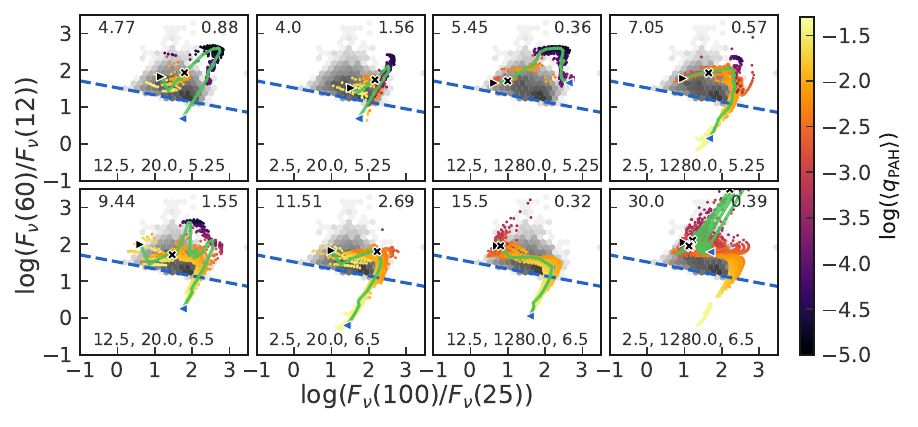}
    \caption{The time evolution of the models at the extremes of the parameter space, as discussed in Sec.~\ref{sect:IRAS plane and parameters}. In each sub-panel, the black triangle marks the start of the evolution, and the blue one represents the end. The dissolution time (in Myr) is given on the top left, a value of 30 corresponds to the shell not being dissolved during the period of evolution. The cross is the point when the shell switches from being pressure-driven to momentum-driven, which for the lowest density cases is also the moment when the shell has swept the entire cloud. The associated time (in Myr) is given on the top right. For systems undergoing multiple shell re-collapse events, only the first switch time is given. The text at the bottom specifies the model parameters in the order $\epsilon_{\mathrm{SF}}[\%],~n_{\mathrm{cl}}[\mathrm{cm^{-3}}],~\log(M_{\mathrm{cl}}/M_{\odot})$. The hexbins represent the histogram for Galactic \HII regions as given in \citet{2018MNRAS.476.3981Y}.}
    \label{fig:IRAS_plane_explanation}
\end{figure*}


In this section we consider the infrared regime to identify trends in our models. This serves as a complimentary diagnostic to the BPT diagram, especially for highly extincted sources. We use the four IRAS bands with wavelengths centered at $12,~25,~60,~100~\micron$ to populate the $F_{\nu}(60)/F_{\nu}(12)$ vs. $F_{\nu}(100)/F_{\nu}(25)$ color-color plane with our models. In the following, we refer to the color-color plane as the IRAS plane, while the flux densities in the bands are simply written as $F_{60},~F_{12},~F_{100}$ and $F_{12}$.
$F_{12}$ represents the PAH emission from the PDRs, and $F_{25}$ comes predominantly from the hot dust in ionized regions within the PAH-containing zones. $F_{60},~F_{100}$ track relatively cooler dust components of the shells. $F_{60}/F_{12}$ could be considered a proxy for $q^{-1}_{\mathrm{PAH}}$ and $F_{100}/F_{25}$ tracks the amount of cooler dust relatively to the hot one.

Figs.~{\ref{fig:Hii_region_IRAScolors_primary}}
and {\ref{fig:Hii_region_IRAScolors_secondary}} show the IRAS plane populated by our models as a function of the primary and secondary variables, respectively. In each sub-panel, the dashed blue line marks the boundary derived by \cite{2018MNRAS.476.3981Y} to distinguish Galactic \HII regions from other sources. 
This criterion suggests that the points lying in the region $\log(F_{60}/F_{12}) \geq (-0.19 \times \log (F_{100}/F_{25}) + 1.52)$ are those containing dust illuminated by young stars.
In a manner akin to the analysis performed for the BPT diagram, we explore the distribution of the models across the IRAS plane in relation to the primary variables, invoking the secondary variables as needed.

The top row in Fig.~{\ref{fig:Hii_region_IRAScolors_primary}} shows the histogram for our models on the IRAS plane. Overall, the models move upward and rightward with increasing metallicity. We attribute this to the dust in the shell/cloud systems becoming colder and the FIR peak moving rightward with increasing metallicity. The evolution of the effective dust temperatures can be seen through the FIR SED peak shown in the second row of Fig.~{\ref{fig:Hii_region_IRAScolors_secondary}}. Apart from this, the $F_{60}/F_{12}$ is impacted by the $F_{12}$ emission which comes predominantly from PAHs. As described in Sec.~\ref{ChemicalAbund.subsect} the maximum value of $q_\mathrm{PAH}$ in our models is $4.6\%$, but it is allowed to vary throughout the dust-containing gas based on its chemical composition, i.e., it is scaled by the atomic hydrogen abundance. At all metallicities, models with low $q_{\mathrm{PAH}}$ tend to separate out, forming a recognizable branch with higher $F_{60}/F_{12}$ values.

The second row of Fig.~{\ref{fig:Hii_region_IRAScolors_primary}} shows the model IRAS colors as a function of the age of the system. 
The evolution of the models' colors with age can be understood by considering the instantaneous incident flux at the inner edge of the shell,
\begin{equation}
    \mathcal{F}_{\mathrm{in}} = \frac{L_{\star,\mathrm{total}}}{r^2_{\mathrm{sh, in}}} \, ,
\label{eqn:logC_equivalent_Fin}
\end{equation} 
which is directly linked to the dust temperature. 
This parameter is shown in the top row of Fig.~{\ref{fig:Hii_region_IRAScolors_secondary}}. $\mathcal{F}_{\mathrm{in}}$ falls with age for systems with no recollapse events as the luminosity of the central cluster falls and the shells expand with age. Broadly speaking, models of this kind initially move up, reach a maximum, and then move to lower values of the $F_{60}/F_{12}$ color as the FIR peak rightward and $F_{60}$ decreases. Furthermore, as the shells expand and the stellar population ages, shells start to diffuse out. The diffused shells irradiated with softer radiation after the death of the most massive stars tend to be rich in atomic hydrogen, thus possessing a $q_{\mathrm{PAH}}$ close to the saturation value of $4.6\%$, which also contributes to lowering the $F_{60}/F_{12}$ value.
The time evolution of the models on the $F_{100}/F_{25}$ axis can also be understood in terms of the rightward movement of the FIR peak.
In contrast, systems that undergo multiple collapse events tend to have, on average, shells that are more compact during their evolution. This results in higher values of $\mathcal{F}_{\mathrm{in}}$, allowing these models to remain above the blue demarcation line as long as the youngest cluster within the system contains massive stars.
Most of these shells start infalling after they have swept through the entire cloud.  In-falling shells are generally high-density, high-cloud mass systems that possess high column densities. 
In the case of higher metallicity systems, the shrinking radii lead to increased $\mathcal{F}_{\mathrm{in}}$ and lower $q_{\mathrm{PAH}}$ as the high gas and dust columns allow for the formation of molecular hydrogen. This leads to higher $F_{60}/F_{12}$ and lower values of $F_{100}/F_{25}$ for the systems undergoing re-collapse forming a distinct region in the IRAS plane for these metallicities (cf. row 5 in Fig.~\ref{fig:Hii_region_IRAScolors_secondary} for shell velocities). On the other hand, low metallicity clusters do not show such a distinct region on the IRAS plane populated by infalling shells.

The third, fourth, and fifth rows in Fig.~{\ref{fig:Hii_region_IRAScolors_primary}} shows the trends on the IRAS plane as a function of $\epsilon_{\mathrm{SF}}$, $n_{\mathrm{cl}}$ and $M_{\mathrm{cl}}$, respectively. The trends on the IRAS plane due to these three parameters can be understood by considering the combinations of their extreme values in the parameter space. 
Fig.~\ref{fig:IRAS_plane_explanation} shows the evolution of these extreme systems for $Z=0.02$. The green line shows the evolution on the IRAS plane for the parameter trio listed at the bottom of each sub-panel. The scatter points are the results from the points adjacent to the mentioned trio on both the lower and higher sides in the parameter space. The values at the top right and top left corners are the times at which the bubble begins the transition to the momentum-driven phase (for the first time if recollapse/s occur), and the dissolution time for the shell, respectively. To facilitate comparison with observational data, the histogram representing the IRAS colors of Galactic \HII regions as compiled by \citet{2018MNRAS.476.3981Y}, is presented as grey hexbins in each sub-panel. 
It can be gathered from Fig.~\ref{fig:IRAS_plane_explanation} that shells formed from low-mass clouds have a tendency to move away from \cite{2018MNRAS.476.3981Y} demarcation and towards higher $F_{60}/F_{12}$ values as they evolve. This is due to the falling $q_{\mathrm{PAH}}$ as these clouds are rapidly thinned out.
Only in the cases where the parent clouds are dense and the star formation efficiency is low, these shells are optically thick to the ionizing radiation throughout their lifetime. 
In contrast, the shells formed out of high-mass clouds tend to remain optically thick throughout their lifetimes except for the shells carved out of the lowest-density clouds containing the highest stellar content. The higher overall $q_{\mathrm{PAH}}$ keeps these kinds of shells from possessing high $F_{60}/F_{12}$ as the shell expands. High $F_{60}/F_{12}$ are only encountered in the case of high-density, low star formation efficiency cases where the system exhibits recollapse events. As these shells live longer than their low-mass counterparts, if the stellar feedback is successful, they are quite diffuse at later times and exhibit low $F_{100}/F_{25}$.
Without a detailed comparison with the observational data, we note that the parameter space of our $Z=0.02$ models is able to capture the variety in the IRAS colors of Galactic \HII regions.

\section{Integrating \texttt{TODDLERS} into \texttt{SKIRT}}
\label{sect:integration_with_SKIRT}

\begin{figure*}
\includegraphics[width=.9\textwidth]{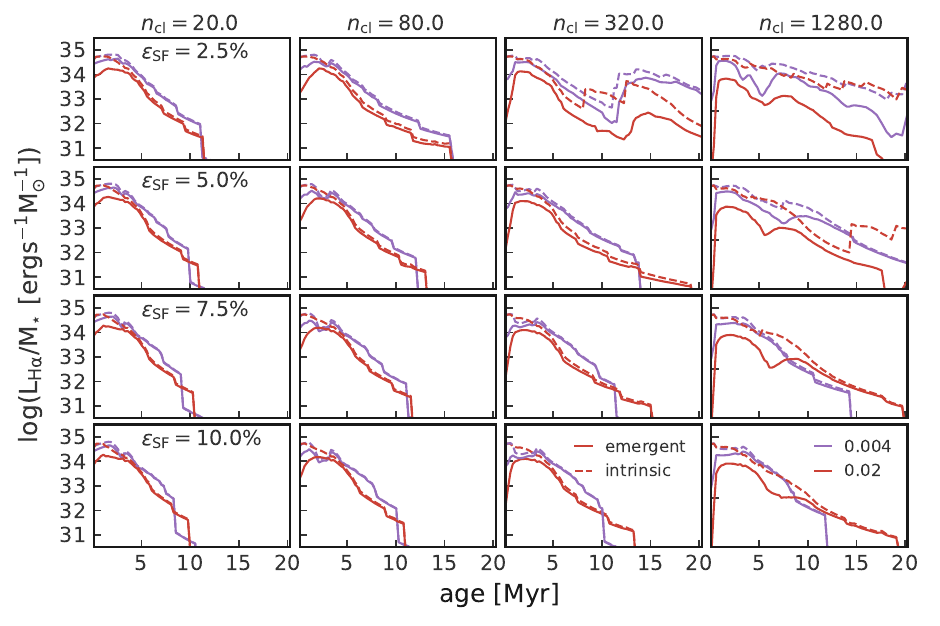}
\caption{The emergent (solid curves) and intrinsic (dashed curves) stellar mass normalized H$\alpha$ luminosity from various star-forming complexes consisting of a population described in Sec.~\ref{sect:Star-forming cloud complex population} with $Z=0.02$ (red), and $Z=.004$ (purple). The cloud density from left to right: $n_{\mathrm{cl}}=20,~80,~320,~1280~\mathrm{cm^{-3}}$, star formation efficiency from the top to bottom: $\epsilon_{\mathrm{SF}}=2.5,~5,~7.5,~10~\%$.}
\label{fig:Halpha_luminosity_HiiComplex}
\end{figure*}


The inclusion of the observables from the \texttt{Cloudy} post-processing in \texttt{SKIRT} consists of two steps: 
\begin{enumerate}[leftmargin=*]
    \item {Normalisation of the data. This is to make the library widely applicable to simulations of varying mass resolutions}. 
    \item {The generation of SEDs including the stellar, nebula, and dust continua, with high resolution around the included emission lines.}
\end{enumerate}
We describe these two steps below.

\subsection{Star-forming cloud complexes}
\label{sect:Star-forming cloud complex population}
A key objective of this work is to generate a library for post-processing simulated galaxies of different resolutions using \texttt{SKIRT}. In order to do so, we generate star-forming complexes consisting of a family of clouds (referred to as the components of a cloud complex) for each point in the parameter space. 

We assume that stellar clusters originate in a cloud population that follows a power-law distribution in masses \citep[see the discussion in][and references therein]{2015ARA&A..53..583H}. This power-law is given as:
\begin{equation}
\frac{\mathrm{d} N_{\mathrm{cl}}}{\mathrm{~d} M} \propto M^{\alpha_{\mathrm{cl}}} \quad \text {with } M \in\left[10^{5},10^{6.75}\right] \mathrm{M}_{\odot} \text{,~and~} \alpha_{\mathrm{cl}} = -1.8
\label{eqn:cloud_distribution}
\end{equation}
We then normalize the observables by the stellar mass present in the system at that time, making the particle mass simply a scaling factor for the assigned SED. By doing so, we aim to conserve the total mass of young stellar particles in a given simulation being post-processed.

{We opted for a power-law population to minimize the models' parameter space while maximizing the utilization of simulation data. This approach allows us to directly extract information on age, metallicity, and star and cloud density from the simulation. The cloud density could potentially be estimated from the cold gas density derived in the sub-grid effective equation of state. However, it is important to note that alternative methods for reducing and averaging over the parameter space could be conceived. For instance, one could use an age average, average over a log-normal density distribution \citep[for example][]{2015ApJ...811L..28B, 2022ApJ...930...76K}, or relate cloud mass and density through Larson's laws \citep{1981MNRAS.194..809L}, among other possibilities.}
This approach leads to a realistic representation of star-forming regions, consisting of several embedded and unembedded sources, including those containing multi-generation stellar populations. We note that although such a cloud complex generation is a physically motivated way of normalizing the SEDs assigned to simulation particles, it effectively reduces the spatial resolution by carrying out a mass-weighted average over individual clouds. An effect of this, for example, is on the emergent H$\alpha$ luminosity from a star-forming complex. As the emergent H$\alpha$ luminosity is going to be dominated by unembedded components, the Balmer decrement-based correction would lead to an underestimation of intrinsic H$\alpha$ for the particle. This effect is similar to the one discussed in \citet{2020MNRAS.498.4205V}. Therefore, individual cloud models could be used for applications requiring higher resolution, which are also available to use in \texttt{SKIRT}.

\begin{figure*}
    \centering
\includegraphics[width=.9\textwidth]{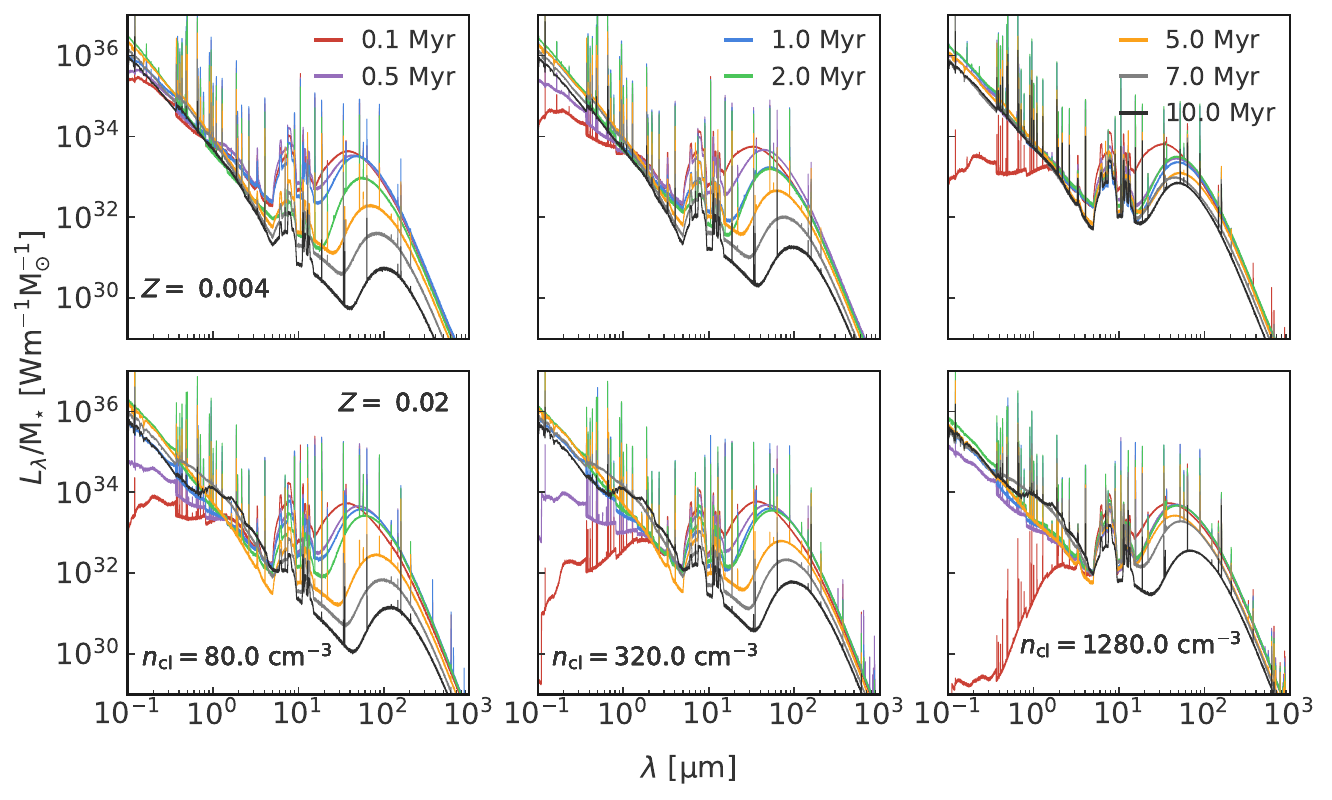}
\caption{UV--mm SED emerging from star-forming complexes described in Sec.~\ref{sect:Star-forming cloud complex population}. Top row: $Z=0.004$, bottom row: $Z=0.02$. From left to right, the densities are $80,~320,~1280~\mathrm{cm^{-3}}$ with the star formation efficiency fixed at $5\%$. The SEDs shown here have a spectral resolution $R=3000$.}
\label{fig:complex_SED_evo}
\end{figure*}

Fig.~\ref{fig:Halpha_luminosity_HiiComplex} shows the stellar mass normalized emergent H$\alpha$ luminosity for a selected part of our parameter space. In any given row, moving from left to right increases the density by a factor of four, while moving top to bottom in the same column increases the star-formation efficiency. Both the emergent and intrinsic H$\alpha$ line luminosities are shown for two metallicities, $Z=0.02$ and $0.004$. 
In Fig.~\ref{fig:Halpha_luminosity_HiiComplex}, increasing the density at a given $\epsilon_{\mathrm{SF}}$ increases the likelihood of re-collapse events and the overall dissolution of the complex takes longer. The bumps in the intrinsic H$\alpha$ luminosity are seen when one of the component shells undergoes a re-collapse event. The emergent H$\alpha$ does not necessarily show an immediate rise in its value due to high $A_{\mathrm{v}}$ values which occur at large shell densities (see Fig.\ref{fig:Hii_region_BPTratios_Nii_Oiii_secondary}) and the presence of the unswept cloud around the shells after they are initially formed. Note that metallicity is directly linked to the amount of dust in our models, and high metallicity models show an overall higher extinction.
Increasing the $\epsilon_{\mathrm{SF}}$ at a fixed density, on average, pushes out the gas more rapidly leading to a quicker dissolution of the component shells. 
The presence of enhanced Ly$\alpha$ pressure in the low metallicity case leads to a faster dissolution of the component clouds, especially at the higher-density and intermediate to high star formation efficiency end (cf. $t_{\mathrm{evo}}$ in Figs.~\ref{fig:R_contours_fix_SFE} and \ref{fig:R_contours_fix_n}).

Fig.~\ref{fig:complex_SED_evo} shows examples of the evolution of the UV--mm SED for $Z=0.004$ (top row) and $Z=0.02$ (bottom row) for the parameters in the second row, Fig.~\ref{fig:Halpha_luminosity_HiiComplex}, excluding the lowest density case of $20~\mathrm{cm^{-3}}$. In the two lowest-density cases where the population of shells does not exhibit any recollapse, the time evolution is dictated by a monotonic expansion of the component shells and the aging cluster population. This lowers the UV extinction and the overall dust temperatures fall shifting the IR peak rightward (see the discussion in Sec.~\ref{sect:IRAS plane and parameters}). 
Increasing the density leads to an overall higher $\mathcal{F_\mathrm{in}}$ due to more compact components and the likelihood of re-collapse, and thus a lower rate of fall of dust temperature. At $Z=0.02$, in the highest density case shown here, all components with $M_{\mathrm{cl}}>10^{6.0}~M_{\odot}$ show a re-collapse, while this threshold moves up to $10^{6.25}~M_{\odot}$ at $Z=0.004$. For the higher metallicity case, the result of the presence of populations younger than $10~$Myr in the highest density case is reflected by the less prominent red supergiant near-IR hump at $10$~Myr.

It is worth noting that different wavelength ends of the UV--mm SED are tracking different mass components of the star-forming cloud complex. The lower mass shells which expel the gas most rapidly and thin out the shells rapidly contribute to the UV end, while the embedded components emit predominantly in the IR. This effect is seen in the UV slope being nearly the same at the same age in all three densities, while the normalizations are different. The highest density case where a significant portion remains embedded in all cases shown shows lower UV.
A small fraction of unembedded clusters can thus have an outsized impact on the UV slope \cite[see the discussion in][]{2017MNRAS.472.2315P}.

\subsection{Line and continuum emission integration}
We include the line emission from the mass normalized star-forming cloud complexes in \texttt{SKIRT} by converting the line luminosities into a continuum SED by assigning a Gaussian profile. 
{The linewidth is selected to ensure that the lines in our list do not blend. The Gaussian profile is truncated at \(4 \sigma\), which means that \(\Delta\lambda = 4 \sigma\). This choice of \(\Delta\lambda\) is made by a comparison with a triangular profile of a resolution \(R = 5 \times 10^{4}\). To resolve the Gaussian profile, \(\Delta\lambda\) is sampled at 37 equidistant points, implying that we achieve a spectral resolution for the lines that exceeds \(5 \times 10^{4}\).
}

This approach of slightly broadening the lines is valid given that the major sources of line broadening in a simulated galaxy are likely to be the bulk motion of a simulated particle and the sub-grid gas motions in a complex. 
The bulk motion of the different particles with respect to each other and with respect to the observer is taken care of within {\tt{SKIRT}}, given the 3D velocity vector of each particle \citep{2020A&C....3100381C}. The second source of Doppler broadening linked to the sub-grid motions of the gas can also be accounted for within \texttt{SKIRT} by assigning a user-defined value for each emitting entity. {Given that the requisite data concerning sub-grid motions of the gas can be straightforwardly derived from the shell velocities computed in Sec.~\ref{sect:evolutionary_model} and incorporated via the \texttt{SKIRT} interface, we opted not to include its broadening effect on the lines in the \texttt{SKIRT} tables.
}

\begin{figure*}
\includegraphics[width=.9\textwidth]{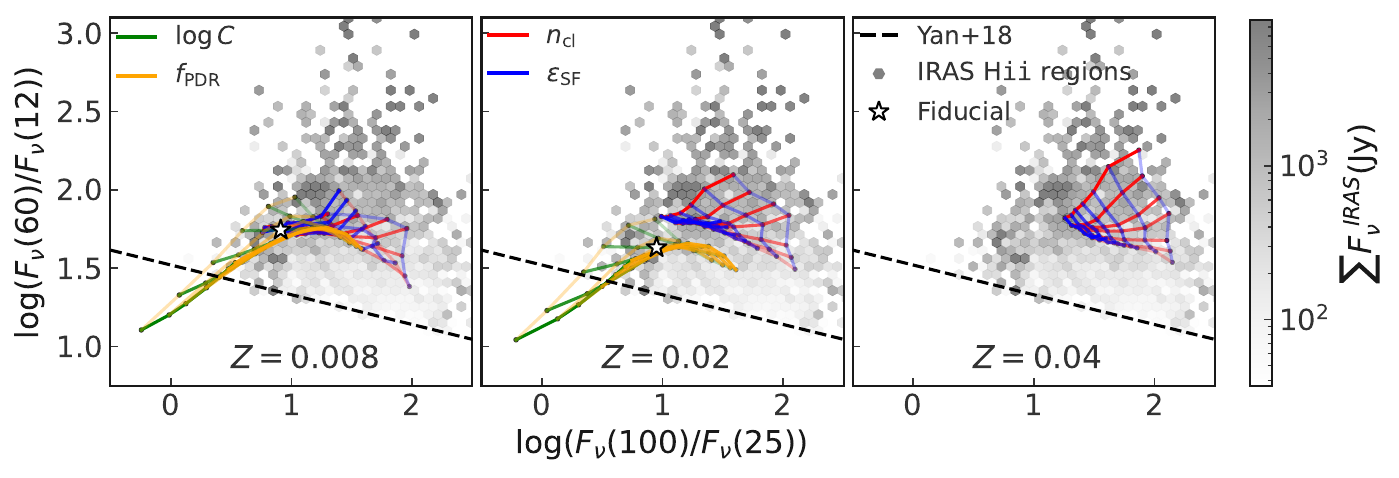}
\caption{The IRAS colour-colour diagram for \texttt{TODDLERS} and \texttt{HiiM3}. The red and blue colors represent the free parameters for the \texttt{TODDLERS} library, $n_{\mathrm{cl}}$ and $\epsilon_{\mathrm{SF}}$, respectively. The green and orange colours represent the free parameters of \texttt{HiiM3}, $\log C$, and $f_{\mathrm{PDR}}$, respectively. 
Increasing opacity of the curves represents increasing values of the variables which are given in Tab.~\ref{Table:TODDLERSvsHiiM3}.
The data is missing at $Z=0.04$ for \texttt{HiiM3}, where the upper limit is $Z=0.025$.
The star labeled as "Fiducial" gives the colors for \texttt{HiiM3} parameters $\log C = 5.0$ and $f_{\mathrm{PDR}} = 0.2$. These numbers have been employed by various authors as their fiducial values when post-processing simulated galaxies using \texttt{HiiM3} \citep{2010MNRAS.403...17J, 2019MNRAS.483.4140R, 2020MNRAS.492.5167V, 2022MNRAS.516.3728T}.
The grey hexbins represent the Galactic \HII regions discussed in  \citet{2018MNRAS.476.3981Y}. The colormap gives the median value of the total flux density in the four IRAS bands. }
\label{fig:TODDLERSvsMAPPINGS_IRAScolors}
\end{figure*}

We add about $150$ lines emanating from various phases of the gas. The list is essentially a merger of the lines' lists\footnote{Available at \url{https://gitlab.nublado.org/cloudy/cloudy/-/blob/master/data/LineList_HII.dat}, \newline \url{https://gitlab.nublado.org/cloudy/cloudy/-/blob/master/data/LineList_PDR_H2.dat}} appropriate for low-density H\textsc{ii} regions, PDR, and the molecular gas supplied with the version of \texttt{Cloudy} used in this work. The consolidated list can be found at \url{www.toddlers.ugent.be}.
We directly use the stellar, nebular, and dust continua reported by \texttt{Cloudy} which have a spectral resolution of $R\approx300$.

\section{Comparison with \texttt{HiiM3}}
\label{sect:comparison_with_HiiM3}

In this section, we compare the \texttt{TODDLERS} library as implemented in \texttt{SKIRT} with \texttt{HiiM3} focusing on their outputs in the MIR--FIR wavelength regime.
A direct comparison between the two libraries is made difficult by the wide variety of differences that exist between the two libraries.
\texttt{TODDLERS} uses physical parameters ($t,\,Z,\,\epsilon_{\mathrm{SF}},\,n_{\mathrm{cl}}$) and considers a finite gas reservoir along with the forces of gravity and shell state-dependent external pressure accounted in the dynamical evolution. The evolution could be pressure or momentum-driven, with a possibility of shell re-collapse in the momentum-driven phase.
On the other hand, the parameters in \texttt{HiiM3} ($Z,\,f_{\mathrm{PDR}},\,\log C,\, P_{\mathrm{ISM}}$) result from the evolution of an adiabatic, pressure-driven bubble evolution without gravity. For such systems, combinations of the cluster's stellar mass and the external gas pressure serve as a scaling on the ionization parameter and control the dust temperature. The combination controlling the dust temperature in the ionized region is $\log C$, 
which is given as:
\begin{equation}
\log C=\frac{3}{5} \log \left(\frac{M_{\mathrm{\star}}}{M_{\odot}}\right)+\frac{2}{5} \log \left(\frac{P_{\mathrm{ISM}} / k}{\mathrm{~cm}^{-3} \mathrm{~K}}\right) \, .
\label{eqn:logC_eqn}
\end{equation}
Increasing $\log C$ at a fixed $P_{\mathrm{ISM}}$ increases the stellar mass in the system and leads to hotter dust, moving the IR peak leftward.
$f_{\mathrm{PDR}}$ is a time-averaged quantity accounting for the absorption of radiation arising from the ionized region {by a surrounding neutral medium (similar to our neutral cloud around the shell)}. Cases in which the PDR entirely surrounds the ionized region correspond to $f_{\mathrm{PDR}}=1$, while completely uncovered ionized regions have $f_{\mathrm{PDR}}=0$. Increasing the value of $f_{\mathrm{PDR}}$ {to 1} leads to higher PAH and cooler dust emission on account of increasing UV absorption. 
The two models also differ in the dust models, e.g., \texttt{TODDLERS} uses a dust mix which has a lower-end cutoff in dust sizes at $0.03~\micron$, while this value is significantly lower in \texttt{HiiM3} at $0.004~\micron$. PAHs in \texttt{TODDLERS} are not associated with ionized or molecular gas, whereas, \texttt{HiiM3} includes them in the molecular cloud covering the ionized region. Further differences in the gas density law also exist, we refer the reader to the discussion in Sec.~\ref{cloudy_methods_library_gen.sec} and that in \cite{2008ApJS..176..438G}. We remark that although this list of differences is useful, it is unlikely to be exhaustive.

\begin{figure*}
\includegraphics[width=.9\textwidth]{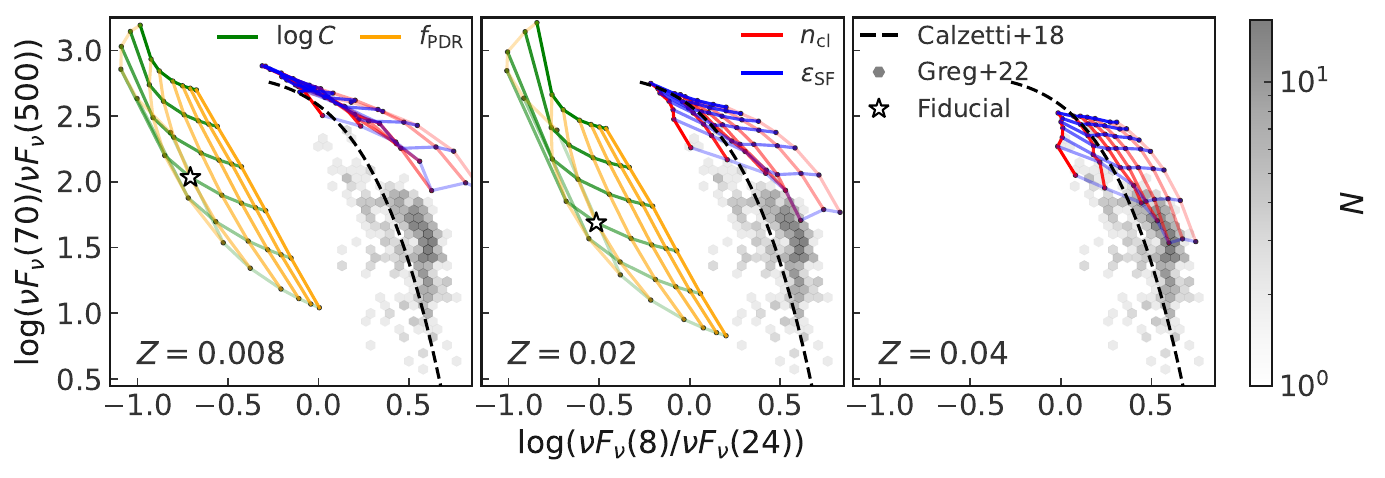}
\caption{Same as Fig.~\ref{fig:TODDLERSvsMAPPINGS_IRAScolors}, but for the MIR--FIR colour-colours discussed in Sec.~\ref{subsect:MIRFIR_colors_of_luminosity_weighted_models}. The grey hexbins represent the colors from the ``Regime-1'' galaxies discussed in \citet{2022ApJ...928..120G}. The hexbin colormap depicts the number of spaxels in a given bin. {The black broken curve in each panel represents the relation from \citet{2018ApJ...852..106C}}.}
\label{fig:TODDLERSvsMAPPINGS_MIRFIR}
\end{figure*}

Due to the wide array of differences that exist between the two libraries, we compare them by simply contrasting a selected set of observables mapped by their input parameters. For each of the two libraries, we scan their respective parameter spaces. The parameter values studied are listed in Tab.~\ref{Table:TODDLERSvsHiiM3}. In the case of \texttt{HiiM3}, We fix $P_{\mathrm{ISM}}$ to a value of $10^{-12}~\mathrm{Pa}$. Changing $P_{\mathrm{ISM}}$ at a fixed $\log C$ does not impact the broadband continua, which is the basis of the comparison here. We also note that $f_{\mathrm{PDR}}$ is only available at the values of $0$ and $1$, and is linearly interpolated at all other values in between.
We restrict the comparison of the two libraries in the metallicity regime that represents massive galaxies in the nearby universe and the Milky Way. This is also the regime where the majority of the past work \citep{2019MNRAS.484.4069B, 2020MNRAS.494.2823T, 2022MNRAS.516.3728T, 2021MNRAS.506.5703K,2022MNRAS.512.2728C} has recognized the shortcomings of \texttt{HiiM3}. Therefore, we consider three metallicities, $Z\in[0.008, 0.02, 0.04]$ to confront the results from the two libraries with observational data.
 We mention that the highest metal fraction available in the case of \texttt{HiiM3} is $0.025$, thus in all the plots where comparison is made with the \texttt{TODDLERS}'s value of $Z=0.04$, \texttt{HiiM3} data is missing. 
As \texttt{HiiM3} considers a luminosity weighted average of various ages between $0-10$~Myr, we generate fluxes for \texttt{TODDLERS} by uniformly sampling the period between $0-30$~Myr. We note that the \texttt{SKIRT} implementation of \texttt{TODDLERS} is normalized by the stellar mass of the system (see Sec.~\ref{sect:integration_with_SKIRT}), while \texttt{HiiM3} is normalized by the SFR. {Thus, for an arbitrary SFR}, we ensure that $\mathrm{SFR}[M_{\odot}\mathrm{yr^{-1}}]\times10^{7}[\mathrm{yr}] = M_{\mathrm{\texttt{TODDLERS}}}$. 
In practice, {we use an SFR of unity} and sample $10^4$ emitting entities with an age lying in the range $0-30$~Myr, each one of their SED is then scaled by the mass $M_{\mathrm{\texttt{TODDLERS}}}/10^{4}$.
We focus on the comparison in the IR regime and use two comparison strategies: 
\begin{enumerate*}
\item The IRAS color plane,  similar to the one discussed in Sec.~\ref{sect:IRAS plane and parameters}.
\item The MIR--FIR colour plane using IRAC~$8\micron$, MIPS~$24~\micron$ and PACS~$70~\micron$, and SPIRE~$500~\micron$ bands.
\end{enumerate*}
We mention that the IR bands used here are affected by the presence of diffuse dust outside the star-forming regions, but it is worthwhile to investigate the behavior of our models in isolation. A detailed comparison using simulated galaxies, including the impact of diffuse dust, is the subject of the second paper in this series. As the UV emission is also subject to attenuation by diffuse dust, we also defer the comparison in the UV regime to that paper.

\begin{table}
\caption{The parameters used for comparing \texttt{TODDLERS} and \texttt{HiiM3}.}
\centering
\renewcommand{\arraystretch}{1.25}
\begin{tabular}{lc}
\hline
Parameter & Values  \\
\hline 
\multicolumn{2}{c}{\texttt{TODDLERS}} \\
$Z$ & 0.008, 0.02, 0.04  \\
$n~[\mathrm{cm^{-3}}]$ & 10.0, 20.0, 80.0, 320.0, 640.0, 1280.0, 2560.0  \\
$\epsilon_{\mathrm{SF}}[\%]$ & 1.0, 2.5, 5.0, 7.5, 10.0, 12.5, 15.0 \\
Age & Uniform sampling between 0-30~Myr \\
\hline
\multicolumn{2}{c}{\texttt{HiiM3}} \\
$Z$ & 0.008, 0.02  \\
$P_{\mathrm{ISM}}~[\mathrm{Pa}]$ & $10^{-12}$  \\
$f_{\mathrm{PDR}}$ & 0.0, 0.1, 0.2, 0.4, 0.6, 0.8, 1.0 \\
$\log C$ & 4.0,4.5, 5.0, 5.5 6.0, 6.5 \\
\hline
\label{Table:TODDLERSvsHiiM3}
\end{tabular}
\renewcommand{\arraystretch}{0.8}
\end{table}

\subsection{IRAS colours}
\label{subsect:IRAS_colors_of_luminosity_weighted_models}
Following the discussion in Sec.~\ref{sect:IRAS plane and parameters}, we use the IRAS plane to populate the luminosity-weighted models as shown in Fig.~\ref{fig:TODDLERSvsMAPPINGS_IRAScolors}. The \texttt{TODDLERS} parameter space is shown in red-blue, while that of \texttt{HiiM3} is shown in orange-green. Higher opacity of the curves is associated with higher variable values, which are listed in Tab.~\ref{Table:TODDLERSvsHiiM3}. The \HII region demarcation is shown along with the colors from individual Galactic \HII regions. 
The individual \HII regions' data is shown as hexbins of the median value of the total flux density in all four IRAS bands, serving as a proxy for the total-IR flux density. 
Note that this dataset is comprised of individual \HII regions, likely composed of different cluster ages, stellar/gas masses, and metallicities. In Fig.~\ref{fig:IRAS_plane_explanation}, we showed that the individual \texttt{TODDLERS} models are able to span the range of colors exhibited by the observational data. In contrast, Fig.~\ref{fig:TODDLERSvsMAPPINGS_IRAScolors} shows time-averaged model data where the colors are dominated by younger, brighter components of the mix. Thus, the observational data is shown simply to give an idea of the parameter space on the IRAS plane. 

The two models cover somewhat different regions on the IRAS plane with the \texttt{HiiM3} data generally exhibiting lower values of both $F_{100}/F_{25}$ and $F_{60}/F_{12}$ than \texttt{TODDLERS}. The \texttt{TODDLERS} models occupy the region marked by bright Galactic \HII regions' with a better coverage of the IRAS plane. The offset between \texttt{HiiM3} and \texttt{TODDLERS} is in part driven by two factors, \begin{enumerate*}
    \item The differences in the lower end of the grain size distribution. The model employed in \texttt{TODDLERS} lacks non-PAH small grains which could emit efficiently at $25~\micron$ (see, e.g., \cite{2012A&A...545A..39R} for dust size dependence of emission in various bands). In contrast, the larger grains in our model emit predominantly at longer wavelengths, shifting the \texttt{TODDLERS}' data upwards and rightwards. 
    \item In \cite{2005ApJ...619..755D}, the mechanical luminosity is reduced by a factor $10$ to {ensure that the bubble expands slowly and the internal pressure remains low}. We suspect that this alteration of dust radius relative to the stellar cluster could also be playing a role in driving the higher dust temperatures in the case of \texttt{HiiM3}, resulting in the different regions of the IRAS color--color plot covered by \texttt{TODDLERS} and \texttt{HiiM3}.
\end{enumerate*}

\subsection{MIR--FIR colours}
\label{subsect:MIRFIR_colors_of_luminosity_weighted_models}
We further examine the shape of the IR SED using the color--color plot employing the IRAC~$8~\micron$, MIPS~$24~\micron$, PACS~$70~\micron$, and SPIRE~$500~\micron$ bands. This is inspired by the work carried out by \cite{2018ApJ...852..106C, 2022ApJ...928..120G}, where this colour--colour plot is generated at a kpc resolution for nearby galaxies. For the rest of this section, we use the notation $F_{n}$ to represent $\nu F_{\nu}(n)$, the flux in a band with pivot wavelength $n$. 

Fig.~\ref{fig:TODDLERSvsMAPPINGS_MIRFIR} shows how the two libraries cover this plane. We also show the relation found by \cite{2018ApJ...852..106C} along with the colors for individual pixels for the ``Regime-1'' galaxies discussed in \cite{2022ApJ...928..120G} which exhibit uniform and high star-formation surface densities. {Note that the curve from \cite{2018ApJ...852..106C} uses $F_{1100}$ instead of $F_{500}$, here we have made the conversion based on the factor given in \cite{2022ApJ...928..120G}.}

As we compare the trends on this color--color plot, it is worth noting that $F_{500}$ and $F_{8}$ are impacted by the dust emission originating outside of the star-forming regions. The observational sample, despite being the one with high star formation density is affected by the presence of cold dust ($F_{500}$) and to some extent, dust heating by old stars ($F_{8}$). {This model data will move downwards and slightly rightwards once the dust emission originating outside of our model star-formation regions is taken into account. This trend is also expected based on the nature of the \cite{2018ApJ...852..106C} curve, where the addition of diffuse, cooler dust moves the data points rightwards and downwards.}

In the models considered in Fig.~\ref{fig:TODDLERSvsMAPPINGS_MIRFIR}, an increase in $n_{\mathrm{cl}}$ and/or $\epsilon_{\mathrm{SF}}$ leads to an increase in the luminosity-weighted $\mathcal{F}_{\mathrm{in}}$ resulting in an increase in $F_{70}/F_{500}$, a trend observed with increasing $\log C$ (and to some extent with increasing $f_{\mathrm{PDR}}$) in the case of \texttt{HiiM3}.
The lower $\epsilon_{\mathrm{SF}}$ systems tend to have a higher number of shell re-collapse events leading to an increasingly narrow range of values in $F_{70}/F_{500}$ as the effective star formation efficiency ($M_{\star}/M_{\mathrm{cl}}$ at the end of 30~Myr) could reach values closer to higher $\epsilon_{\mathrm{SF}}$ models.
The \texttt{HiiM3} parameter, $f_{\mathrm{PDR}}$, is completely free from the evolution of the bubbles and serves as a geometrical factor on the non-ionizing UV absorbed by the molecular cloud assuming a fixed column depth. At a fixed $\log C$, this leads to models moving rightward and downward on the MIR--FIR plane due to the increasing contribution of PAHs and cold dust. In contrast, \texttt{TODDLERS} calculate IR emission in a single model, and column depths are affected by the state of the component shells of the star-forming complexes. The youngest and the most massive components tend to have the highest impact on the IR SED. These components are ones that are still embedded in their birth cloud and have high dust column depths. If we consider $F_{8}/F_{24}$ a proxy for $q_{\mathrm{PAH}}$, at lower densities, the unswept clouds are more diffused and contain neutral Hydrogen. This leads to a higher luminosity-weighted PAH abundance, moving the models rightward with decreasing $n_{\mathrm{cl}}$ at a fixed $\epsilon_{\mathrm{SF}}$. Increasing $\epsilon_{\mathrm{SF}}$ at a fixed $n_{\mathrm{cl}}$ tends to decrease the PAH fraction, leading to leftward movement on the MIR--FIR plane.  

On this MIR--FIR color plane the two libraries cover markedly different regions, with the \texttt{TODDLERS} library showing a closer match to the observational colors. 
We confirm that this trend is also seen in the colors generated with simulated galaxies accounting for dust outside the star-forming regions (subject of Paper 2 in this series).
We attribute the large differences in the $F_{8}/F_{24}$ to the differences in the dust models used in the two libraries, in particular, the dust model employed in \texttt{HiiM3} appears to be more emissive around the $24\,\micron$ band.

\section{Summary and outlook}
\label{sect:conclusions}
In this work, we have presented a new emission library \texttt{TODDLERS} with the primary aim of producing time-dependent emission diagnostics from gas and dust around young stars for simulated galaxies. 
To achieve this, we have run a large suite of semi-analytic calculations that allow us to infer the gas--star geometry as the gas evolves under stellar feedback. The calculations assume a cloud with a constant density profile and a finite mass evolving under the influence of the feedback of a central cluster. This idealized approach allows us to sweep a large parameter space while accounting for complex feedback physics in a simplified fashion. 
The dominant stellar feedback channel evolves as a function of metallicity. At high metallicities, the gas is pushed predominantly by stellar winds and the subsequent SNe, whereas, the dominant feedback channel at metallicities below $Z\lesssim 0.004$ is the Ly$\alpha$ radiation pressure due to multiple resonant scatterings. This is especially the case for the higher end of cloud densities and masses where there is ample neutral gas for prolonged periods. At higher metallicities, the role of Ly$\alpha$ radiation pressure is subdominant but non-negligible. The clues from this idealized Ly$\alpha$ radiation pressure setup point to the importance of this feedback channel in young star-forming clouds warranting more detailed studies, even at higher metallicities.

The semi-analytic calculations are passed on to \texttt{Cloudy} to carry out calculations involving detailed chemistry enabling us to produce time-dependent UV--mm observables. 
The tabulated observables include emission lines originating from ionized, photo-dissociation, and molecular regions along with the nebular, stellar, and dust continuum emission. We have used the BPT diagram and the IRAS color-color diagram to map the parameter space of the models onto the observable space demonstrating that they populate the expected regions when compared to observational data from nearby galaxies and the Milky Way.
We have integrated these observables into \texttt{SKIRT} to use these data for post-processing simulations where star-forming regions are not resolved by assuming a power-law cloud population.
A comparison focused on the IR colors produced using the aforementioned \texttt{TODDLERS} implementation to that of the currently only available star-forming regions' library in \texttt{SKIRT} was carried out. When confronted with observations, \texttt{TODDLERS}' IR colors are in better agreement with observations in comparison to \texttt{HiiM3}, which until now was the only option in \texttt{SKIRT} to incorporate star forming regions' emission. In a companion paper, we use \texttt{TODDLERS} to produce UV-submm diagnostics using simulated galaxies. By doing this, we can make a direct and detailed comparison between broadband and line-emission data from simulated galaxies with those from recent observational studies, such as \citet{2022ApJ...928..120G} and \citet{2023MNRAS.520.4902G}.

This work serves as a proof of concept where we followed the evolution of homogeneous clouds using a single set of stellar templates and the observable generation employs a single chemical abundance set and dust model. Thus, several areas remain open for exploration within the framework employed in this work.
As far as the evolutionary model is concerned, the cloud density profile dictates its binding energy, modifying which could lead to significant changes in the system's evolution \citep{2019MNRAS.483.2547R}. {For example, consider two clouds with identical mean density and mass. One exhibits a Bonnor-Ebert profile with a high-density core \citep{1955ZA.....37..217E, 1956MNRAS.116..351B}, while the other has a constant density. Comparatively, the cloud with the Bonnor-Ebert profile possesses a greater gravitational binding energy than its constant-density counterpart, making it more resistant to destruction.}

We have assumed that the unswept cloud is not dynamically affected by stellar feedback, which, for example, lowers the effective outward momentum deposition at low metallicities when the ionization front lies outside the shell. We have also neglected the density and velocity gradients that would appear as a result of the Ly$\alpha$ feedback which could lower the gas columns seen by Ly$\alpha$ photons in non-trivial ways. 
While the problem of how feedback affects the gas in star-forming regions is intrinsically 3D, relaxing these assumptions in spherical symmetry would be an interesting iteration of the current work \cite[see, for example,][]{2023MNRAS.521.5686K}.

We have shown that one of the key parameters that determine the emission properties of a nebula, $U$, is critically dependent on the properties resulting directly from the stellar templates ($Q_{\mathrm{H}}$ and $F_{\mathrm{ram}}$).
At the same time, the dust temperature is a function of $\mathcal{F}_{\mathrm{in}}$, which depends on the stellar feedback. Such dependencies demand a thorough investigation of the results presented here while changing the stellar library. 
Modifying the stellar library could mean, among other things, varying the initial elemental abundances, the physical processes employed in the stellar models, and its IMF.  
\citet{2021ApJ...908..241G} have shown the need to use abundance sets beyond the metallicity-scaled solar abundances for massive stars, especially due to their effects on the line emission at the lower metallicity end. 
Similarly, the presence of stellar rotation, binarity, and changes in the IMF play important roles in controlling the ionizing photon production rate and mass loss from the stellar clusters \citep{2014ApJS..212...14L, 2019A&A...621A.105S}. 
Such model choices will have an impact on the various feedback channels incorporated here, leading to interesting effects on the emission lines and dust emission.
Thus considering variations of the stellar templates based on the abundance sets, physical processes, and IMF represent interesting avenues to explore.

Another intriguing prospect is to consider variations in the dust-to-metal ratio as a function of metallicity and environment, which are expected based on numerical modeling and observations \citep{2013EP&S...65..213A, 2013A&A...557A..95R, 2016MNRAS.457.1842S, 2019A&A...623A...5D}.
The computational efficiency offered by the 1D calculations makes it feasible to generate models with the aforementioned variations, we plan to pursue this in the future.

\section{Data Availability}
The \texttt{TODDLERS} library for post-processing galaxy formation simulations is available for download along with the \texttt{SKIRT} code.
All other data used in this work, including the \texttt{Cloudy} output for individual models, are publicly available at \url{www.toddlers.ugent.be}.

\section{Acknowledgements}
We thank Ilse De Looze, Jérémy Chastenet, and Brian Van Den Noortgaete for useful discussions.
We thank Benjamin Gregg for providing the MIR-FIR colors' data from the KINGFISH sample.
We thank Qing-Zeng Yan for providing the IRAS colors' data.
AUK, AN, MB acknowledge the financial support of the Flemish Fund for Scientific Research (FWO-Vlaanderen).
The simulations carried out for this work used the Tier-2 facilities of the Flemish Supercomputer Center (https://www.vscentrum.be/) located at Ghent University.
We are grateful to the members, particularly Gary Ferland and Peter van Hoof, for addressing numerous questions on the \texttt{Cloudy} online forum.
We also thank the anonymous referee for their valuable comments and suggestions.


\FloatBarrier
\bibliographystyle{mnras}
\bibliography{bibliography}

\FloatBarrier


\appendix
\section{Comparison with WARPFIELD-2}
\label{appendix:comparison_warpfield}
\begin{figure}
    \centering \hspace{-0.75cm}\includegraphics[width=.8\columnwidth]{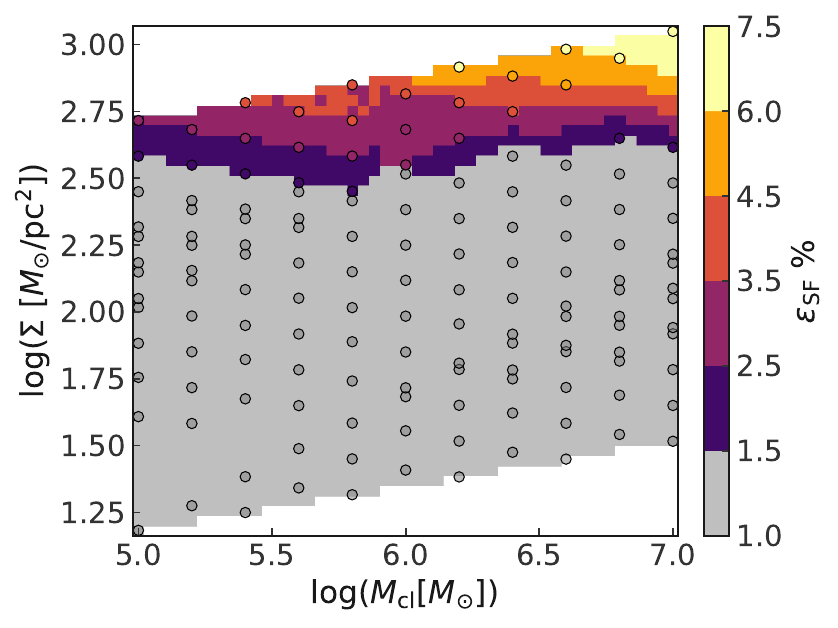}
    \caption{Minimum Star-formation efficiency $\epsilon_{\mathrm{SF}}$ across the cloud mass and surface density parameter space for comparison to Fig.~5 from \citet{2019MNRAS.483.2547R}. The colormap represents an interpolation over the data points shown.}
    \label{fig:benchmarking}
\end{figure}
To compare our implementation of the evolutionary model with the work carried out by \cite{2019MNRAS.483.2547R}, we compare the minimum star formation efficiencies needed as a function of cloud surface density and cloud mass for a homogeneous cloud. 
For this comparison, we turn off the Lyman-$\alpha$ feedback. We also use the same stellar templates as the ones used in that work.  The terminal velocities for SNe ejecta are fixed to $v_{\mathrm{sn}}=10^{4}~{\mathrm{m/s}}$ as mentioned in \cite{2017MNRAS.470.4453R}, and we assume that the same is done in \cite{2019MNRAS.483.2547R}. 
 The parameter space used for this comparison is given in Table~\ref{Table:benchmarking_parameter_space}
Our results for the minimum $\epsilon_{\mathrm{SF}}$ required to
disrupt a homogeneous cloud are shown in Fig. \ref{fig:benchmarking}. These results
are in very good agreement with those in \cite{2019MNRAS.483.2547R}. A more detailed comparison was not possible as WARPFIELD-2 is not publicly available.
\begin{table}
\caption{The $n_{\mathrm{cl}},~M_{\mathrm{cl}},~Z,~\epsilon_{\mathrm{SF}}$ parameter space for the comparison with WARPFIELD-2.}
\centering
\renewcommand{\arraystretch}{1.25}
\begin{tabular}{lccc}
\hline
Parameter & Min & Max & Step size \\
\hline
 $\log n_{\mathrm{cl}}~(\mathrm{~cm}^{-3})$ & $0.9$ & $2.7$ & $0.15$ \\
$\log M_{\mathrm{cl}}/M_{\odot}$ & $5$ & 7 & $0.2$ \\
$Z$ & $0.014$ & $-$ & $-$  \\
$\epsilon_{\mathrm{SF}}~\%$ & \multicolumn{3}{l}{1, 1.5, 2.5, 3.5, 4.5, 6.0, 7.5} \\
\hline
\label{Table:benchmarking_parameter_space}
\end{tabular}
\renewcommand{\arraystretch}{0.8}
\end{table}

{\section{Determining molecular fraction in the shell}
\label{appendix:krumholz2013}
The model described in \citet{2013MNRAS.436.2747K} calculates $f_{\mathrm{mol}}$ as:}

{\begin{equation}
\begin{aligned}
f_{\mathrm{mol}}=
\begin{cases}
1-3 s /(4+s) & \text { for } s<2 \\
0 & \text { for } s \geqslant 2
\end{cases}, \\
s \equiv \frac{\ln \left(1+0.6 \chi+0.01 \chi^2\right)}{0.6 \tau_{\mathrm{c}}} , \\
\chi \equiv 7.2 U_{\mathrm{MW}}\left(\frac{n_{\mathrm{CNM}}}{10 \mathrm{~cm}^{-3}}\right)^{-1},
\end{aligned}
\end{equation}}

{where $U_{\mathrm{MW}}$ is the UV radiation field relative to the solar neighborhood's average interstellar radiation, with a flux of $3.43 \times 10^{-8}$ photons s$^{-1}$ cm$^{-2}$ Hz$^{-1}$ at $1000$\,\AA~\citep{1978ApJS...36..595D}, and $n_{\mathrm{CNM}}$ is the cold neutral medium's number density. $\tau_{\rm{c}}$ represents the cloud's optical depth. We adapt this model to determine the shell's depth where the gas becomes molecular, using the stellar SED around $1000$\,\AA~to compute the flux at a specific shell location and employing $n_{\rm{sh}}$ and $\tau_{\rm{d}}$ from Eqns.~\eqref{eqn:Shell_structure_density_neutral}, \eqref{eqn:Shell_structure_dust_optical_depth_neutral} as $n_{\mathrm{CNM}}$ and $\tau_{\rm{c}}$, respectively.}

{\section{Atomic Hydrogen column density trends}
\label{appendix:Atomic_column_trends}
\begin{figure}
    \centering
    \includegraphics[width=\columnwidth]{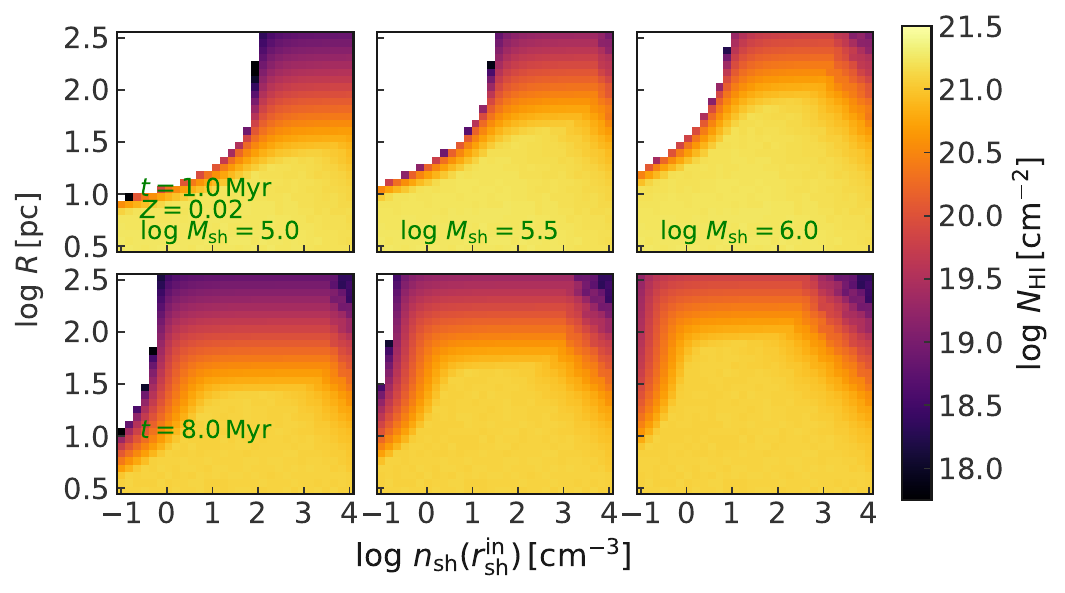}
    \caption{Atomic Hydrogen column density as a function of the density at the inner face of the shell and its radius calculated using the equations in Sec.~\ref{sect:shell_structure} for $Z=0.02$. The top row shows the results when the shells are irradiated by a stellar cluster of age 1 Myr, while the bottom row uses a cluster of age 8 Myr. The columns vary the mass of the shell as listed in the columns of the top row. The mass of the stellar cluster irradiating the shells is $5\,\times\,10^{4}\,\rm{M_{\odot}}$ in all cases. Regions that appear empty correspond to scenarios with low inner-edge density and large radii, which are conditions under which the shells are optically thin to ionizing radiation. The impact of the presence of molecular Hydrogen can be seen in the reduction in the atomic columns for large radii (lower UV flux) and high density (lower dissociation of H$_{2}$) cases, see Appendix~\ref{appendix:krumholz2013}.}
    \label{fig:NHI_Z02}
\end{figure}
\begin{figure}
    \centering
    \includegraphics[width=\columnwidth]{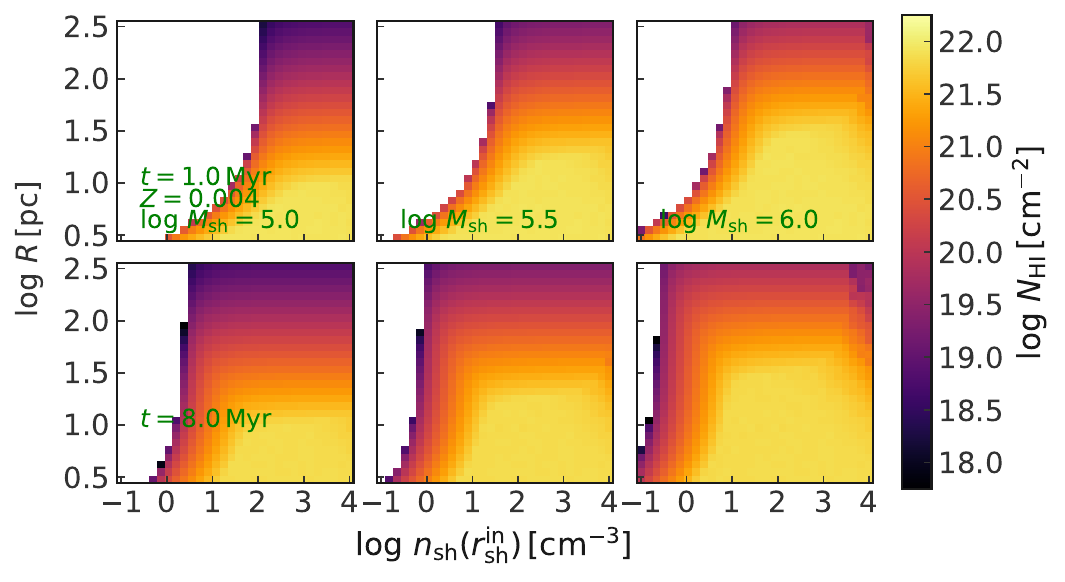}
    \caption{Same as Fig.~\ref{fig:NHI_Z02}, but for $Z=0.004$.}
    \label{fig:NHI_Z004}
\end{figure}
To understand the influence of various parameters on atomic Hydrogen column densities, we apply the equations detailed in Sec.~\ref{sect:shell_structure} while treating the shell's inner edge density, radius, and mass as independent parameters. This method provides a generalized version of our models. Additionally, to examine the influence of cluster age, we analyze a young system (1 Myr) and a system where the most massive stars have died (8 Myr), keeping the stellar mass irradiating the system constant at $5\,\%$ of the largest shell considered, i.e., $M_{\star}=5\,\times\,10^{4}\,\rm{M_{\odot}}$.}

{Fig.~\ref{fig:NHI_Z02} shows the atomic Hydrogen column density as a function of the density at the inner face of the shell and its radius calculated for $Z=0.02$, while Fig.~\ref{fig:NHI_Z004} shows the same for $Z=0.004$.}

{It is clear that higher column densities of neutral Hydrogen are associated with clouds which promote lower shell radii, higher masses, and higher inner edge densities.
In our models, while these parameters are interdependent, specific inferences can still be made for certain types of clouds. For instance, during the early phases of expansion, clouds with higher density tend to have shells with smaller radii, higher swept-up mass, and higher inner edge density compared to their low-density counterparts. This leads to higher atomic Hydrogen column densities (if any, depending on the swept-up mass and the ionizing radiation strength) at earlier stages, which in turn results in a stronger coupling with Ly$\alpha$ radiation. Such a dynamic can significantly impact the evolution of a system, particularly in environments with lower metallicities. 
Likewise, clouds (and, consequently, shells) with higher masses can result in high atomic Hydrogen column densities, even when the shell radii are moderately large and the inner edge densities are relatively low.}

{
\section{Absorption of ionizing radiation by dust}
\label{appendix:Draine_fion}
We estimate the fraction of ionizing photons that escape absorption by dust, denoted as $f_{\rm{ion}}$ (or $1 - f^{\rm{ion}}_{\rm{abs,\,dust}}$), using the fitting function from \cite{2011ApJ...732..100D}. This analysis is valid for hydrostatic shells, as is appropriate for our \texttt{Cloudy} models.
\begin{equation}
f_{\text {ion }}\left(\beta, \gamma, \tau_{d 0}\right) \approx \frac{1}{1+(2 / 3+A B) \tau_{d 0}}+\frac{A B \tau_{d 0}}{1+B \tau_{d 0}}, 
\end{equation}
with
\begin{align}
A &= \frac{1}{1+0.75 \gamma^{0.65} \beta^{-0.44}}, \\
B &= \frac{0.5}{1+0.1(\gamma / \beta)^{1.5}}, \\
\beta &\equiv \frac{L_{\mathrm{n}}}{L_{\mathrm{i}}}, \\
\gamma &= 11.2 \,T_4^{1.83}\left(\frac{18 \mathrm{eV}}{\langle h \nu\rangle_{\mathrm{i}}}\right)\left(\frac{\sigma_d}{10^{-21} \mathrm{~cm}^2}\right).
\end{align}
Here, $\tau_{d 0}$ denotes the dust optical depth within the ionized region, $\langle h \nu\rangle_{\mathrm{i}}$ represents the mean energy of the ionizing photons, and $T_{4}$ is the temperature in the \HII region in units of $10^{4}\,$K.}

{We utilize the output from \texttt{Cloudy} to determine $\tau_{d 0}$. In our calculations, the ionizing region is delineated by zones where the electron fraction exceeds $0.5$. We adopt an absorption cross section consistent with the \HII regions' dust model used in the \texttt{Cloudy} models as $\sigma_{d}$ here. Additionally, the mass-weighted temperature is directly derived from these models. Both $\langle h \nu\rangle_{\mathrm{i}}$ and $\beta$ are extracted directly from the stellar templates.}

{
\section{Line ratios of highly attenuated objects}
\label{appendix:line_ratios_of_highest_Av_objects}
\begin{figure}
    \centering \hspace{-0.75cm}\includegraphics[width=.8\columnwidth]{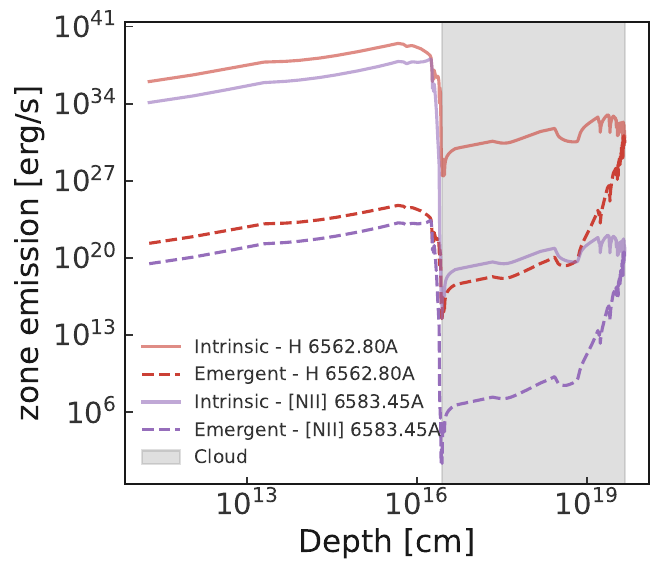}
    \caption{Zone by zone emission from a model where the shell is deeply embedded in the cloud.}
    \label{fig:emission_structure_highest_Av}
\end{figure}
In Fig.~\ref{fig:Hii_region_BPTratios_Nii_Oiii_secondary}, we observe that highly attenuated objects inhabit a unique area on the BPT diagram. This behavior is further understood by analyzing the outputs from \texttt{Cloudy}. Specifically, \texttt{Cloudy} provides both intrinsic and emergent line-luminosities. The intrinsic line luminosities factor in dust grains across all processes that influence line production without taking into account foreground attenuation. In contrast, emergent luminosities incorporate this foreground attenuation. Notably, when we use intrinsic line luminosities, this set of models reverts to its customary position on the BPT diagram.}

{
Delving deeper, Fig.~\ref{fig:emission_structure_highest_Av} offers a zone-by-zone emission analysis for an early age model, at a stage where the shell remains deeply embedded within its birth cloud. Both H$\alpha$ and \NII emission originating from the \HII region are considerably attenuated—a fact underscored when one contrasts the intrinsic with the emergent zone luminosities. Concurrently, H$\alpha$ possesses a secondary non-recombination contribution from outside the \HII region. In an environment devoid of dust, where the \HII region is clearly visible, this contribution would go unnoticed. However, with the \HII region being significantly attenuated, this secondary contribution becomes the primary driver of the H$\alpha$ line luminosity emerging from the cloud. For the \NII emission, however, the emergent contribution primarily stems from the \HII region itself.}

{This phenomenon results in such models being positioned in the distinctive area of the BPT diagram, as illustrated in both Fig.~\ref{fig:Hii_region_BPTratios_Nii_Oiii_primary} and \ref{fig:Hii_region_BPTratios_Nii_Oiii_secondary}. A similar explanation could also be given for the line ratios encountered in the case of infalling shells.
However, due to the pronounced attenuation, such models are unlikely to be detected.}

\section{Examples of the evolution of key shell properties}
\label{appendix:examples_shell_prop_evolution}
\begin{figure*}
        \begin{subfigure}[b]{0.29\textwidth}
        \centering
        \includegraphics[width=\textwidth]{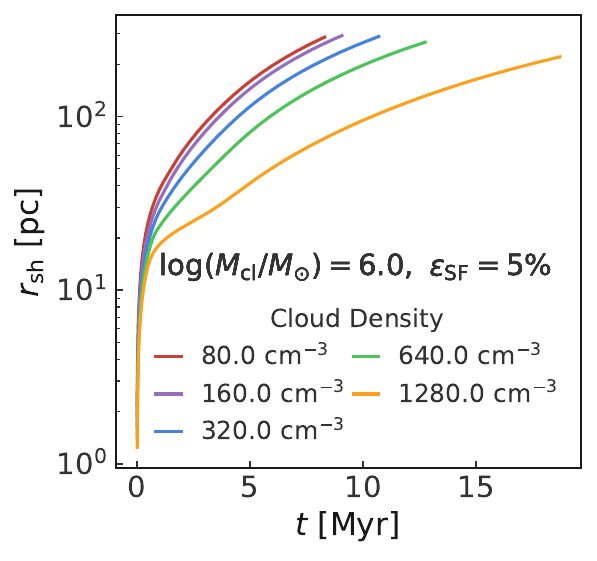}
        \phantomcaption{}  
      \label{fig:var_Ncl_radius_evolution_Z02}
    \end{subfigure}
    \begin{subfigure}[b]{0.29\textwidth}  
        \centering 
        \includegraphics[width=\textwidth]{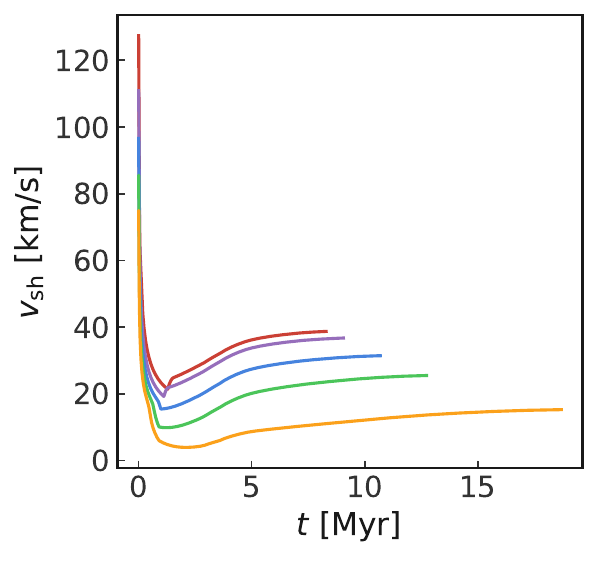}
        \phantomcaption{}     
        \label{fig:var_Ncl_velo_evolution_Z02}
    \end{subfigure}
    \begin{subfigure}[b]{0.2985\textwidth}   
        \centering 
        \includegraphics[width=\textwidth]{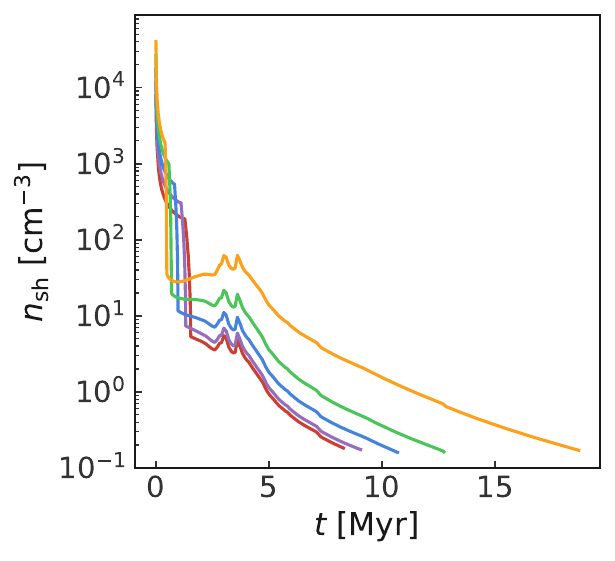}
        \phantomcaption{}      
        \label{fig:var_Ncl_nInner_evolution_Z02}
    \end{subfigure}

    \begin{subfigure}[b]{0.29\textwidth}
        \centering
        \includegraphics[width=\textwidth]{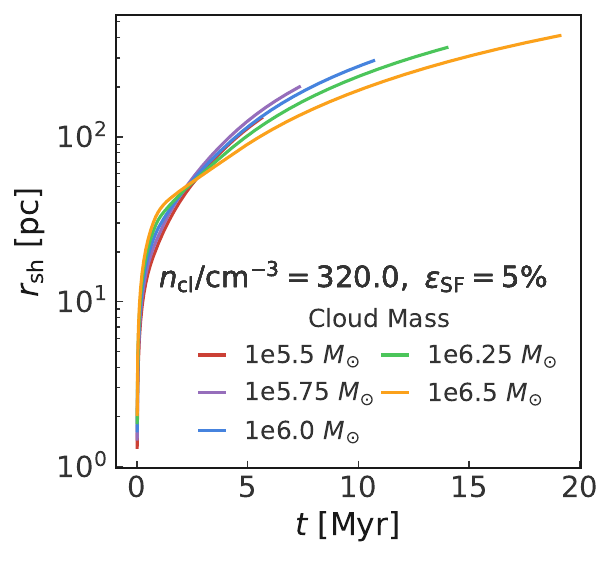}
        \phantomcaption{}      
        \label{fig:var_logM_radius_evolution_Z02}
    \end{subfigure}
    \begin{subfigure}[b]{0.29\textwidth}  
        \centering 
        \includegraphics[width=\textwidth]{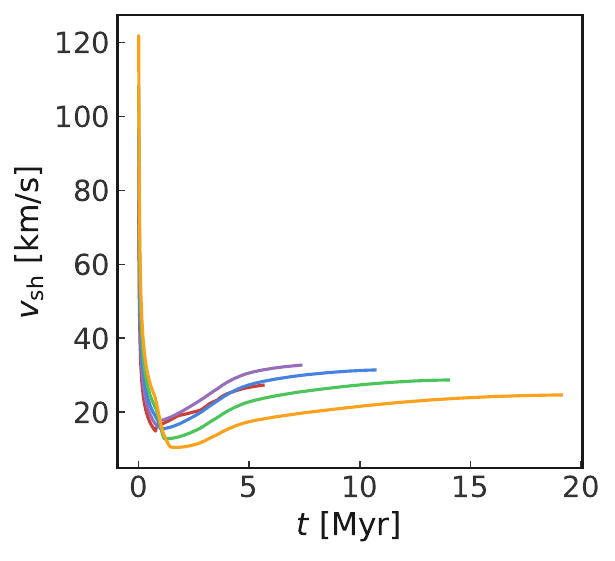}
        \phantomcaption{}    
        \label{fig:var_logM_nShell_evolution_Z02}
    \end{subfigure}
    \begin{subfigure}[b]{0.2985\textwidth}   
        \centering 
        \includegraphics[width=\textwidth]{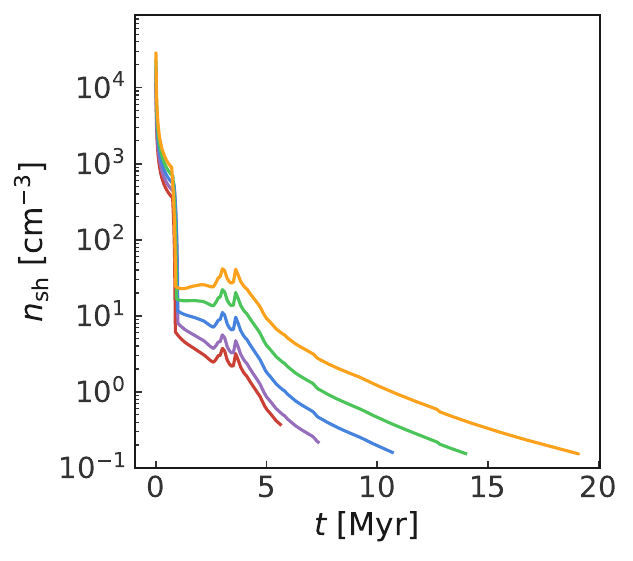}
        \phantomcaption{}     
     \label{fig:var_logM_etaCoupling_evolution_Z02}
    \end{subfigure}

        \begin{subfigure}[b]{0.29\textwidth}
        \centering
        \includegraphics[width=\textwidth]{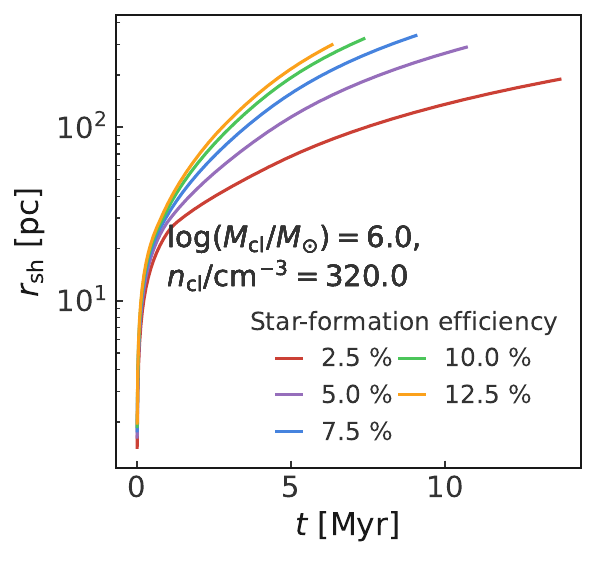}
        \phantomcaption{}     
        \label{fig:var_SFE_radius_evolution_Z02}
    \end{subfigure}
    \begin{subfigure}[b]{0.29\textwidth}  
        \centering 
        \includegraphics[width=\textwidth]{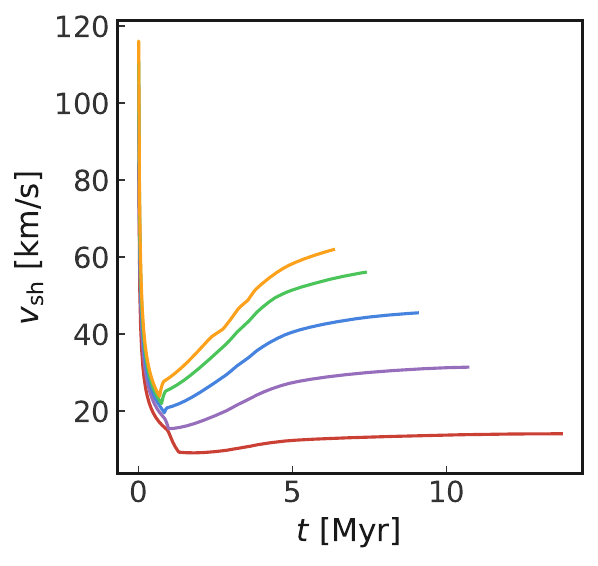}
        \phantomcaption{}     
        \label{fig:var_SFE_velo_evolution_Z02}
    \end{subfigure}
    \begin{subfigure}[b]{0.2985\textwidth}   
        \centering 
        \includegraphics[width=\textwidth]{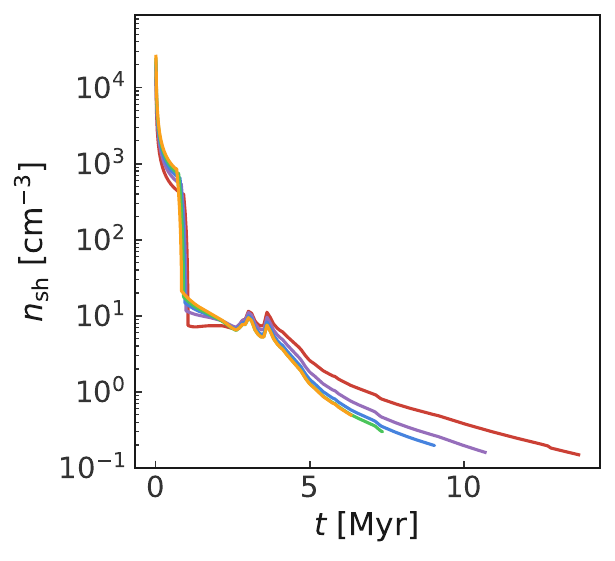}
        \phantomcaption{}     
        \label{fig:var_SFE_nShell_evolution_Z02}
    \end{subfigure}
    \caption[]{Evolution of key shell properties as a function of the parameters of the model for $Z=0.02$: 
    The top row shows the effect of changing the cloud density of models with $\epsilon_{\mathrm{SF}}= 5.0\%$, and $\log(M_{\mathrm{cl}}/M_{\odot})=6.0$. The second row shows the effect of changing the cloud mass of a system with $\epsilon_{\mathrm{SF}}= 5.0\%$, and  $n_{\mathrm{cl}} = 320.0~\mathrm{cm^{-3}}$. The last row depicts the impact of changing the star-formation efficiency for a system with $n_{\mathrm{cl}} = 320.0~\mathrm{cm^{-3}}$, and $\log(M_{\mathrm{cl}}/M_{\odot})=6.0$.  The first, second, and third columns show the shell radius, velocity, and inner-face density for the shells, respectively. The fixed parameters of each row are listed in bold in the first column, while the legend title gives the parameter being varied.} 
\label{fig:evolution_vary_all_params_Z02}
\end{figure*}

\begin{figure*}
        \begin{subfigure}[b]{0.2905\textwidth}
        \centering
        \includegraphics[width=\textwidth]{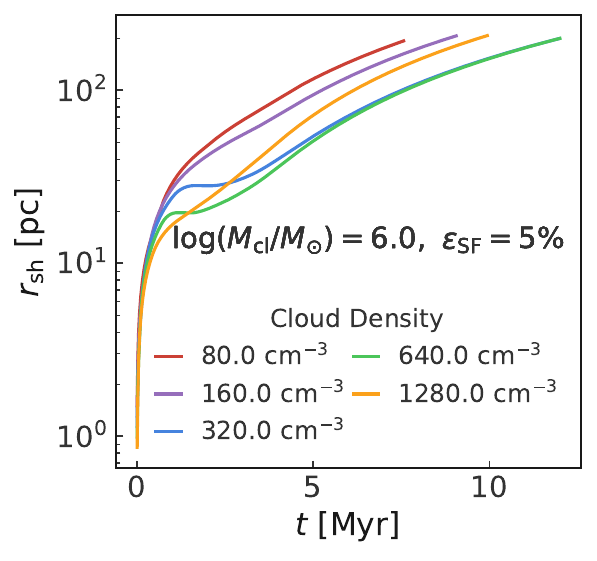}
        \phantomcaption{}  
      \label{fig:var_Ncl_radius_evolution_Z004}
    \end{subfigure}
    \begin{subfigure}[b]{0.2825\textwidth}  
        \centering 
        \includegraphics[width=\textwidth]{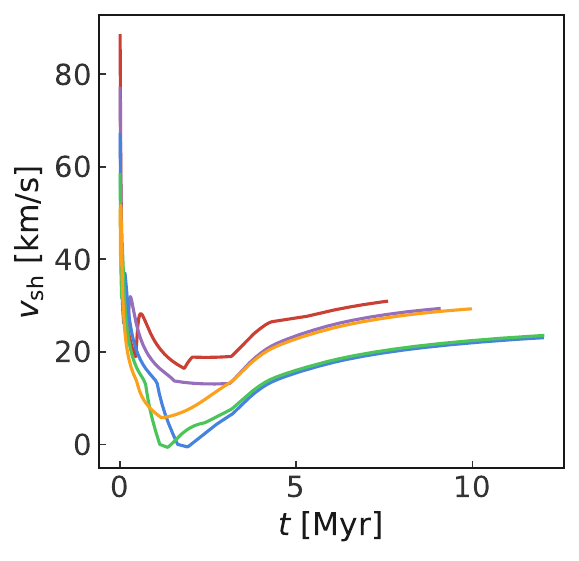}
        \phantomcaption{}     
        \label{fig:var_Ncl_velo_evolution_Z004}
    \end{subfigure}
    \begin{subfigure}[b]{0.2985\textwidth}   
        \centering 
        \includegraphics[width=\textwidth]{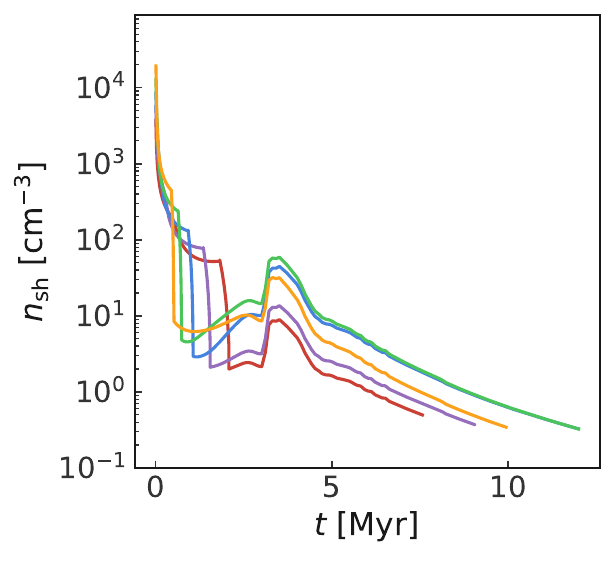}
        \phantomcaption{}      
        \label{fig:var_Ncl_nInner_evolution_Z004}
    \end{subfigure}
    \begin{subfigure}[b]{0.2905\textwidth}
        \centering
        \includegraphics[width=\textwidth]{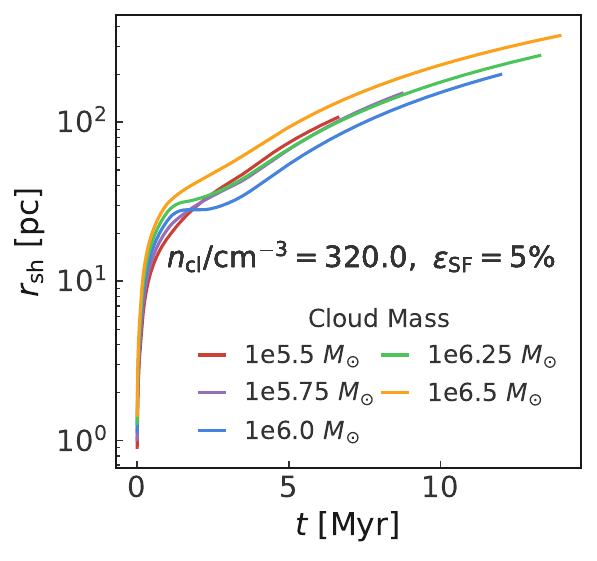}
        \phantomcaption{}      
        \label{fig:var_logM_radius_evolution_Z004}
    \end{subfigure}
    \begin{subfigure}[b]{0.2825\textwidth}  
        \centering 
        \includegraphics[width=\textwidth]{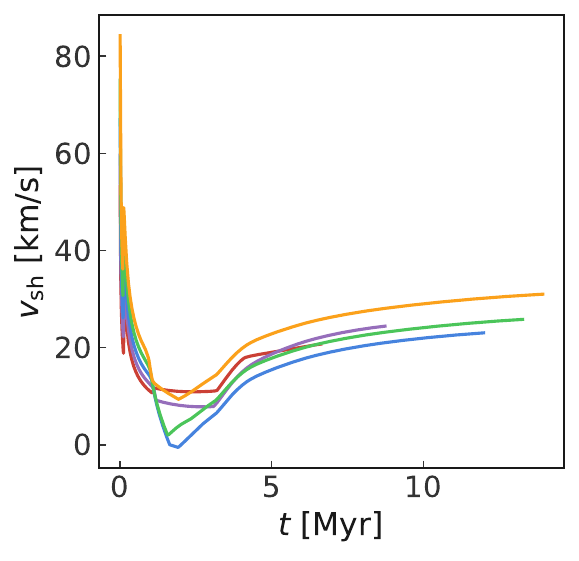}
        \phantomcaption{}    
        \label{fig:var_logM_nShell_evolution_Z004}
    \end{subfigure}
    \begin{subfigure}[b]{0.2985\textwidth}   
        \centering 
        \includegraphics[width=\textwidth]{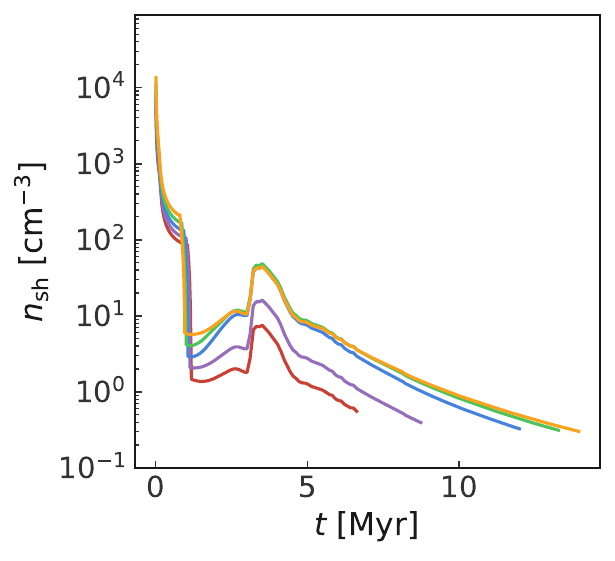}
        \phantomcaption{}     
     \label{fig:var_logM_etaCoupling_evolution_Z004}
    \end{subfigure}
        \begin{subfigure}[b]{0.2905\textwidth}
        \centering
        \includegraphics[width=\textwidth]{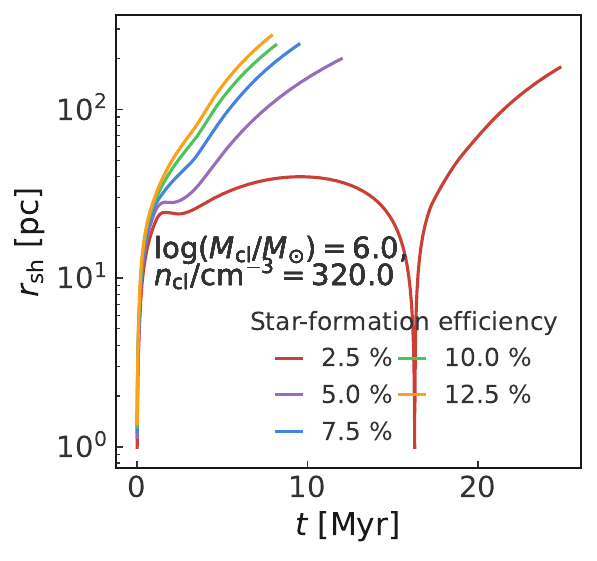}
        \phantomcaption{}     
        \label{fig:var_SFE_radius_evolution_Z004}
    \end{subfigure}
    \begin{subfigure}[b]{0.2825\textwidth}  
        \centering 
        \includegraphics[width=\textwidth]{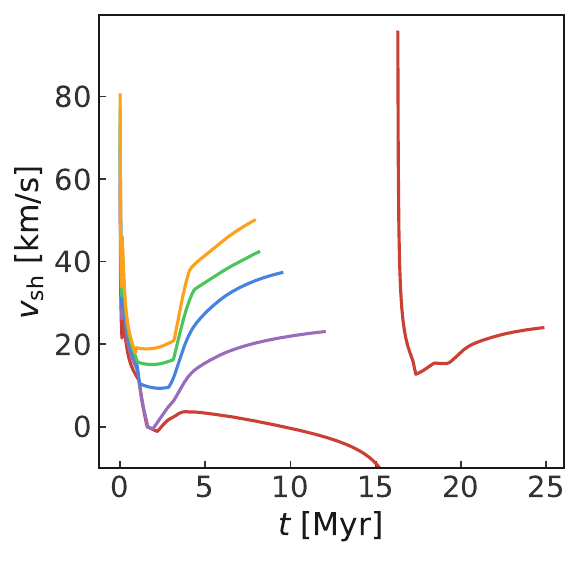}
        \phantomcaption{}     
        \label{fig:var_SFE_velo_evolution_Z004}
    \end{subfigure}
    \begin{subfigure}[b]{0.2985\textwidth}   
        \centering 
        \includegraphics[width=\textwidth]{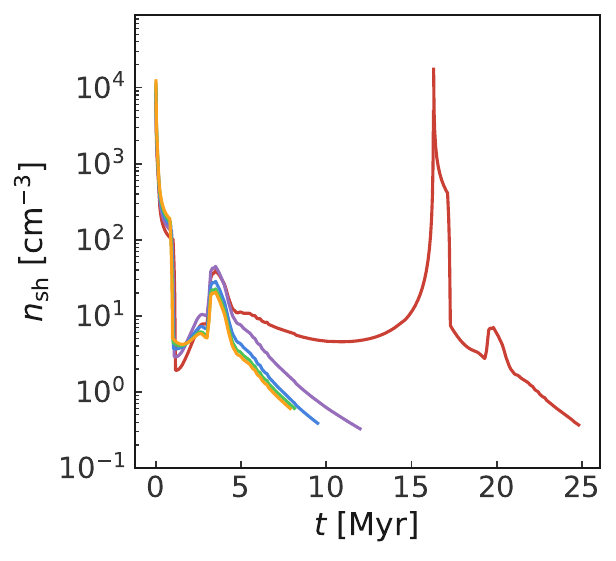}
        \phantomcaption{}     
        \label{fig:var_SFE_nShell_evolution_Z004}
    \end{subfigure}
    \caption[]{Same as Fig.~\ref{fig:evolution_vary_all_params_Z02}, but for $Z=0.004$.} 
\label{fig:evolution_vary_all_params_Z004}
\end{figure*}

\bsp	
\label{lastpage}
\end{document}